\documentclass{emulateapj}
\usepackage{amssymb,amsmath,graphicx,longtable,subfigure,wrapfig}
\usepackage{rotating}
\usepackage[colorlinks,linkcolor=blue,anchorcolor=green,citecolor=blue]{hyperref}
\usepackage{natbib}

\shorttitle{GMRT imaging of X-shaped radio galaxies}
\shortauthors{Lal et~al.}

\begin{document}

\def\func#1{\mathop{\rm #1}\nolimits}
\def\unit#1{\mathord{\thinspace\rm #1}}


\title{GMRT Low-frequency Imaging of an Extended Sample of X-shaped Radio Galaxies}


\author{Dharam V. Lal\altaffilmark{1}, Biny Sebastian\altaffilmark{1}, C. C. Cheung\altaffilmark{2}, A. Pramesh Rao\altaffilmark{1}}

\altaffiltext{1}{National Centre for Radio Astrophysics - Tata Institute of Fundamental Research, Post Box 3, Ganeshkhind P.O., Pune 41007, India}
\altaffiltext{2}{Space Science Division, Naval Research Laboratory, Washington, DC 20375-5352, USA}
\email{dharam@ncra.tifr.res.in}

\begin{abstract}
We present a low-frequency imaging study of an extended sample of X-shaped radio sources using the Giant Metrewave radio telescope (GMRT) at two frequencies (610 and 240 MHz). 
The sources were drawn from a Very Large Array FIRST-selected sample and extends an initial GMRT study at the same frequencies, of 12 X-shaped radio galaxies predominantly from the 3CR catalog \citep{2007MNRAS.374.1085L}. 
Both the intensity maps and spectral index maps of the 16 newly observed sources are presented.
With the combined sample of 28 X-shaped radio sources, we found no systematic differences in the spectral properties of the higher surface brightness, active lobes versus the lower surface brightness, off-axis emission.
The properties of the combined sample are discussed, including the possible role of a twin active galactic nuclei model in the formation of such objects.
\end{abstract}


\section{Introduction}
\label{intro}

Morphologically, a small ($\sim$10\%) fraction of the general radio galaxy population is comprised of a special class of objects known as  X-shaped radio galaxies \citep{1992ersf.meet..307L,1984MNRAS.210..929L}. 
These objects appear to possess two pairs of radio lobes extending along two distinct axes, unlike the majority of radio galaxies where most of the radio emission is well aligned along a single, primary axis. 
Thus, a defining characteristic is the presence of synchrotron emitting plasma off the primary axis of the source \citep{2018ApJ...852...47R}.
The more diffuse and lower surface brightness of the two pair of lobes are often termed `wings' or secondary lobes, whereas the higher surface brightness ones are termed active or primary lobes.

A consensus has not been reached regarding the formation mechanism of X-shaped radio sources. 
There are several models that were proposed for the formation of these objects in the literature.
(a) The jet material backflow from the terminal hotspots, therefore they may stream back towards the host galaxy to form the wings of X-shaped sources \citep{1984MNRAS.210..929L}.
(b) The lobes of a radio galaxy have a lower density than the surrounding medium \citep{2005ApJ...622..149K}, therefore buoyancy may have an impact on the large-scale morphology of the radio lobes and can form such X-shaped structures \citep{2011ApJ...733...58H,2009ApJ...695..156S,1995ApJ...449...93W}.
\cite{2011ApJ...733...58H,Capettietal2002} invoked a combination of (a, b), first the over-pressured cocoon model for initial expansion and subsequently the buoyancy to explain the observed morphology of X-shaped sources.
\cite{Capettietal2002} also suggests that jets propagate in an asymmetric enviroment, where the cocoon grows faster in the
direction of maximum pressure gradient, generating the wings -- this accounts
for the observed alignement between the axis of the wings and the minor axis
of the host \citep[see also,][]{Gilloneetal2016,Rossietal2017}.
(c) \cite{Parmaetal1985} suggested precession of the central engine as the formation mechanism of X-shaped radio sources.
(d) A reorientation of the jet axis due to a minor merger may also form X-shaped structures \citep{2002MNRAS.330..609D,2002Sci...297.1310M}.
(e) \cite{2005MNRAS.356..232L,2007MNRAS.374.1085L} used the low-frequency spectra from 610/240 MHz data at different locations in the radio structures to understand the nature of these sources and suggested that they could consist of two pairs of jets, which are associated with unresolved binary active galactic nuclei (AGN) system.

Broadly, the above formations models can be categorized as two scenarios, ``environmental'' (a, b), or due to the central engine (c, d, e). 
Recently, \cite{2018ApJ...852...48S} and \cite{2015ApJS..220....7R} suggested that the X-shaped morphology arises due to a combination of both scenarios, suggesting an interplay of the reorientation of the jet
axis and the backflow models. 
\cite{2007MNRAS.374.1085L} studied all the known X-shaped radio sources, predominantly from the 3CR sample and had also pointed out that various X-shaped radio galaxies need not form via a unique formation mechanism, but instead could form via a variety of scenarios which in-turn could explain the differences in the spectral properties of different X-shaped radio galaxies.
The earlier studies were based on a small number of objects and \cite{2007MNRAS.374.1085L} noted the necessity to improve the statistics.

\begin{table*}[tbph]
        \centering
        \caption{Summary of sources and observations.}
        \label{obs-log}
 \begin{tabular}{lccccccccccc}
\hline \hline
Object name     &  Redshift & R.~A. (J2000)   &  Decl. (J2000)  & Obs. date   & Flux Cal. & Phase Cal.  & t$_{int}$ (hrs)\\
\multicolumn{1}{c}{(1)} & (2) & (3) & (4) & (5) & (6) & (7) & (8) \\
\hline
J0113$+$0106 & 0.281 & 01:13:41.08 & $+$01:06:09.0 & 2006 Dec 25 & 3C48  & 3C468.1,    & 2.91/2.97 \\
               &       &           &                &             &       & 0025$-$260, &         &        \\
               &       &           &                &             &       & 0116$-$208  &         &        \\
J0115$-$0000 & 0.381 & 01:15:27.33 & $-$00:00:01.1 & 2006 Dec 25 & 3C48  & 3C468.1,     & 3.05/3.05 \\
               &       &           &                &             &       & 0025$-$260, &         &         \\
               &       &           &                &             &       & 0116$-$208  &         &         \\
J0702$+$5002 & 0.094 & 07:02:47.92 & $+$50:02:05.3 & 2007 Nov 22 & 3C48  & 0542$+$498  & 3.82/3.91 \\
J0859$-$0433 & 0.356 & 08:59:50.19 & $-$04:33:06.90 & 2006 Dec 26 & 3C48, & 0834$+$555  & 2.04/0.58 \\
               &       &           &                &             & 3C286 &             &         &         \\
J0914$+$1715 & 0.520 & 09:14:05.19 & $+$17:15:54.4 & 2006 Dec 26 & 3C48, & 1021$+$219  & 2.08/0.67 \\
               &       &           &                &             & 3C286 &             &         &         \\
J0917$+$0523 & 0.591 & 09:17:44.33 & $+$05:23:09.4 & 2007 Nov 22 & 3C48  & 1021$+$219  & 3.34/3.33 \\
J0924$+$4233 & 0.2274 & 09:24:47.01 & $+$42:33:47.4 & 2007 Nov 23 & 3C147 & 0834$+$555, & 2.88/2.68 \\
               &       &           &                &             &       & 1123$+$055  &         &         \\
J1055$-$0707 & $-$   & 10:55:52.56 & $-$07:07:19.1 & 2007 Nov 24 & 3C286 & 1123$+$055  &  2.62/2.62 \\
J1130$+$0058 & 0.1325 & 11:30:21.42 & $+$00:58:22.9 & 2006 Dec 26 & 3C48, & 1123$+$055  & 2.56/2.98 \\
               &       &           &                &             & 3C286 &             &         &         \\
J1218$+$1955 & 0.424 & 12:18:59.16 & $+$19:55:28.0 & 2006 Dec 26 & 3C48, & 1123$+$055  & 2.91/3.07 \\
               &       &           &                &             & 3C286 &             &         &         \\
J1309$-$0012 & 0.419 & 13:09:49.65 & $-$00:12:35.2 & 2006 Dec 24 & 3C147,& 3C468.1,       & 3.52/3.68 \\
               &       &           &                &             & 3C286 & 1419$+$064  &         &         \\
J1339$-$0016 & 0.1452 & 13:39:34.26 & $-$00:16:35.8 & 2007 Nov 24 & 3C147,& 1419$+$064  & 2.63/2.76 \\
               &       &           &                &             & 3C286 &             &         &         \\
J1406$-$0154 & 0.641 & 14:06:48.63 & $-$01:54:17.3 & 2006 Dec 24 & 3C147,& 3C468.1,      & 3.46/3.34 \\
               &       &           &                &             & 3C286 & 1419$+$064  &         &        \\
J1430$+$5217 & 0.3671 & 14:30:17.32 & $+$52:17:35.3 & 2007 Nov 24 & 3C147,& 1419$+$064  & 2.54/2.76 \\
               &       &           &                &             & 3C286 &             &         &         & \\
J1600$+$2058 & 0.1735 & 16:00:38.94 & $+$20:58:51.9 & 2006 Dec 25 & 3C48, & 1021$+$219, & 2.78/1.28 \\
               &       &           &                &             & 3C147 & 1419$+$064, &         &      \\
               &     &             &                &             &       & 1822$-$096  &         &         \\
J1606$+$0000 & 0.059 & 16:06:12.71 & $+$00:00:27.1 & 2006 Dec 25 & 3C48, & 1021$+$219, & 2.81/2.03 \\
               &       &           &                &             & 3C147 & 1419$+$064, &         &         \\
               &     &             &                &             &       & 1822$-$096  &         &         \\
\hline
 \end{tabular}
\tablecomments{
Col.~1: Source name as used in this paper.\\
Col.~2: Redshifts are from \citet[][and references therein]{2009ApJS..181..548C}, plus a subsequently determined value for J0914$+$1715 \citep{2017ApJS..233...25A}.\\
Col.~3: Right ascension (J2000). Col.~4: Declination (J2000).\\
Col.~5: Date of observation.\\
Col.~6: Flux density calibrator, also the bandpass calibrator. Col.~7: Phase calibrator.\\
Col.~8: On-source integration times at 240/610 MHz after flagging bad data.
}
 \end{table*}

In this paper, we extend the Giant Metrewave radio telescope (GMRT) spectral-morphological study of \cite{2007MNRAS.374.1085L}, who studied all the known winged$/$X-shaped radio sources at the time, predominantly those known from the 3CR catalog.
The new targets consist 16 more sources, taken from a sample of 100 candidate X-shaped radio sources using the Very Large Array (VLA) FIRST \cite{1995ApJ...450..559B} survey database \citep{2007AJ....133.2097C}.
We also include results from the original study of 12 sources \citep{2007MNRAS.374.1085L}, thereby increasing the sample of X-shaped sources to 28 sources, in order to understand the formation scenarios of these sources in a statistical manner.  
This large sample would be used to understand the relationship, if any, among three categories, (i) where wings have flatter spectra than primary lobes, (ii) where wings and primary lobes have comparable spectral indices and (iii) where wings have steeper spectra than the primary lobes, of X-shaped radio sources.

The paper is organized as follows.
The sample selection and GMRT observational details of the 16 sources are given in Sections ~\ref{sec.sample} and~\ref{gmrt-obs}, respectively.
The high-resolution GMRT 610 MHz total intensity maps and the spectral index maps constructed at the lower-resolution corresponding to the 240 MHz data are given in Section ~\ref{gmrt-result}.
The Appendix includes the 240 MHz images together with the lower-resolution versions of the 610 MHz data.
Our findings are briefly discussed and summarized in Section~\ref{xs-discuss}.
Throughout, positions are given in J2000 coordinates.

\section{Sample}
\label{sec.sample}

\citet{2007AJ....133.2097C} compiled a sample of 100 candidate winged and X-shaped radio sources drawn from the VLA FIRST survey \citep{1995ApJ...450..559B} database. 
For our GMRT study, we selected sources amongst these candidates which have (i) characteristic `X' shape based on the $\sim$5$^{\prime\prime}$ resolution FIRST 1.4 GHz images,
(ii) both set of lobes passing symmetrically through the center of the associated host galaxy, and 
(iii) an angular size greater than 1.2$'$ so as to allow more than 7$-$8 synthesized beams across the whole source at the observing frequency (240 MHz) utilized.
We additionally observed J1130$+$0058, an X-shaped radio galaxy similarly discovered through its FIRST image \citep{wan03}, and satisfied these criteria. 
The above criteria provided us with a sample comprised of 16 targets listed in Table~\ref{obs-log}.

\begin{figure*}
\begin{center}
\begin{tabular}{rrrr}
\includegraphics[height=4cm]{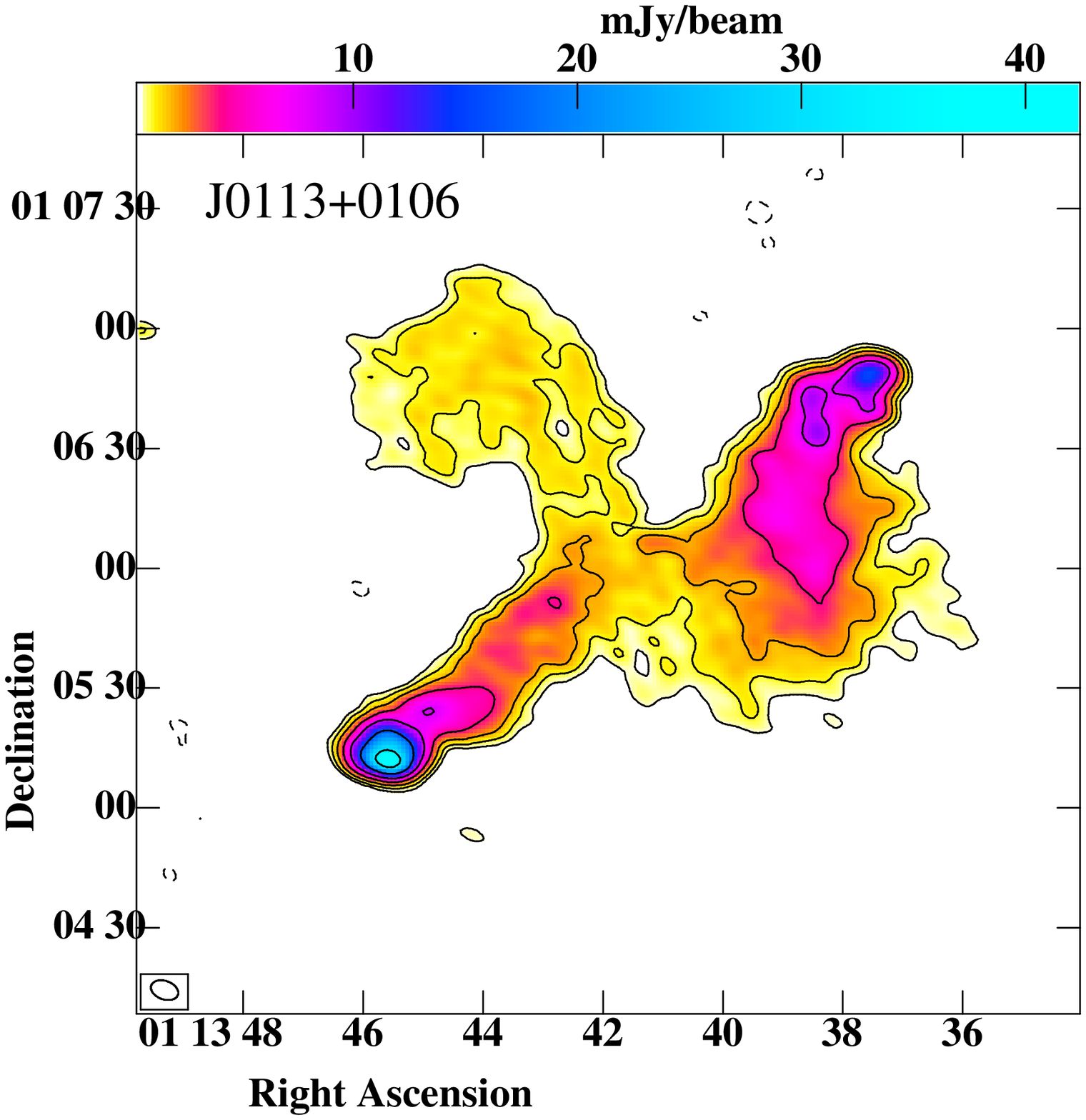} &
\includegraphics[height=4cm]{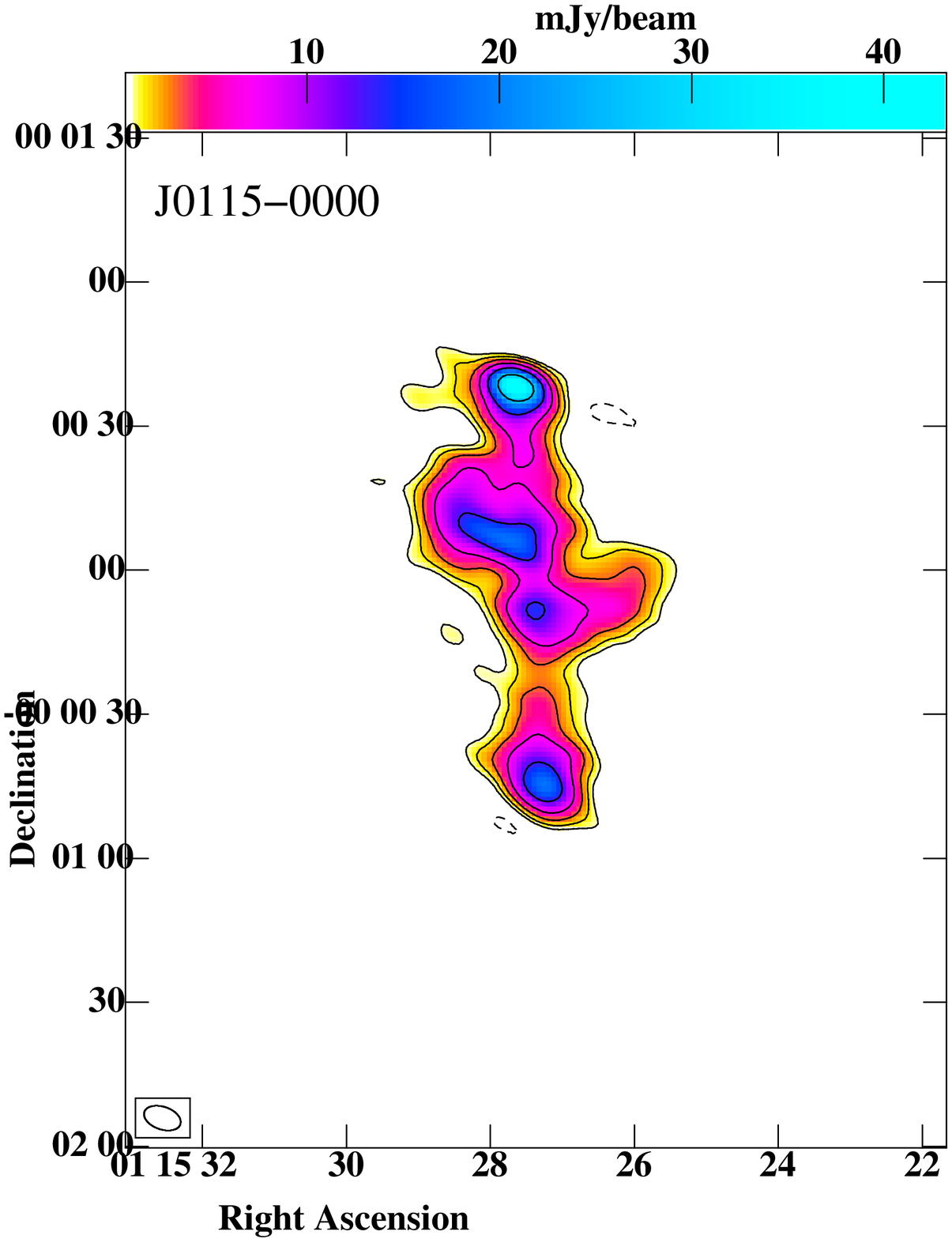} &
\includegraphics[height=4cm]{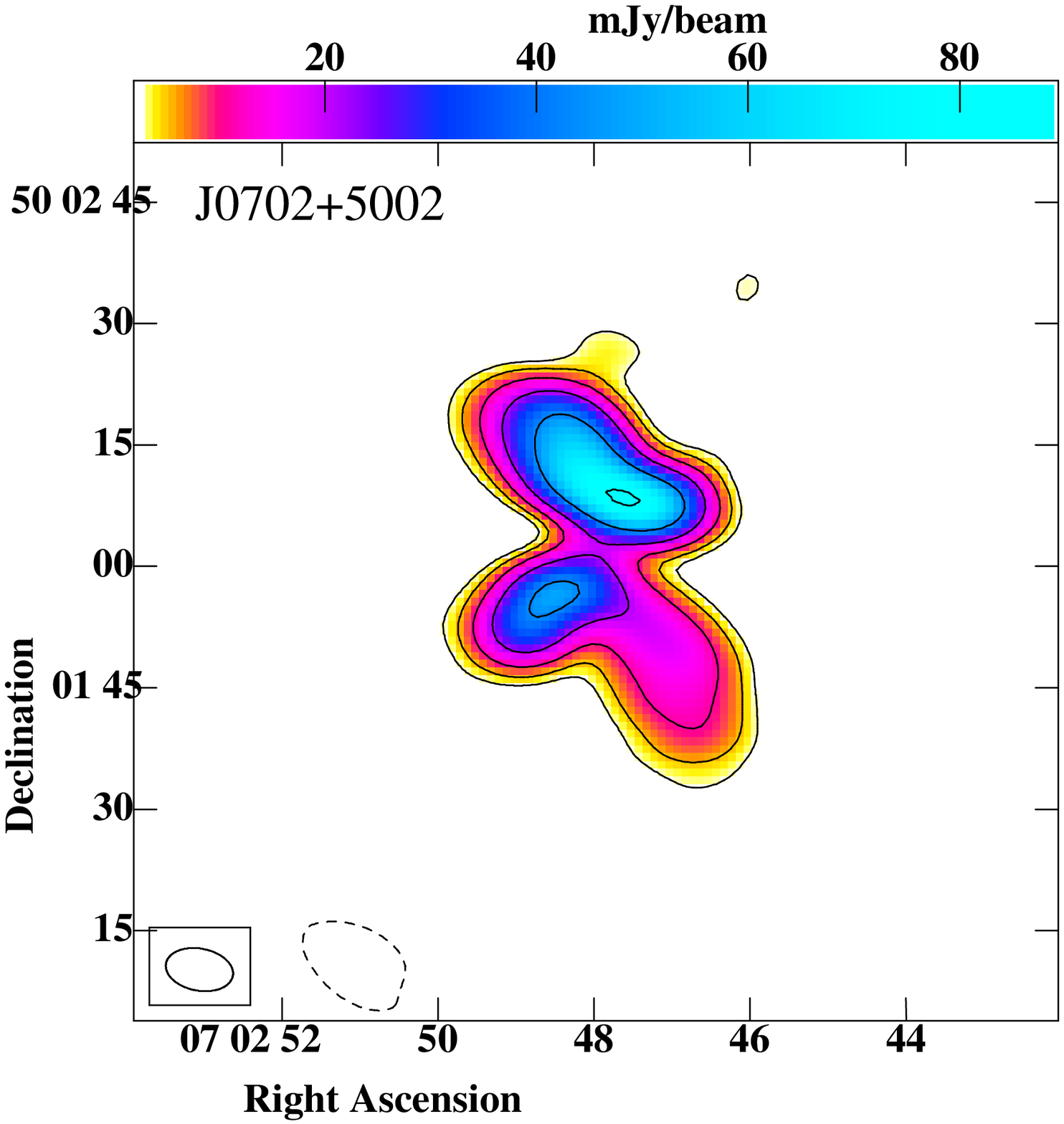} &
\includegraphics[height=4cm]{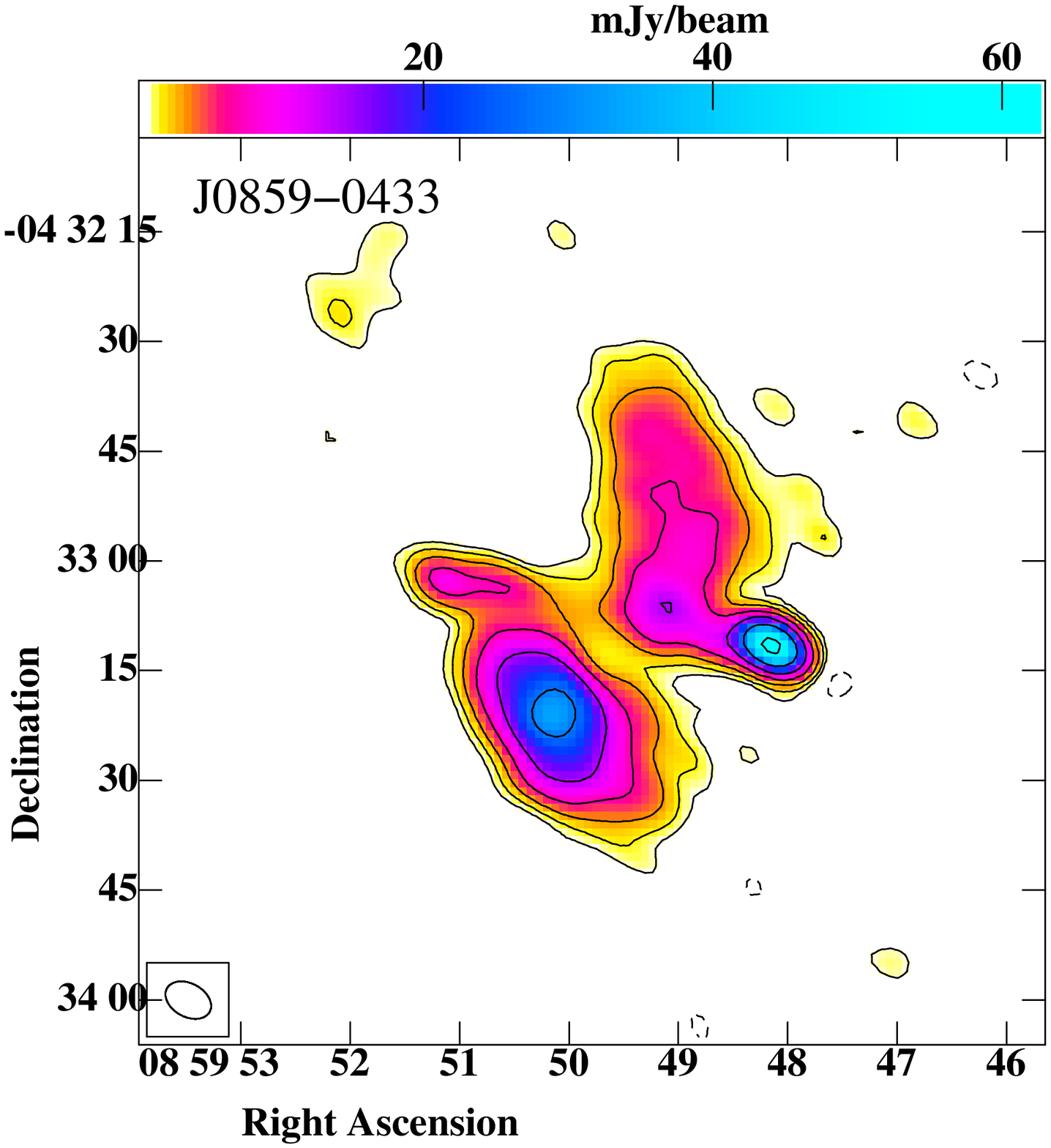} \\
\includegraphics[height=4cm]{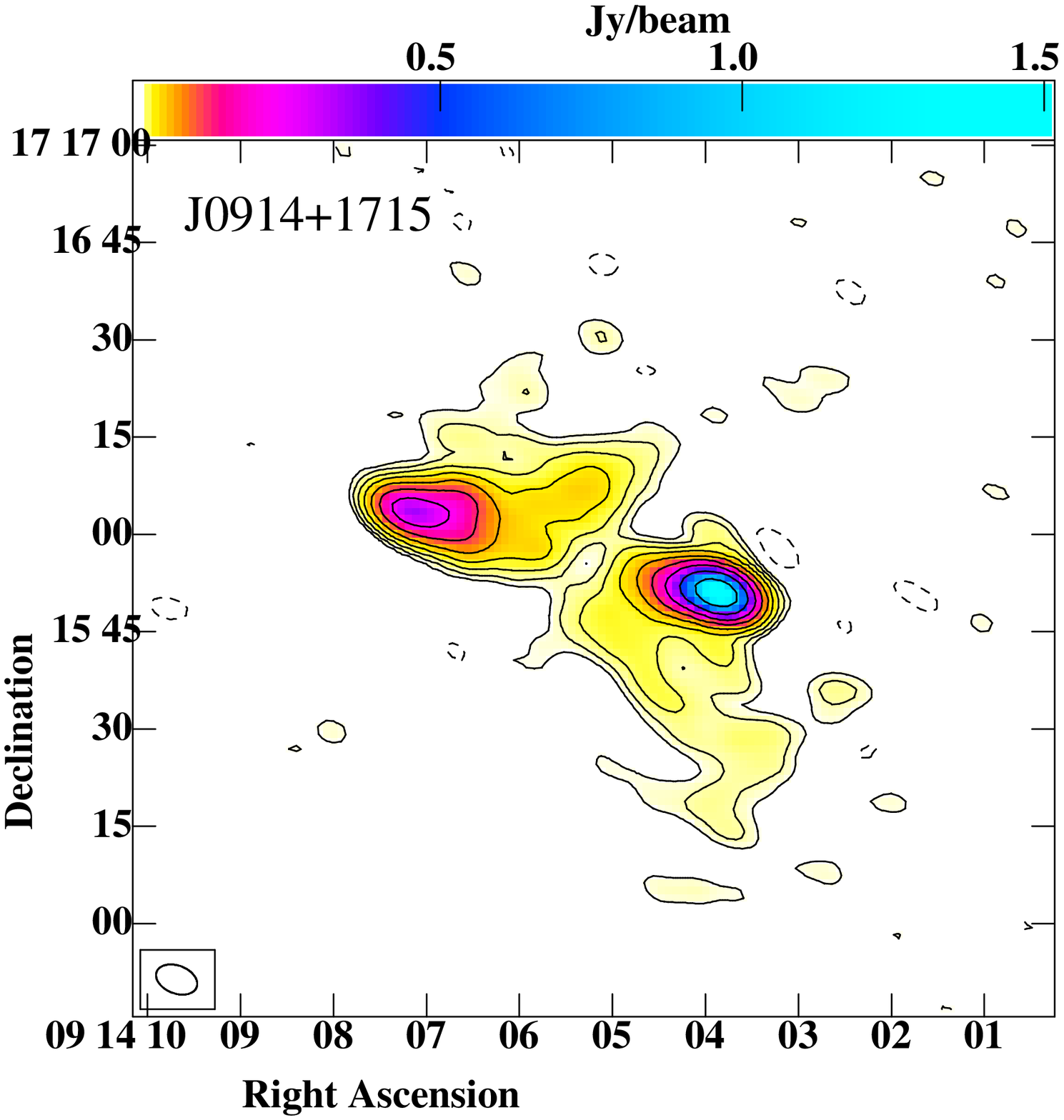} &
\includegraphics[height=4cm]{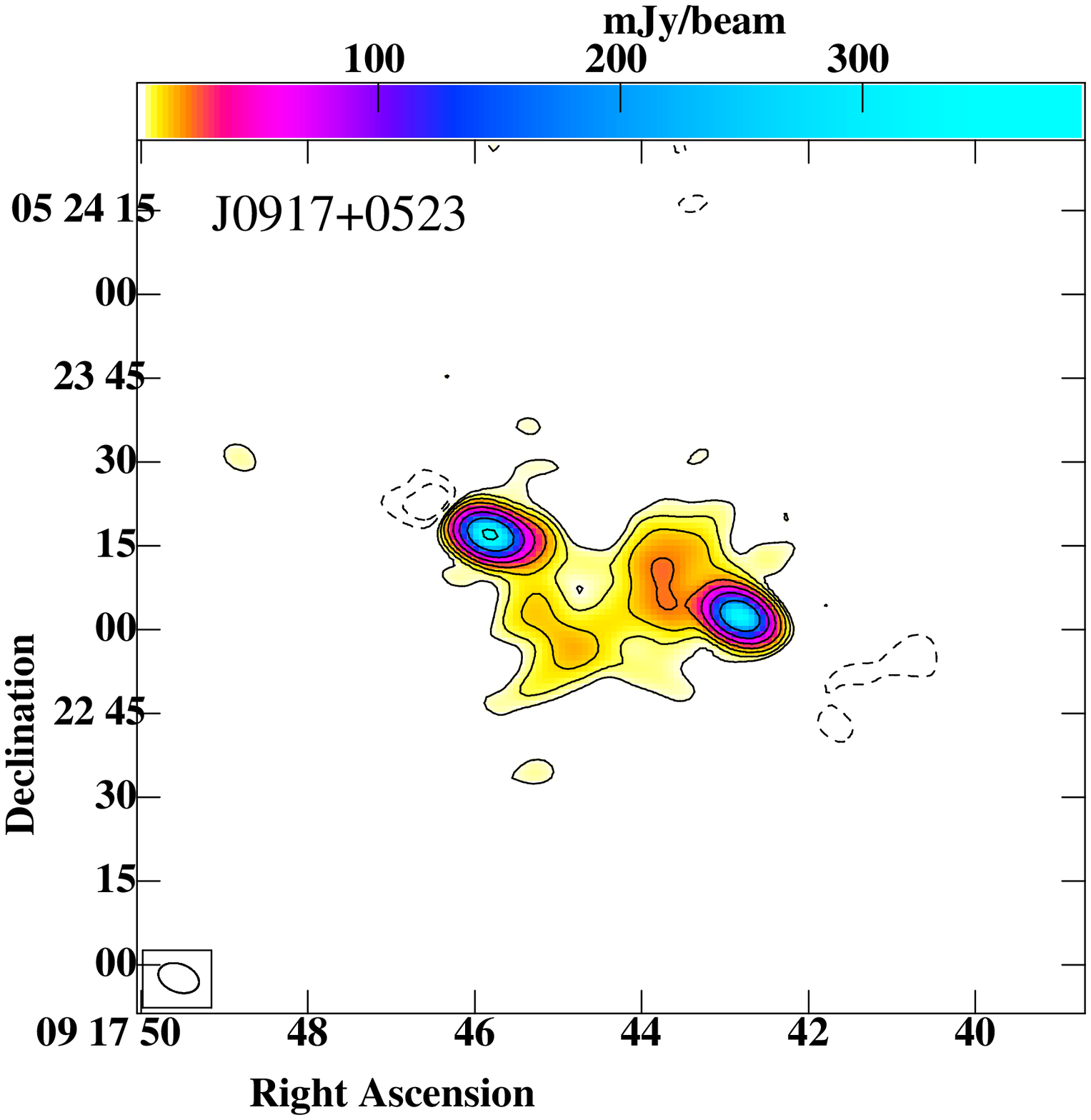} &
\includegraphics[height=4cm]{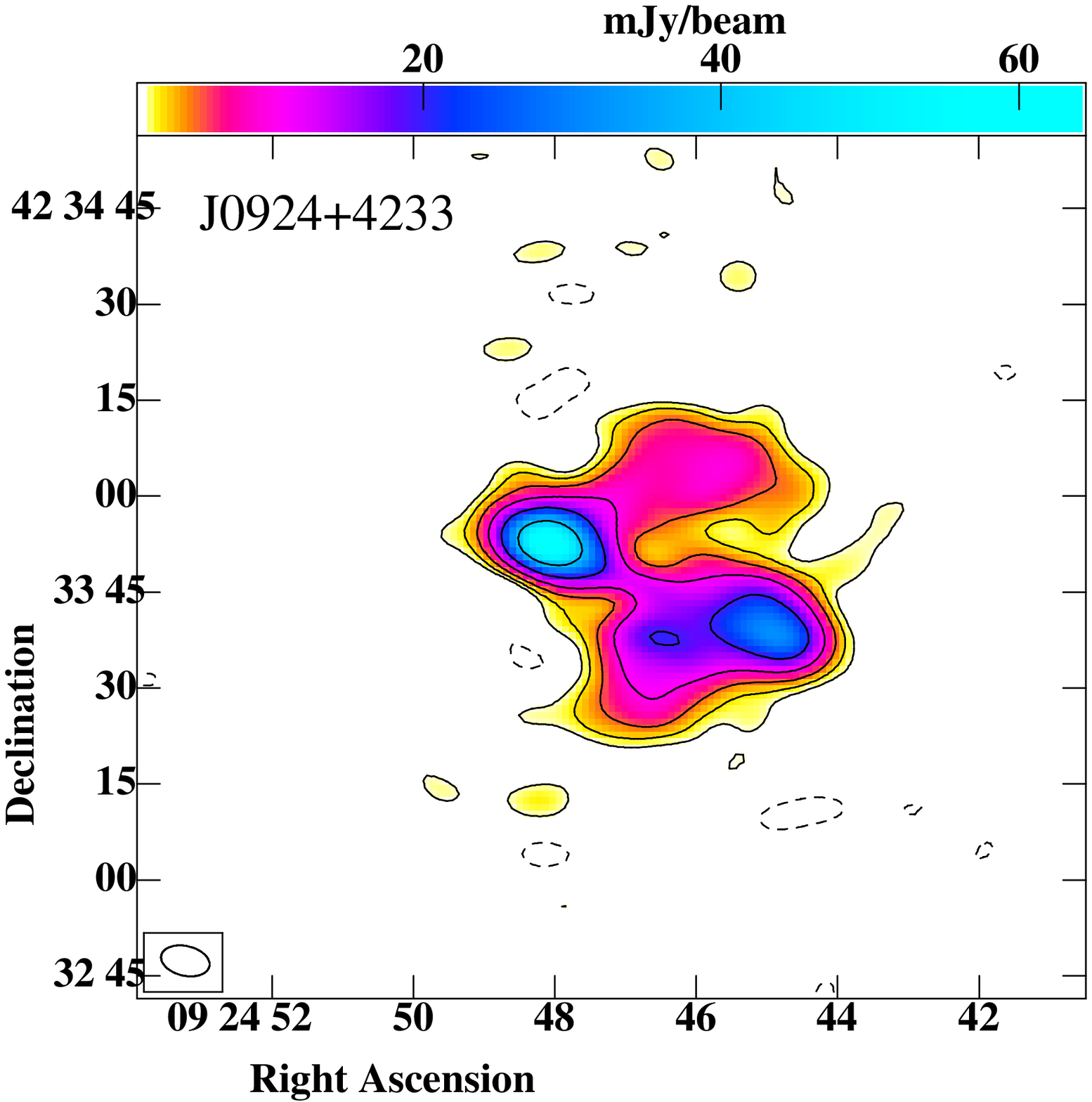} &
\includegraphics[height=4cm]{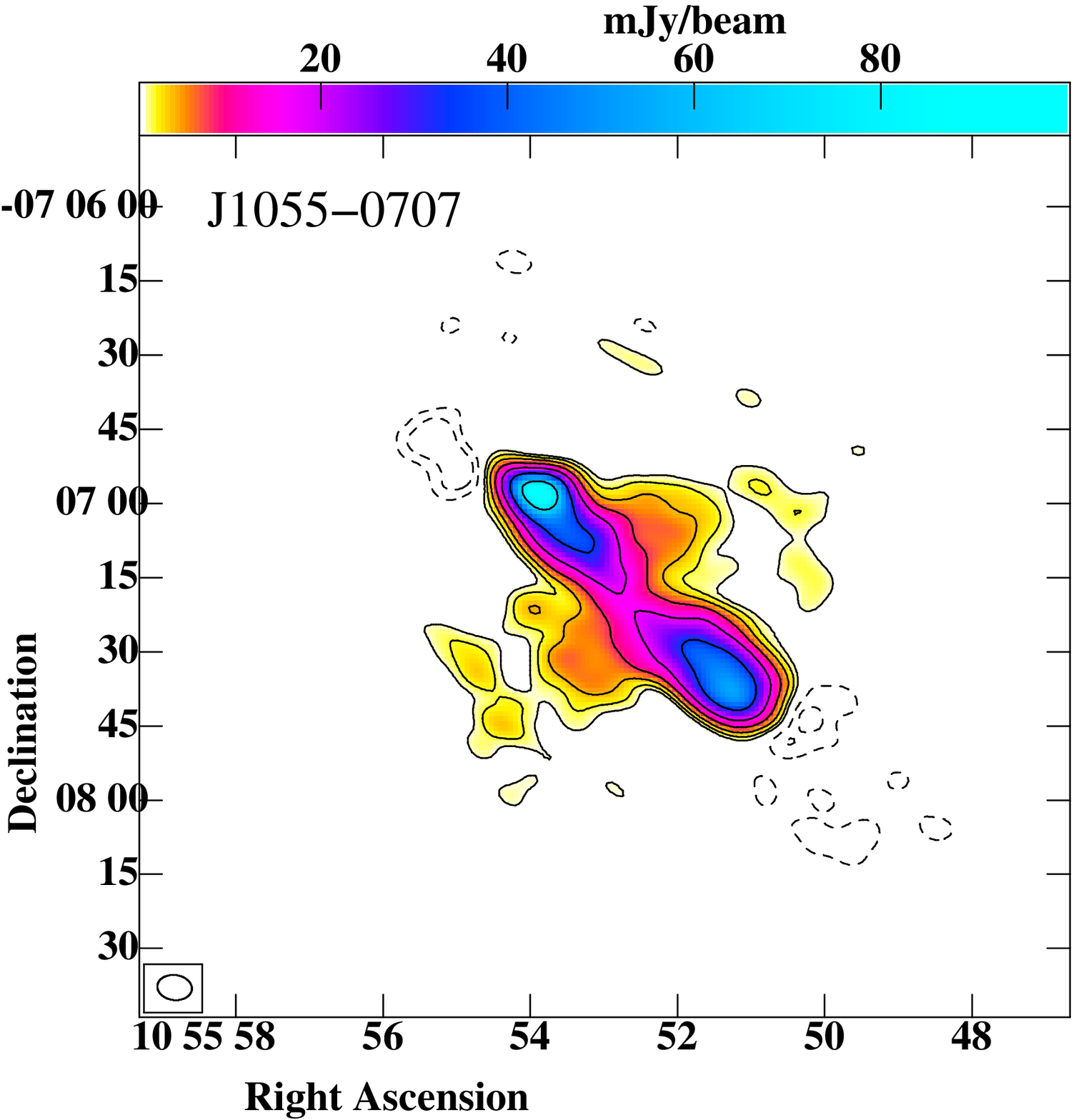} \\
\includegraphics[height=4cm]{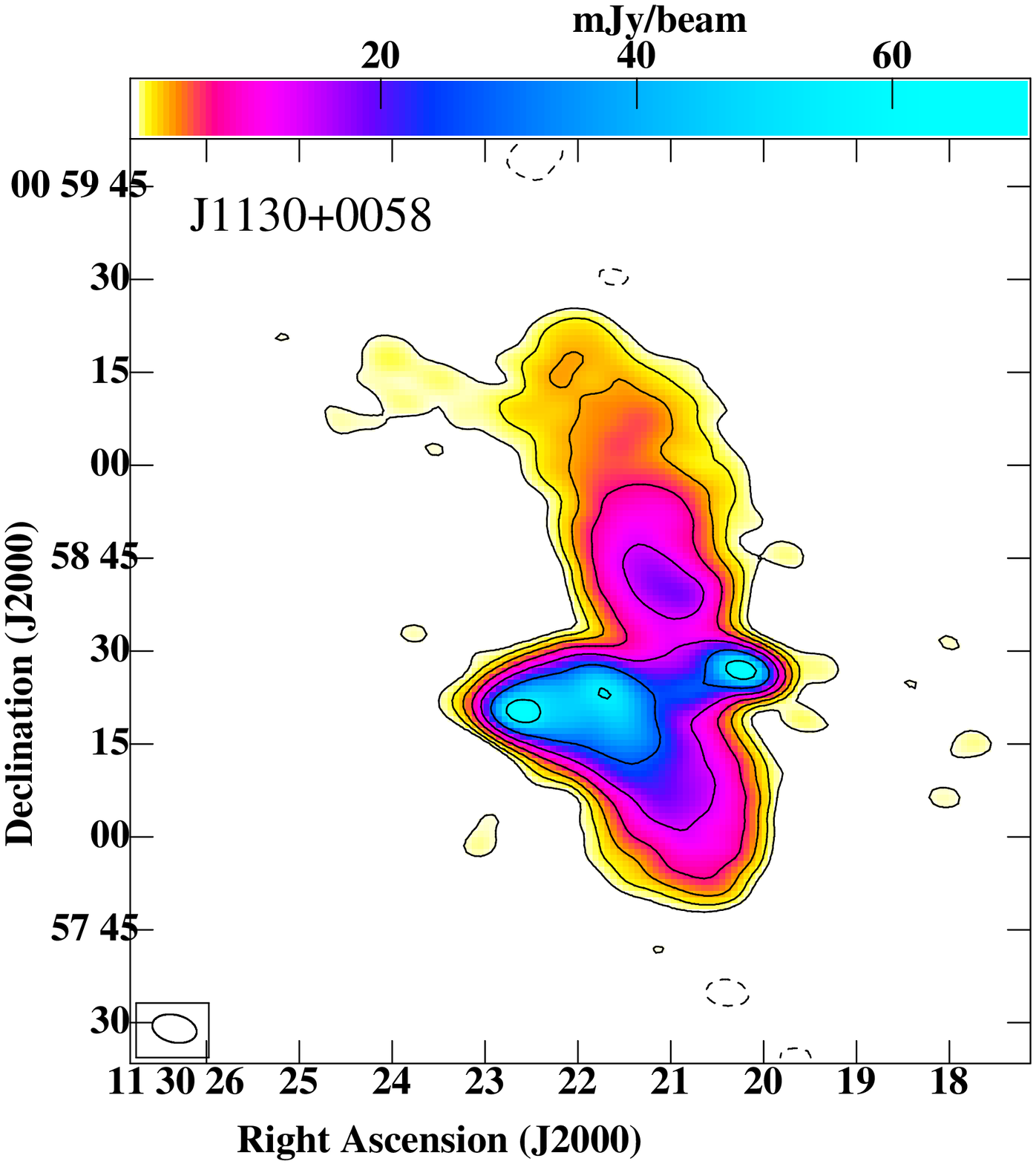} &
\includegraphics[height=4cm]{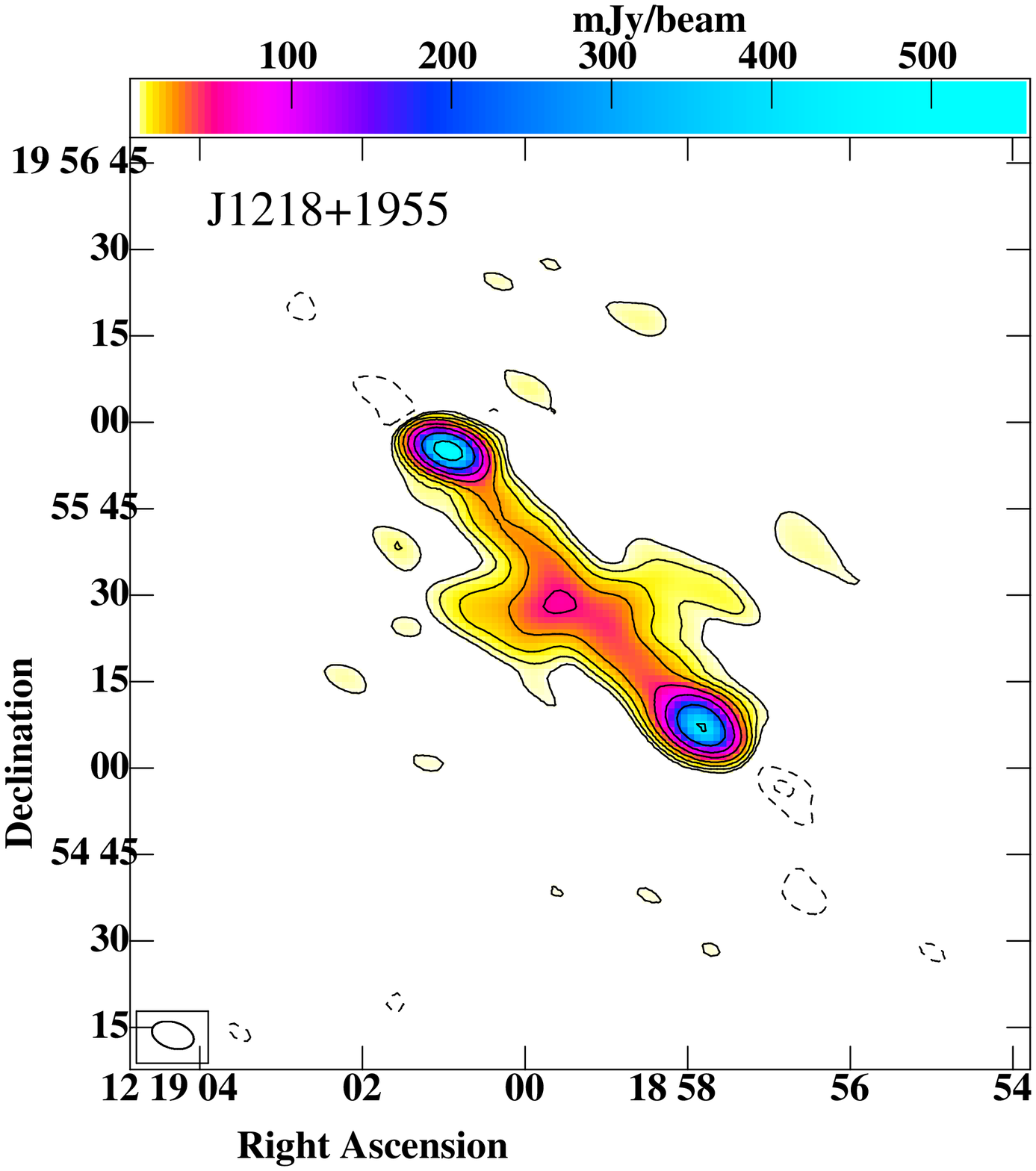} &
\includegraphics[height=4cm]{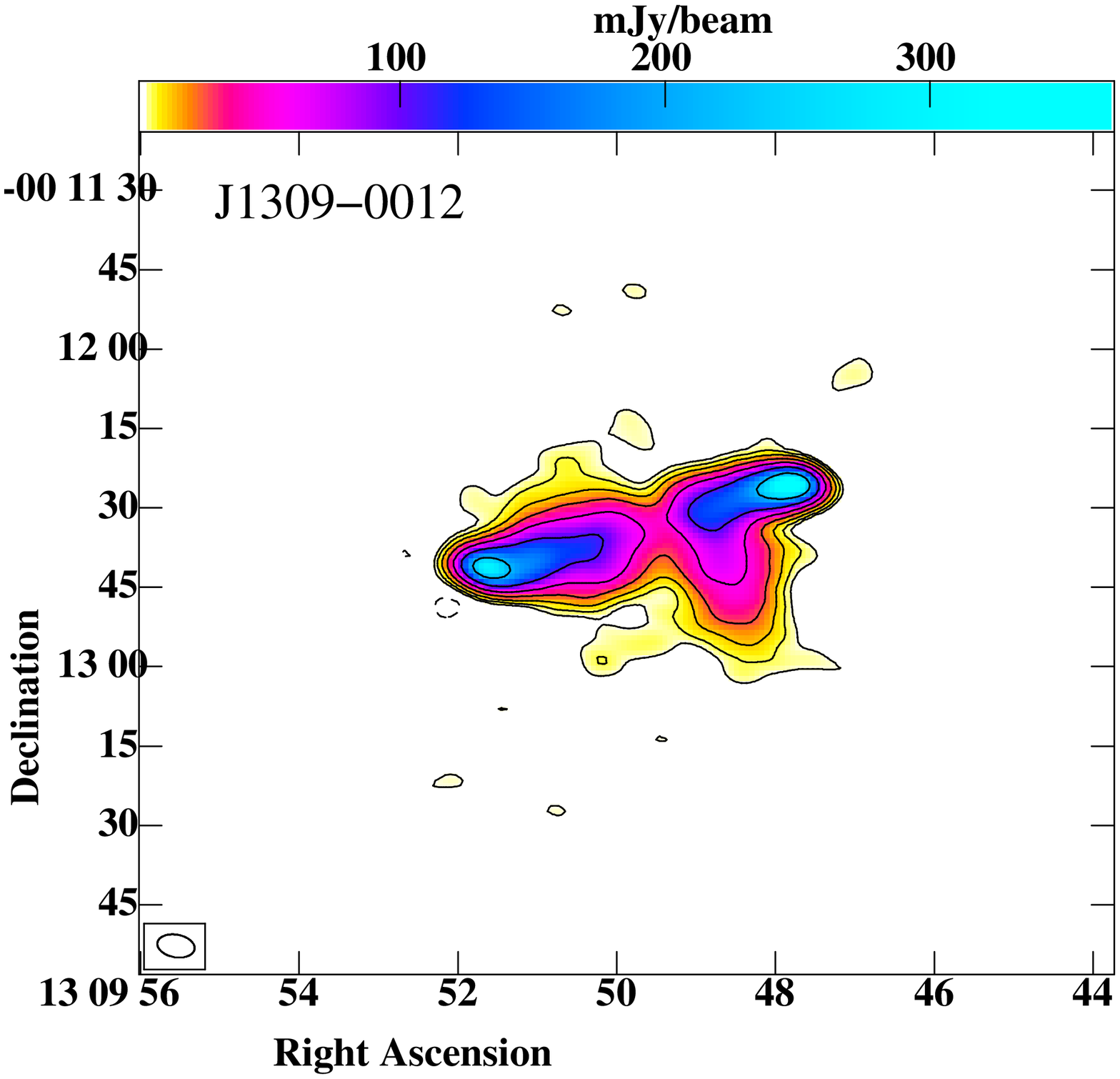} &
\includegraphics[height=4cm]{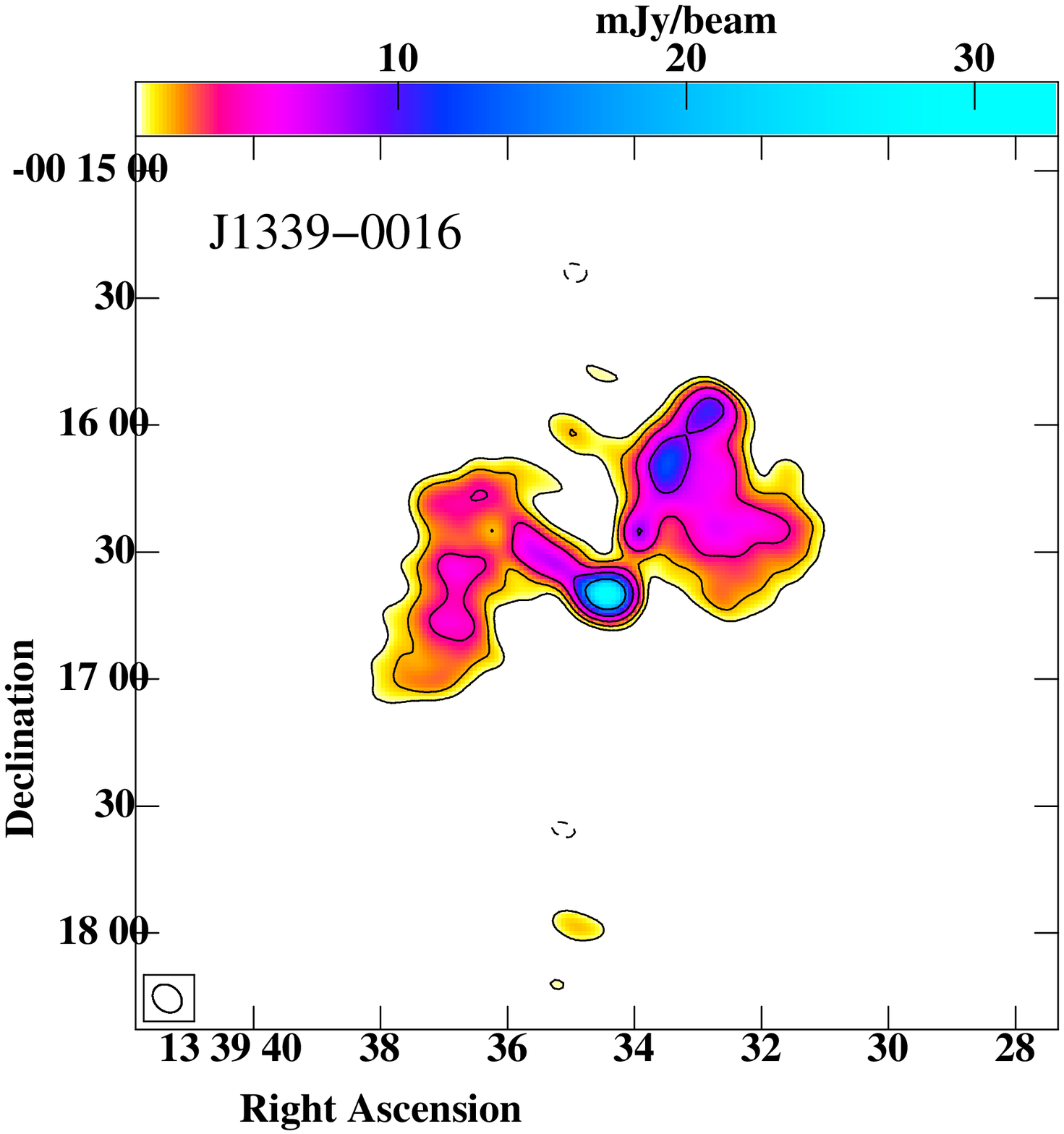} \\
\includegraphics[height=4cm]{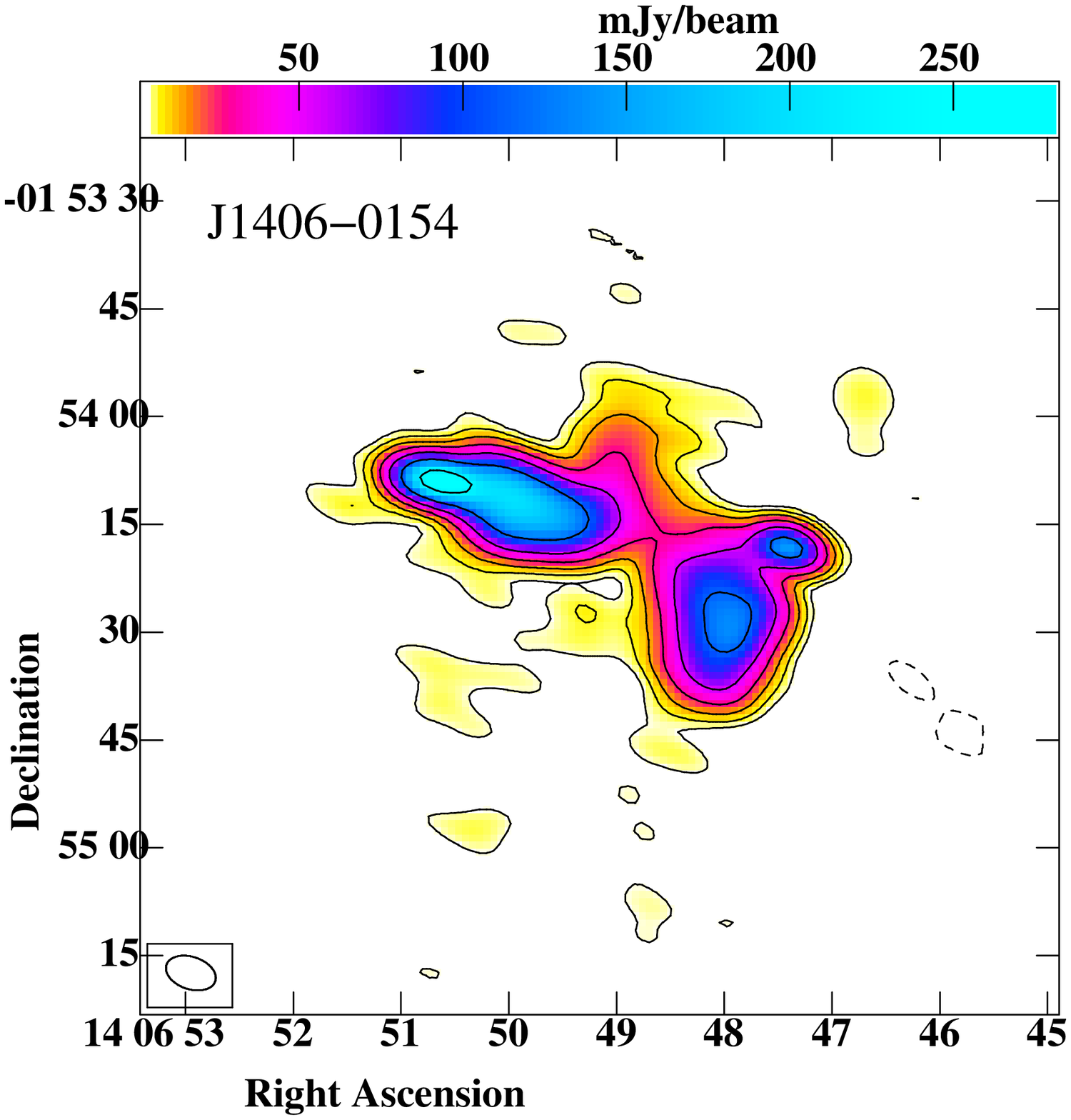} &
\includegraphics[height=4cm]{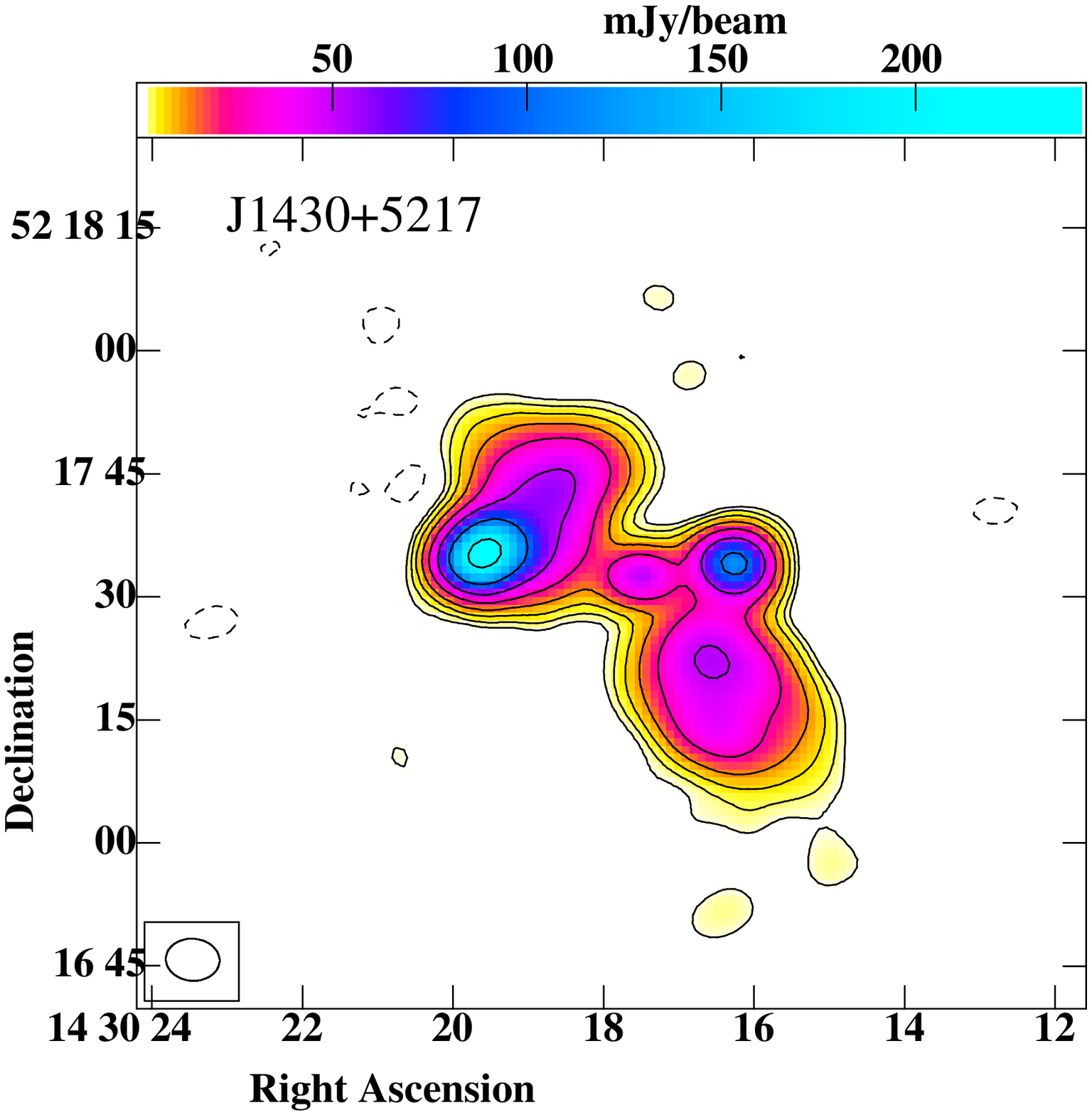} &
\includegraphics[height=4cm]{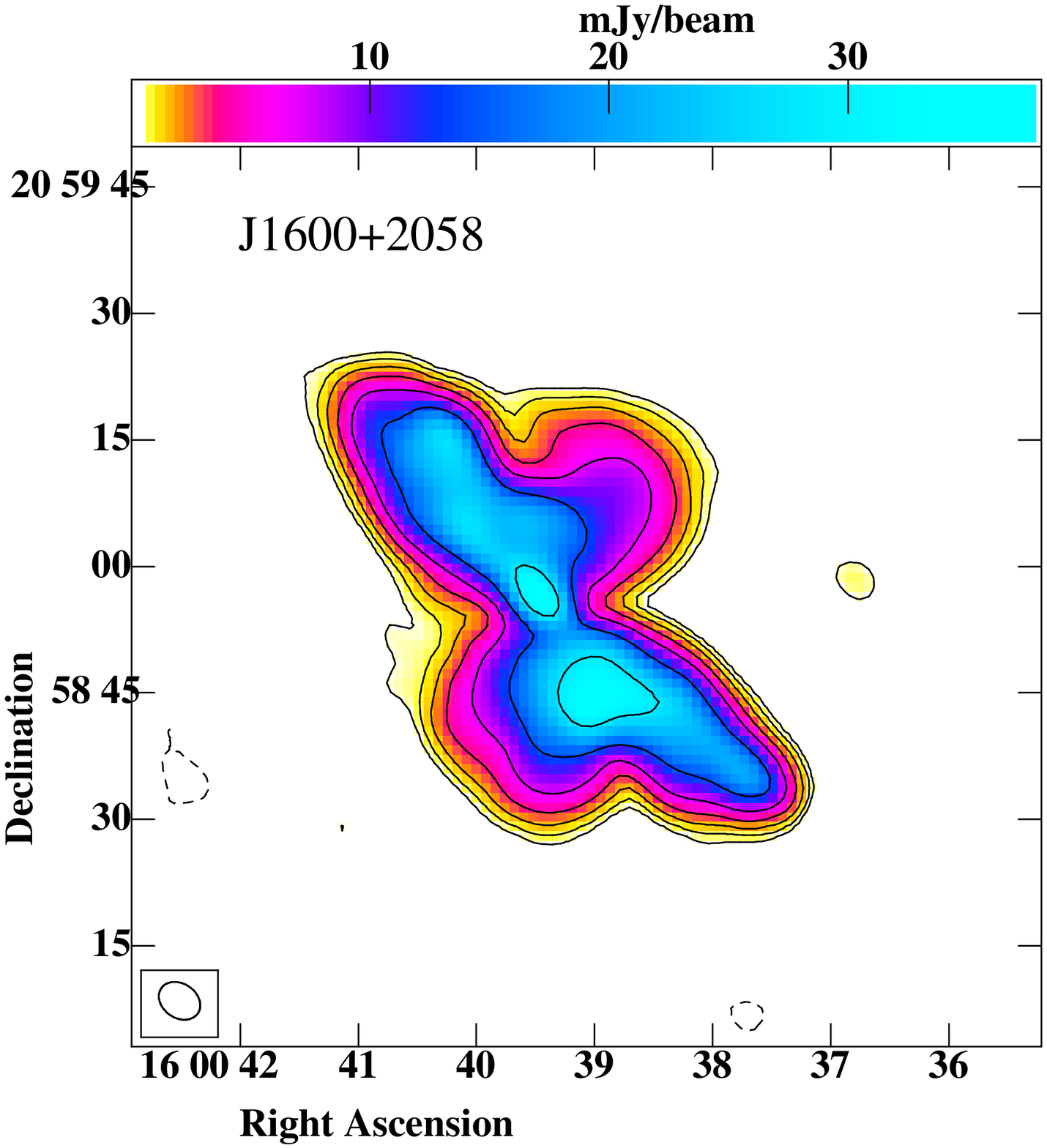} &
\includegraphics[height=4cm]{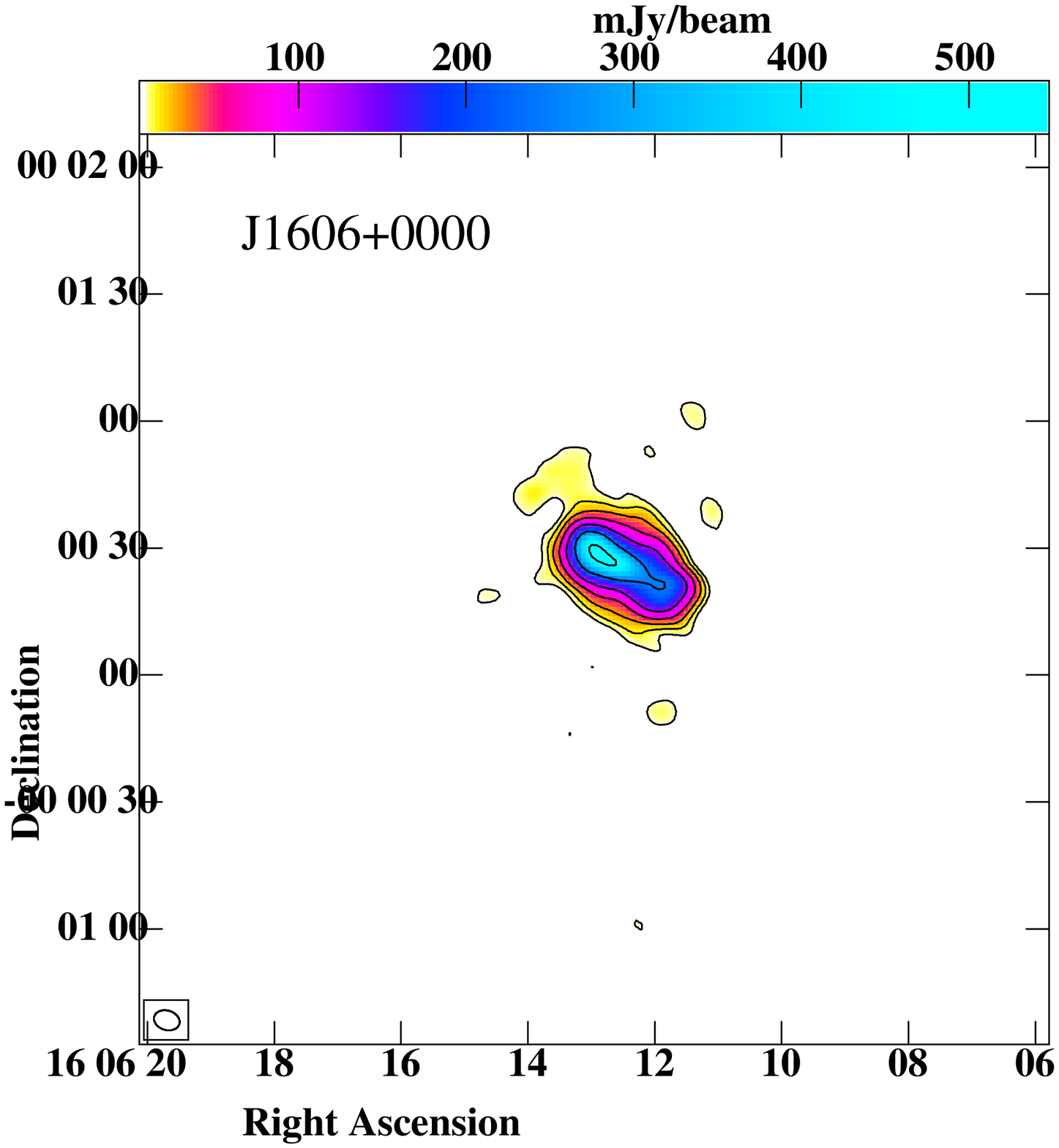}
\end{tabular}
\caption{Full resolution GMRT 610 MHz contour maps for the 16 sample sources, ordered in increasing right ascension from left to right, and then down (see also Table~\ref{obs-log}). 
The contour levels and surface brightness peaks are listed in Table~\ref{610-mhz-maps}.
The boxed ellipse in the lower left-hand corner shows the shape of the FWHM synthesized beam in each case.}
\label{610-sample}
\end{center}
\end{figure*}

\begin{table*}[tbph]
   \centering
   \caption{High-resolution GMRT 610 MHz map parameters presented in Figure~\ref{610-sample}.}
   \label{610-mhz-maps}
\begin{tabular}{lccccc}
\hline \hline
Object name & Beam & P.A.   & Peak & rms noise & Contour levels  \\
\cline{4-6}
            &      & (deg.) & \multicolumn{3}{c}{(mJy~beam$^{-1}$)} \\
\hline
J0113$+$0106 & 7.1$^{\prime\prime}$ $\times$ 14.2$^{\prime\prime}$ & 70.2 &   42.3 & 0.10 & 0.5 $\times$ ($-$1, 1, 2, 4, ...) \\
J0115$-$0000 & 7.9$^{\prime\prime}$ $\times$ 12.0$^{\prime\prime}$ & 70.9 &   43.0 & 0.19 & 0.8 $\times$ ($-$1, 1, 2, 4, ...) \\
J0702$+$5002 & 8.4$^{\prime\prime}$ $\times$ 5.2$^{\prime\prime}$  & 79.4 &   88.6 & 0.39 & 3.0 $\times$ ($-$1, 1, 2, 4, ...) \\
J0859$-$0433 & 6.7$^{\prime\prime}$ $\times$ 4.6$^{\prime\prime}$  & 61.1 &   62.4 & 0.15 & 0.9 $\times$ ($-$1, 1, 2, 4, ...) \\
J0914$+$1715 & 6.6$^{\prime\prime}$ $\times$ 4.3$^{\prime\prime}$  & 68.9 & 1505.2 & 0.39 & 3.6 $\times$ ($-$1, 1, 2, 4, ...) \\
J0917$+$0523 & 7.5$^{\prime\prime}$ $\times$ 5.0$^{\prime\prime}$  & 69.6 &  389.6 & 0.25 & 2.8 $\times$ ($-$2, $-$1, 1, 2, 4, ...) \\
J0924$+$4233 & 7.7$^{\prime\prime}$ $\times$ 13.9$^{\prime\prime}$ & 77.0 &   64.0 & 0.18 & 1.6 $\times$ ($-$1, 1, 2, 4, ...) \\
J1055$-$0707 & 7.0$^{\prime\prime}$ $\times$ 11.2$^{\prime\prime}$ & 83.0 &   99.8 & 0.13 & 1.1 $\times$ ($-$2, $-$1, 1, 2, 4, ...) \\
J1130$+$0058 & 7.3$^{\prime\prime}$ $\times$ 4.4$^{\prime\prime}$  & 77.0 &   70.3 & 0.14 & 0.9 $\times$ ($-$1, 1, 2, 4, ...) \\
J1218$+$1955 & 7.5$^{\prime\prime}$ $\times$ 11.3$^{\prime\prime}$ & 73.1 &  557.6 & 0.30 & 3.2 $\times$ ($-$2, $-$1, 1, 2, 4, ...) \\
J1309$-$0012 & 7.1$^{\prime\prime}$ $\times$ 13.2$^{\prime\prime}$ & 79.9 &  367.6 & 0.37 & 3.6 $\times$ ($-$1, 1, 2, 4, ...) \\
J1339$-$0016 & 7.5$^{\prime\prime}$ $\times$ 6.0$^{\prime\prime}$  & 49.5 &   32.7 & 0.10 & 1.0 $\times$ ($-$1, 1, 2, 4, ...) \\
J1406$-$0154 & 7.2$^{\prime\prime}$ $\times$ 12.1$^{\prime\prime}$ & 71.5 &  280.2 & 0.30 & 3.6 $\times$ ($-$1, 1, 2, 4, ...) \\
J1430$+$5217 & 6.5$^{\prime\prime}$ $\times$ 5.1$^{\prime\prime}$  & 85.9 &  241.9 & 0.23 & 1.7 $\times$ ($-$1, 1, 2, 4, ...) \\
J1600$+$2058 & 5.3$^{\prime\prime}$ $\times$ 4.1$^{\prime\prime}$  & 55.2 &   37.7 & 0.06 & 0.4 $\times$ ($-$1, 1, 2, 4, ...) \\
J1606$+$0000 & 6.2$^{\prime\prime}$ $\times$ 4.6$^{\prime\prime}$  & 73.0 &  545.2 & 0.84 & 7.6 $\times$ ($-$1, 1, 2, 4, ...) \\
\hline
\end{tabular}
\end{table*}

\section{GMRT data and analysis}
\label{gmrt-obs}

The 16 sources were observed with the GMRT during two 3-day sessions in December 2006 and November 2007 (see Table~\ref{obs-log} for a summary). 
All observations used the dual frequency mode observing simultaneously at 610 MHz and 240 MHz with bandwidths of 16 MHz and 8 MHz, respectively.
To improve ($u,v$) coverage, target scans were obtained over a range of hour angles, interleaved with observations of phase and flux density calibrators.
We aimed for $\sim2-3$ hrs of total exposure obtained per source, before editing.

The data were calibrated as in \cite{2007MNRAS.374.1085L}. 
In summary, basic editing of the data, including flagging and calibration was carried out using the \textsc{flagcal} package \citep{2012ExA....33..157P} with the rest of the analysis performed using the NRAO {\tt AIPS} package \citep{bri94}. 
Typically $\lesssim$20\% of the data were flagged (predominantly due to radio frequency interference) for each source.
Wide fields of view were imaged using facets to map each of the two frequencies for every target.
After three to four rounds of phase-only self-calibration, a final self-calibration of both amplitude and phase was performed to obtain the final images.
Finally, the facets were stitched and corrected for the primary beam of the GMRT antennas.

\section{Results}
\label{gmrt-result}

\subsection{Radio Morphology}

The high resolution, high sensitivity radio maps of all our 16 X-shaped radio galaxies at 610 MHz are shown in Figure~\ref{610-sample} with map parameters given in Table~\ref{610-mhz-maps}.
The root mean square (rms) noise measurements were taken in source-free regions away from phase-centre, and are occasionally lower than the local noise levels in the vicinity of our targets.  
Hence, the first contour-levels are 3--5~$\times$ the rms noise in the vicinity of our targets.

The features observed in these high-resolution GMRT 610 MHz maps have similar resolution and sensitivity to the 5$^{\prime\prime}$ resolution VLA-FIRST 1.4 GHz images \citep[see][]{2007AJ....133.2097C}.
In most cases, the radio morphology of the primary lobes can be characterized as having edge-brightened \citep[FR\,II type;][]{FanaroffRiley} radio structures, confirmed in higher-resolution VLA images \citep{2015ApJS..220....7R,2018ApJ...852...47R}.  
The two exceptions are J0702$+$5002 and J1600$+$2058 with morphology better classified as FR\,I.  
J0702$+$5002 is absent of clear hotspots, even in the high-resolution image at 610 MHz in the source, and we take the lobes pointed towards the north-east and south-west direction that have higher peak surface brightness as the primary lobes.
\citet{2018ApJ...852...48S} suggested J1600$+$2058 may be an FRII, but it shows evidence for higher intensity close to the radio core, indicative of an FR\,I radio galaxy in line with the original classification \citep{2007AJ....133.2097C}.  We define the north-east and south-west direction as the axis of the primary lobes.

Note that the wings in J1606$+$0000 are barely detected in our GMRT images, although they are clearly seen in the VLA FIRST survey image \citep{2007AJ....133.2097C}. 
Similarly, J1339$-$0016 shows `W'-shaped morphology in our 610 MHz map, possibly because the two jets and hence the two primary lobes are bent because the source is moving through the dense gas in the cluster environment \citep[e.g.,][]{hao10}
We define the northern, high surface brightness regions close to the radio core as the active lobes and southern, low surface brightness regions far away from the radio core as the wings.
Both these sources are included in our sample and the inclusion of these two sources do not affect the overall spectral trends we observe over the total sample of 28 objects.

It is also worth mentioning that J0115$-$0000 is a peculiar source and stands out from the rest of X-shaped radio sources studied. In higher-resolution VLA images \citep{2018ApJ...852...47R}, the edge-brightened inner structure becomes more apparent, particularly in the polarization structure, indicating this is a possible `winged double-double' radio galaxy. Interestingly, the wings identified by \cite{2007AJ....133.2097C} are apparently connected with the inner lobes in our maps, giving the inner structure an overall `Z'-shaped symmetry \citep{gop03} as well.

The lower resolution maps at 240 MHz are shown for the 16 individual sources in the Appendix together with similar (lower) resolution versions of the 610 MHz data (Figure~\ref{low-res-collage}). 
These data were used to create spectral index images, where the final calibrated ($u,v$) data at 610 MHz were re-mapped using a ($u,v$) taper corresponding to the 240-MHz data, giving the beam sizes at these two frequencies that are nearly identical.
Both maps were then restored using the circular/symmetric synthesized beam corresponding to the major axis of the beam-shape of the 240 MHz map and corrected for misalignment, if any.
Pixels below 5 $\times$ rms were clipped at both frequencies and the spectral index maps were constructed. 
We use the ratio of log($S_{\rm 240~MHz} (x, y)$/$S_{\rm 610~MHz} (x, y)$) and log(240 MHz/610 MHz) to construct the spectral index distribution between maps, $S_{\rm 240~MHz} (x, y)$ and $S_{\rm 610~MHz} (x, y)$.

\begin{figure*}
\begin{center}
\begin{tabular}{r}
\includegraphics[width=17.8cm]{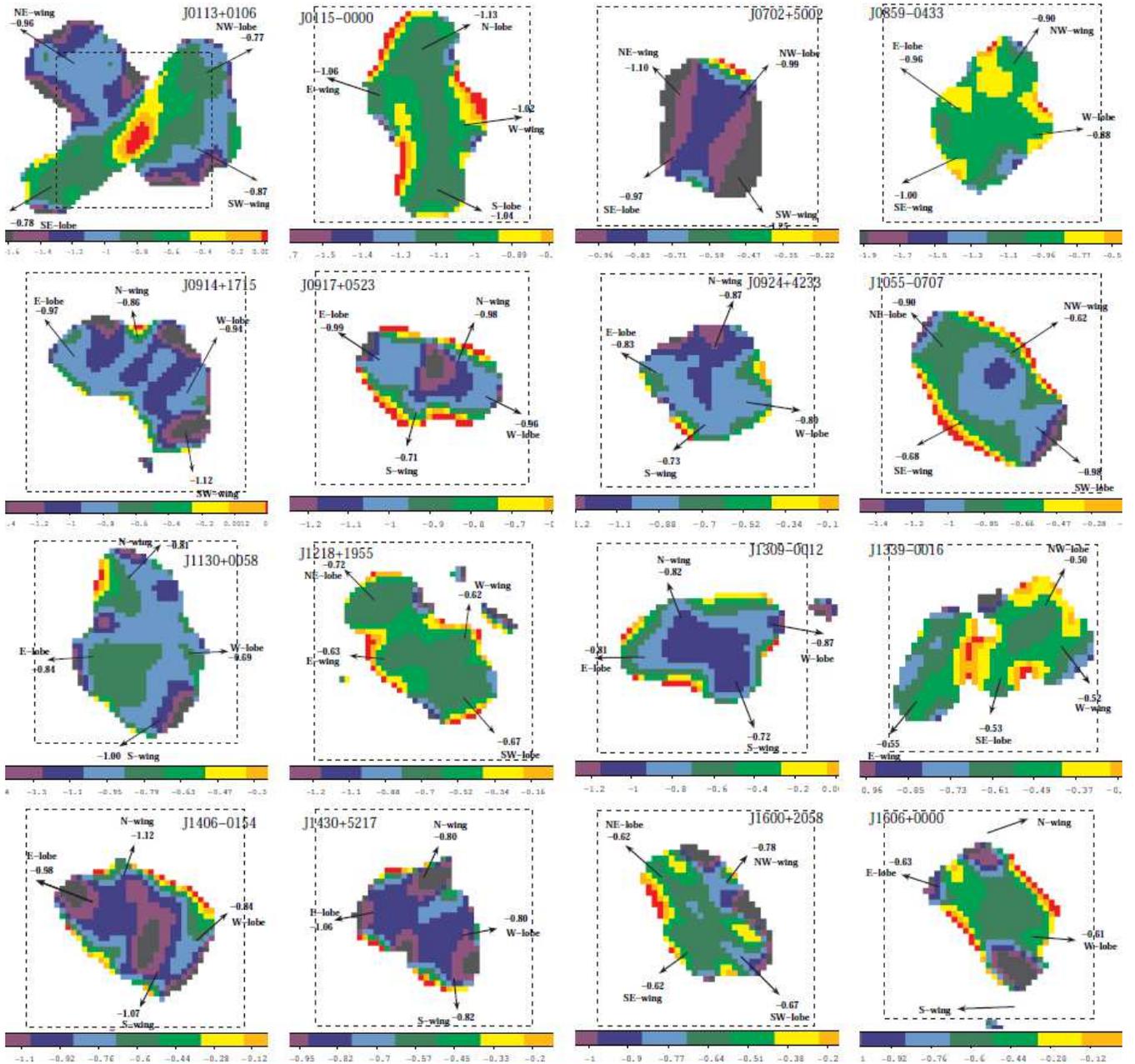}
\end{tabular}
\caption{The maps of our 16 sample sources showing the spatial distribution of the spectral indices between 240 and 610 MHz (color bars), ordered in increasing right ascension as in Figure~\ref{610-sample}.
Dashed lines in each panel indicate 2$^\prime$-sided squares.}
    \label{sp-in-sample}
\end{center}
\end{figure*}

\subsection{Global Spectral Index Properties}
\label{global-sp-in}

The two-frequency spectral index maps for the 16 newly observed X-shaped radio sources are shown in Figure~\ref{sp-in-sample}. 
They show remarkable variation across each of the sources.
We have shown only a small range of spectral indices for clarity, though the full range is large in each case.

We follow the recipe of \cite{2007MNRAS.374.1085L} to determine flux densities at 610 MHz and at 240 MHz and hence the spectral index for the active lobes and the wings using regions defined at their respective locations. 
This was done for our 16 newly observed X-shaped sources, thereby making this sample consistent
in terms of data analysis with the known sample of X-shaped sources observed by \cite{2007MNRAS.374.1085L} (maps shown in the lower-right panels of Figures~ 1--11 therein).
Hereagain, the measurements are integrated over the region, which is at least four times the beam size (either a circular region of $\sim$6 pixel radius or a polygon shaped region centered at the position of the tail) and above their 5$\sigma$ surface brightness contour levels to reduce statistical errors (Table~\ref{flux-den-region}).
Appropriate care was taken when determining the flux density values in the wings and in the active lobes so as to avoid contamination as much as possible. 
We also tried regions of various sizes and shapes, including irregular polygons, in order to
avoid contamination, if any and our results presented here do not change.
Unlike \cite{2007MNRAS.374.1085L}, here we are more conservative with the flux density error values and assume a 5\% calibration error, which were added to the rms of the map in quadrature. 
The calibration error is dominant as compared to error contributed by the rms of the map itself.
We recalculate the errors in these maps as well as in \cite{2007MNRAS.374.1085L} maps and quote
spectral index values after accounting for the calibration errors.
This translates to a maximum probable error of 0.06 on the spectral index values. 

\begin{figure}[ht]
\begin{center}
\begin{tabular}{c}
\includegraphics[width=8cm]{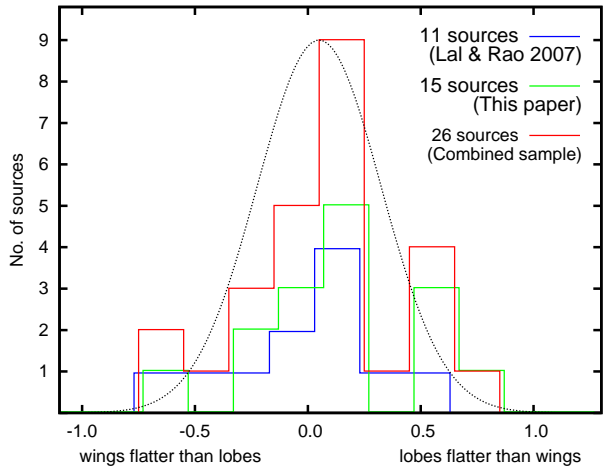} 
\end{tabular}
\end{center}
\caption{
The distribution of $\Delta{\alpha}$, difference between the averaged spectral indices in the two primary active lobes and the two secondary wings.
The three histograms (for clarity, shifted slightly with respect to one another) are for the present sample of 16 candidate X-shaped sources (green), the earlier \cite{2007MNRAS.374.1085L} sample (blue), and for the combined sample (red).
Also shown is the Gaussian function fit to the combined sample of 28 sources, with a standard deviation, $\sigma$ ($\simeq \pm$0.27).
The sources in the central two bins, whose differences in spectral indices are not significant, i.e., $-1\sigma$ $<$ $\Delta{\alpha}$ $<$ $+1\sigma$, are classified as category 
(ii) sources showing comparable spectral indices for the primary active lobes and the secondary wings.
The sources in the left-bins, $\Delta{\alpha}$ $\le$ $-1\sigma$ are sources with secondary wings to be flatter than the primary active lobes, 
category (i) and in the right-bins, $\Delta{\alpha}$ $\ge$ $+1\sigma$ are sources with secondary wings to be steeper than the primary active lobes, category (iii).}
    \label{fig:hist}
\end{figure}

The radio lobe emission and its spectrum for each of the sample sources is typical of 
radio lobes of classical FR\,II source \citep[e.g.,][]{2008MNRAS.390.1105L}.
Though the emission from
diffuse wings is similar in surface brightness to the diffuse emission seen in classical
FR\,II sources, but their the spectra are different.  Diffuse wings occassionally show
flat spectrum in some sources as compared to steep spectra, typically seen in FR\,II sources.
We determine the difference between the averaged spectral indices in the two active (primary) lobes and in the two wings (secondary lobes), $\Delta{\alpha} = \alpha_{\rm primary} - \alpha_{\rm secondary}$, for each source in our earlier observed 12 sources in \cite{2007MNRAS.374.1085L} and the 16 X-shaped sources in this paper.
Following \cite{2007MNRAS.374.1085L}, the sample sources can again be categorized into
following three groups:
\begin{itemize}
\item[(i)] the secondary, low-surface brightness wings have flatter spectra as compared to the primary active lobes, with  $\Delta{\alpha}$ $\le$ $-1\sigma$, \\ [-0.7cm]
\item[(ii)] the secondary, low-surface brightness wings  and primary active lobes have comparable spectral indices, with $-1\sigma$ $<$ $\Delta{\alpha}$ $<$ $+1\sigma$, and \\ [-0.7cm]
\item[(iii)] the secondary, low-surface brightness wings have steeper spectra as compared to the primary active lobes, with $\Delta{\alpha}$ $\ge$ $+1\sigma$, \\ [-0.6cm]
\end{itemize}
\noindent
where, $\sigma$ ($\simeq$ 0.27; see also Figure~\ref{fig:hist}) is the standard deviation, assuming a normal distribution for $\Delta{\alpha}$.
Note that we do not include sources J1606$+$0000 and B2\,0828$+$28 from our current sample of candidate X-shaped galaxies and from Lal \& Rao (2007) known sample of X-shaped galaxies, respectively where we do not have detections for the radio wings.

\begin{table*}[ht]
\centering
\caption{Flux densities of the distinct regions, and the spectral indices along with its error estimates in lobes and wings.}
\label{flux-den-region}
\begin{tabular}{llrrr}
\hline \hline
 Object name& Region & \multicolumn{2}{c}{$S_\nu$ (mJy)} & \multicolumn{1}{c}{$\alpha_{\rm 240 MHz}^{\rm 610 MHz}$} \\
        &  & {610} & {240} & \\
\hline
\multicolumn{1}{l}{J0113$+$0106} & SE-lobe &   131.4  &   272.0  & $-$0.78  $\pm$0.06 \\
                                 & NW-lobe &    75.1  &   154.1  & $-$0.77  $\pm$0.04 \\
                                 & NE-wing &    45.7  &   111.2  & $-$0.96  $\pm$0.05 \\
                                 & SW-wing &    95.9  &   215.9  & $-$0.87  $\pm$0.10 \\
\multicolumn{1}{l}{J0115$-$0000} &  N-lobe &    51.8  &   148.6  & $-$1.13  $\pm$0.06 \\
                                 &  S-lobe &    44.4  &   117.1  & $-$1.04  $\pm$0.05 \\
                                 &  E-wing &    21.6  &    58.1  & $-$1.06  $\pm$0.06 \\
                                 &  W-wing &    18.3  &    47.4  & $-$1.02  $\pm$0.03 \\
\multicolumn{1}{l}{J0702$+$5002} & SE-lobe &    14.6  &    36.1  & $-$0.97  $\pm$0.02 \\
                                 & NW-lobe &    17.6  &    44.4  & $-$0.99  $\pm$0.02 \\
                                 & NE-wing &    16.1  &    44.9  & $-$1.10  $\pm$0.06 \\
                                 & SW-wing &    25.2  &    80.8  & $-$1.25  $\pm$0.10 \\
\multicolumn{1}{l}{J0859$-$0433} &  E-lobe &    44.7  &   109.5  & $-$0.96  $\pm$0.04 \\
                                 &  W-lobe &    41.4  &    94.1  & $-$0.88  $\pm$0.06 \\
                                 & NW-wing &    32.1  &    74.3  & $-$0.90  $\pm$0.06 \\
                                 & SE-wing &    18.5  &    46.9  & $-$1.00  $\pm$0.10 \\
\multicolumn{1}{l}{J0914$+$1715} &  E-lobe &   570.3  &  1409.5  & $-$0.97  $\pm$0.03 \\
                                 &  W-lobe &  1593.8  &  3830.3  & $-$0.94  $\pm$0.05 \\
                                 &  N-wing &    72.3  &   161.3  & $-$0.86  $\pm$0.07 \\
                                 & SW-wing &    42.2  &   120.0  & $-$1.12  $\pm$0.08 \\
\multicolumn{1}{l}{J0917$+$0523} &  E-lobe &   376.1  &   947.0  & $-$0.99  $\pm$0.03 \\
                                 &  W-lobe &   364.9  &   893.4  & $-$0.96  $\pm$0.04 \\
                                 &  N-wing &     9.7  &    24.2  & $-$0.98  $\pm$0.07 \\
                                 &  S-wing &    20.3  &    39.4  & $-$0.71  $\pm$0.04 \\
\multicolumn{1}{l}{J0924$+$4233} &  E-lobe &   136.8  &   296.7  & $-$0.83  $\pm$0.02 \\
                                 &  W-lobe &    98.2  &   207.1  & $-$0.80  $\pm$0.04 \\
                                 &  N-wing &    36.2  &    81.5  & $-$0.87  $\pm$0.03 \\
                                 &  S-wing &    16.0  &    31.6  & $-$0.73  $\pm$0.08 \\
\multicolumn{1}{l}{J1055$-$0707} & NE-lobe &   233.8  &   541.3  & $-$0.90  $\pm$0.07 \\
                                 & SW-lobe &   221.1  &   551.6  & $-$0.98  $\pm$0.01 \\
                                 & NW-wing &    12.5  &    22.3  & $-$0.62  $\pm$0.04 \\
                                 & SE-wing &    24.8  &    46.8  & $-$0.68  $\pm$0.04 \\
\multicolumn{1}{l}{J1130$+$0058} &  E-lobe &   189.1  &   414.0  & $-$0.84  $\pm$0.05 \\
                                 &  W-lobe &   106.5  &   202.7  & $-$0.69  $\pm$0.06 \\
                                 &  N-wing &    61.1  &   130.1  & $-$0.81  $\pm$0.04 \\
                                 &  S-wing &    55.3  &   140.6  & $-$1.00  $\pm$0.07 \\
\multicolumn{1}{l}{J1218$+$1955} & NE-lobe &   552.6  &  1081.7  & $-$0.72  $\pm$0.04 \\
                                 & SW-lobe &   654.0  &  1221.8  & $-$0.67  $\pm$0.03 \\
                                 &  E-wing &    22.0  &    39.6  & $-$0.63  $\pm$0.03 \\
                                 &  W-wing &    13.9  &    25.0  & $-$0.62  $\pm$0.04 \\
\multicolumn{1}{l}{J1309$-$0012} &  E-lobe &   789.2  &  1680.1  & $-$0.81  $\pm$0.05 \\
                                 &  W-lobe &   883.8  &  1989.8  & $-$0.87  $\pm$0.05 \\
                                 &  N-wing &    22.8  &    49.0  & $-$0.82  $\pm$0.03 \\
                                 &  S-wing &    71.9  &   140.7  & $-$0.72  $\pm$0.04 \\
\multicolumn{1}{l}{J1339$-$0016} & SE-lobe &    42.7  &    70.0  & $-$0.53  $\pm$0.07 \\
                                 & NW-lobe &    70.7  &   112.7  & $-$0.50  $\pm$0.06 \\
                                 &  E-wing &    31.0  &    51.8  & $-$0.55  $\pm$0.05 \\
                                 &  W-wing &    36.9  &    59.9  & $-$0.52  $\pm$0.03 \\
\multicolumn{1}{l}{J1406$-$0154} &  E-lobe &   804.6  &  2007.2  & $-$0.98  $\pm$0.03 \\
                                 &  W-lobe &   548.4  &  1200.6  & $-$0.84  $\pm$0.06 \\
                                 &  N-wing &    42.9  &   122.0  & $-$1.12  $\pm$0.11 \\
                                 &  S-wing &    27.3  &    74.1  & $-$1.07  $\pm$0.09 \\
\multicolumn{1}{l}{J1430$+$5217} &  E-lobe &   429.9  &  1155.6  & $-$1.06  $\pm$0.04 \\
                                 &  W-lobe &   240.9  &   508.1  & $-$0.80  $\pm$0.02 \\
                                 &  N-wing &    13.1  &    27.6  & $-$0.80  $\pm$0.04 \\
                                 &  S-wing &    28.1  &    60.4  & $-$0.82  $\pm$0.08 \\
\multicolumn{1}{l}{J1600$+$2058} & NE-lobe &   152.8  &   272.5  & $-$0.62  $\pm$0.04 \\
                                 & SW-lobe &   148.6  &   277.6  & $-$0.67  $\pm$0.05 \\
                                 & NW-wing &    12.1  &    25.0  & $-$0.78  $\pm$0.08 \\
                                 & SE-wing &    20.3  &    36.2  & $-$0.62  $\pm$0.07 \\
\multicolumn{1}{l}{J1606$+$0000} &  E-lobe &  1915.2  &  3446.9  & $-$0.63  $\pm$0.03 \\
                                 &  W-lobe &   275.9  &   487.3  & $-$0.61  $\pm$0.09 \\
                                 &  N-wing & $<$ 35.6 &  $<$51.0 &                    \\
                                 &  S-wing & $<$158.4 &  $<$18.5 &                    \\
\hline
\end{tabular}
\end{table*}

The resultant distribution of $\Delta{\alpha}$, shown in Figure~\ref{fig:hist}.
The bin-width of the histograms showing distribution of $\Delta{\alpha}$ is sufficiently large, which is larger than the error estimates on the spectral indices.
In the 16 newly observed sources (green histogram), 3/15, 8/15, and 4/15 fall into categories (i), (ii) and (iii), respectively.
Similarly, results from the earlier sample by \cite{2007MNRAS.374.1085L},  3/11, 6/11, and 2/11 are in categories (i), (ii) and (iii), respectively (blue histogram).
Note that there is a marginal difference than the earlier quoted statistics \citep{2007MNRAS.374.1085L} and is because of the averaged spectral properties as against the spectral errors for each of the primary active lobes and the secondary wings.
Finally, combining both these sample, we find 6/26, 14/26, and 6/26 in categories (i), (ii), and (iii), respectively, as shown in Figure~\ref{fig:hist} (red histogram).
Note that excluding J1339$-$0016, source showing `W'-shaped morphology 
does not change the statistics.
The spectra of active lobes is flatter than wings in the former, whereas, using the limits for spectral indices, the wings seem to be flatter than the active lobes in the latter.

This statistical difference in three categories of sources is significant and we do not believe it being due to different ($u,v$) coverages at 610 MHz and 240 MHz.
Unlike the array configurations of VLA, GMRT is not a scaled array. The GMRT has a good ($u,v$) coverage. The sample sources are small, 2$^\prime$--3$^\prime$, which is much smaller than the short baseline lengths of $\sim$35$^{\prime}$ ($\simeq$100 wavelengths) at 610 MHz and $\sim$100$^{\prime}$ ($\simeq$35 wavelengths) at 240 MHz.
Furthermore it is also unlikely that the uncertainties, if any in flux density measurements along with difficulty in mapping the diffuse low-surface brightness emission would make secondary wings to have flatter spectra than primary active lobes.
This is because we rely on the median value for the difference in spectral indices of the sample, 0.07 $\pm$0.06 rather than the difference in spectral indices of individual sources.

\section{Discussion and Summary}
\label{xs-discuss}

Several models for the formation of X-shaped radio sources outlined in the Section~\ref{intro} give differing predictions for the expected spectral differences between the radio emission in active lobes and wings.  
More importantly, the low-frequency GMRT observations are below the typically observed spectral breaks \citep[e.g.,][]{2013MNRAS.433.3364H,2007A&A...462...43M}, and thus more closely represents the injection index of the power-law population of relativistic electrons.
Additionally, as pointed out earlier by \cite{2007MNRAS.374.1085L}, the injection spectral index may be varying \citep{Palmaetal2000}, there is a likely gradient in the magnetic field strength across the source, and the electron energy spectrum is likely curved \citep{2000AJ....119.1111B}. These differences in the intrinsic source parameters could explain each of these X-shaped sources individually, but a single model presently seems implausible to explain a large sample of X-shaped sources.
Furthermore, the kpc-scale morphology of X-shaped radio galaxies is not
in general \citep[e.g., 3C\,223.1 and 3C\,403;][]{2007MNRAS.374.1085L} compatible with the simple
superposition of two radio sources \citep[unlike the case of 3C 75;][]{owen85},
as would be expected for the two unresolved AGNs model.
Clearly, high-resolution, multifrequency, phase-referenced very long baseline
interferometry imaging of X-shaped sources is needed \citep{2007MNRAS.374.1085L} in order to map two pairs
of radio jets on pc-scales corresponding to the two unresolved AGNs that are embedded well
into the base of the radio core.  This would be a crucial test to determine the
recent active jet and investigate if these sources are indeed examples of
resolved binary AGN systems \citep{SudouIguchiMurata,2006ApJ...646...49R}.

The formation models and the expected spectral dependencies, were discussed in detail in \citet[][Section 6.2 therein]{2007MNRAS.374.1085L}. 
For instance, in the \cite{1984MNRAS.210..929L} model, the wings are the outcome of back flow driven emission from the lobes, and we expect that the wings will have steeper spectra because they derive from an older electron population. 
Similarly, in the buoyancy \citep{1995ApJ...449...93W} and conical precision \citep{Parmaetal1985} models, one would expect to find a random distribution of the angles between the primary and the secondary lobes
of X-shaped sources with no dependence on their spectral indices.
Finally, in the reorientation of jet axis due to a minor merger model \citep{2002Sci...297.1310M} one expects the wings to have steeper spectra than the active lobes.  
This is because the wings would be comprised of an older electron population as compared to the active lobes, which has younger, energetic electron population.

We therefore examined the distribution of the difference between the averaged spectral indices in the two active lobes and the two wings, $\Delta{\alpha}$ for each source (Figure~\ref{fig:hist}).
The histogram has a median of 0.07 $\pm$0.06 and a peak at $\simeq$~0.05, meaning there is no systematic trend in whether the primary lobes have flatter spectra than the wings, or vice versa.
Based on the GMRT 610/240 MHz spectral results for the initial 12 X-shaped radio sources studied, \cite{2007MNRAS.374.1085L} concluded that earlier models do not explain the formation scenario for the X-shaped sources. 
Instead, they suggested an `alternative' viable twin AGN model, the X-shaped sources consist of two pairs of jets, which are associated with two unresolved AGNs, where the spectra of the primary lobes and secondary wings are expected to be uncorrelated.
The new observations of 16 sample sources gleaned from VLA FIRST survey images \citep[see also Section~\ref{sec.sample},][]{2007AJ....133.2097C} help to bolster \cite{2007MNRAS.374.1085L,2005MNRAS.356..232L} results with a more extensive sample.

Both the median and the peak values of the distribution of $\Delta{\alpha}$, suggest that a majority of the sources in our sample have comparable spectral indices for the wings and active lobes, implying a smaller radiative age difference between them \citep{2002MNRAS.330..609D}.
\cite{1980Natur.287..307B} first suggested the possibility that AGN might contain a super-massive binary black hole.
In addition, observationally we know, galaxies often merge and dual super-massive radio cores scales $<$100 kpc \citep{2012ApJ...746L..22K} and even at parsec scale separation have been discovered, e.g., in 0402$+$379 \citep{2006ApJ...646...49R} and NGC\,7674 \citep{2017NatAs...1..727K}.
The inferred rate of X-shaped radio galaxies, $\sim$7\% \citep{1992ersf.meet..307L} and $\sim$1.3\% \citep{2015ApJS..220....7R} is small.  This rate is comparable to
the low rates inferred from optical observations, e.g., 0.1\% of quasar pairs \citep{1999MNRAS.309..836M} and $\sim$1.5\% of AGN pairs \citep{2011ApJ...737..101L} with typical separations of a few tens of kpc.
Therefore it is not surprising that X-shaped sources could be associated with two unresolved AGNs, in particular when several sample sources show the wings and the lobes to be embedded well into the base.

However, it is not only hard to explain the comparable spectral indices for the wings and active lobes, but also the steeper spectra in the primary lobes as compared to the wings for a significant number of sources.
Quantitatively, this large sample of 28 sources show seven sources with wings having flatter spectral index than active lobes, and 14 sources having comparable spectral indices for the wings and active lobes, contrary to our understanding of the evolution of radio sources.
Assuming that the whole class of objects form via a single mechanism of formation, it is hard to find support for any other formation model than the `alternative' twin AGN model, i.e., the X-shaped sources consist of two pairs of jets, which are associated with two unresolved AGNs.

\section*{Acknowledgments}

We thank the anonymous referee for detailed comments on the manuscript.
We also thank Sanjay Bhatnagar for participating in the early stages of this work.
Work by C.C.C. at NRL is supported in part by NASA DPR S-15633-Y.
We thank the staff of the GMRT who made these observations possible. The GMRT is run by the National Centre for Radio Astrophysics of the Tata Institute of Fundamental Research. This research has made use of the NED, which is operated by the Jet Propulsion Laboratory, Caltech, under contract with the NASA, and NASA's Astrophysics Data System.

\appendix

\begin{table*}[ht]
   \centering
   \caption{Map parameters of the GMRT 240 MHz and 610 MHz maps presented in Figure~\ref{low-res-collage}.}
   \label{maps-appendix}
\begin{tabular}{lcccccl}
\hline \hline
Object name & Frequency & Beam & P.A.   & Peak & rms noise & Contour levels \\
\cline{5-7}
            &(MHz)&      &(deg.)  & \multicolumn{3}{c}{(mJy~beam$^{-1}$)} \\
\hline
J0113$+$0106& 240 & 17.7$^{\prime\prime}$ $\times$ 14.2$^{\prime\prime}$ &  68.5 & 219.5 & 1.35  & 6.2 $\times$ ($-$1, 1, 2, 4, ...) \\
            & 610 & 17.7$^{\prime\prime}$ $\times$ 17.7$^{\prime\prime}$ &       & 110.5 & 1.87  & 2.3 $\times$ ($-$1, 1, 2, 4, ... ) \\
J0115$-$0000& 240 & 16.2$^{\prime\prime}$ $\times$ 12.0$^{\prime\prime}$ &  77.5 & 191.6 & 1.5   & 6.1 $\times$ ($-$1, 1, 2, 4, ...) \\
            & 610 & 16.4$^{\prime\prime}$ $\times$ 16.4$^{\prime\prime}$ &       & 283.7 & 1.7   & 1.6 $\times$ ($-$1, 1, 2, 4, ... ) \\
J0702$+$5002& 240 & 19.0$^{\prime\prime}$ $\times$ 11.6$^{\prime\prime}$ &  81.4 & 462.8 & 1.52  & 7.6 $\times$ ($-$1, 1, 2, 4, ...) \\
            & 610 & 19.4$^{\prime\prime}$ $\times$ 19.4$^{\prime\prime}$ &       & 266.0 & 1.54  & 7.8 $\times$ ($-$1, 1, 2, 4, ... ) \\
J0859$-$0433& 240 & 15.3$^{\prime\prime}$ $\times$ 11.0$^{\prime\prime}$ &  66.7 & 200.4 & 1.33  & 7.0 $\times$ ($-$1, 1, 2, 4, ...) \\
            & 610 & 15.0$^{\prime\prime}$ $\times$ 15.0$^{\prime\prime}$ &       &  98.0 & 0.6   & 2.0 $\times$ ($-$1, 1, 2, 4, ... ) \\
J0914$+$1715& 240 & 14.9$^{\prime\prime}$ $\times$ 10.1$^{\prime\prime}$ &  71.2 &4010.3 & 2.76  &30.1 $\times$ ($-$1, 1, 2, 4, ...) \\
            & 610 & 14.9$^{\prime\prime}$ $\times$ 14.9$^{\prime\prime}$ &       &1602.5 & 0.76  & 7.0 $\times$ ($-$1, 1, 2, 4, ... ) \\
J0917$+$0523& 240 & 16.2$^{\prime\prime}$ $\times$ 13.0$^{\prime\prime}$ &  80.0 &1111.5 & 1.64  &14.0 $\times$ ($-$1, 1, 2, 4, ...) \\
            & 610 & 16.2$^{\prime\prime}$ $\times$ 16.2$^{\prime\prime}$ &       & 427.3 & 0.62  & 4.3 $\times$ ($-$2, $-$1, 1, 2, 4, ... ) \\
J0924$+$4233& 240 & 14.9$^{\prime\prime}$ $\times$ 13.9$^{\prime\prime}$ &  89.6 & 327.3 & 1.59  & 7.8 $\times$ ($-$1, 1, 2, 4, ...) \\
            & 610 & 14.8$^{\prime\prime}$ $\times$ 14.8$^{\prime\prime}$ &       & 136.7 & 1.84  & 3.8 $\times$ ($-$1, 1, 2, 4, ... ) \\
J1055$-$0707& 240 & 20.6$^{\prime\prime}$ $\times$ 11.2$^{\prime\prime}$ &  53.6 & 581.8 & 1.02  &11.3 $\times$ ($-$1, 1, 2, 4, ...) \\
            & 610 & 20.8$^{\prime\prime}$ $\times$ 20.8$^{\prime\prime}$ &       & 270.0 & 2.9   & 5.3 $\times$ ($-$1, 1, 2, 4, ... ) \\
J1130$+$0058& 240 & 16.8$^{\prime\prime}$ $\times$ 12.7$^{\prime\prime}$ &$-$80.8& 511.8 & 1.38  & 6.3 $\times$ ($-$1, 1, 2, 4, ...) \\
            & 610 & 16.5$^{\prime\prime}$ $\times$ 16.5$^{\prime\prime}$ &       & 257.5 & 0.89  & 3.1 $\times$ ($-$1, 1, 2, 4, ... ) \\
J1218$+$1955& 240 & 15.6$^{\prime\prime}$ $\times$ 11.3$^{\prime\prime}$ &  60.9 &1279.0 & 2.22  &12.9 $\times$ ($-$2, $-$1, 1, 2, 4, ...) \\
            & 610 & 15.4$^{\prime\prime}$ $\times$ 15.4$^{\prime\prime}$ &       & 629.4 & 4.41  & 3.8 $\times$ ($-$2, $-$1, 1, 2, 4, ... ) \\
J1309$-$0012& 240 & 16.9$^{\prime\prime}$ $\times$ 13.2$^{\prime\prime}$ &  76.3 &1733.9 & 3.18  &21.7 $\times$ ($-$1, 1, 2, 4, ...) \\
            & 610 & 16.7$^{\prime\prime}$ $\times$ 16.7$^{\prime\prime}$ &       & 804.1 &12.43  & 7.6 $\times$ ($-$1, 1, 2, 4, ... ) \\
J1339$-$0016& 240 & 15.7$^{\prime\prime}$ $\times$ 14.2$^{\prime\prime}$ &$-$4.0 &  70.7 & 1.15  & 8.1 $\times$ ($-$1, 1, 2, 4, ...) \\
            & 610 & 15.5$^{\prime\prime}$ $\times$ 15.5$^{\prime\prime}$ &       &  48.1 & 0.3   & 2.1 $\times$ ($-$1, 1, 2, 4, ... ) \\
J1406$-$0154& 240 & 16.7$^{\prime\prime}$ $\times$ 12.1$^{\prime\prime}$ &  64.9 &1753.3 & 4.34  &19.3 $\times$ ($-$1, 1, 2, 4, ...) \\
            & 610 & 16.4$^{\prime\prime}$ $\times$ 16.4$^{\prime\prime}$ &       & 746.2 &10.48  & 9.0 $\times$ ($-$1, 1, 2, 4, ... ) \\
J1430$+$5217& 240 & 14.2$^{\prime\prime}$ $\times$ 12.4$^{\prime\prime}$ &  82.4 & 881.5 & 1.53  & 7.7 $\times$ ($-$1, 1, 2, 4, ...) \\
            & 610 & 14.2$^{\prime\prime}$ $\times$ 14.2$^{\prime\prime}$ &       & 412.6 & 0.84  & 3.2 $\times$ ($-$1, 1, 2, 4, ... ) \\
J1600$+$2058& 240 & 15.5$^{\prime\prime}$ $\times$ 12.6$^{\prime\prime}$ &  43.1 & 343.3 & 2.9   & 7.0 $\times$ ($-$1, 1, 2, 4, ...) \\
            & 610 & 15.4$^{\prime\prime}$ $\times$ 15.4$^{\prime\prime}$ &       & 211.3 & 0.37  & 2.3 $\times$ ($-$1, 1, 2, 4, ... ) \\
J1606$+$0000& 240 & 15.7$^{\prime\prime}$ $\times$ 13.0$^{\prime\prime}$ &  60.6 &2917.0 & 4.36  &23.2 $\times$ ($-$1, 1, 2, 4, ...) \\
            & 610 & 15.8$^{\prime\prime}$ $\times$ 15.8$^{\prime\prime}$ &       &1851.1 & 1.86  &12.3 $\times$ ($-$1, 1, 2, 4, ... ) \\
\hline
\end{tabular}
\end{table*}

\section{Matched Resolution GMRT maps}

Here, we present the GMRT 240 MHz maps at native resolution side-by-side with lower-resolution 
versions of the 610 MHz maps presented in Figure~\ref{610-sample} made by tapering the ($u,v$) data.
The map parameters are given in Table~\ref{maps-appendix}. 
The maps are grouped in two, ordered in increasing right ascension (see also Table~\ref{obs-log}).
These data were ultimately used to produce the spectral index maps at matched resolution (Figure~\ref{sp-in-sample}).

\begin{figure*}[ht]
\begin{center}
\begin{tabular}{cccc}
\includegraphics[height=4cm]{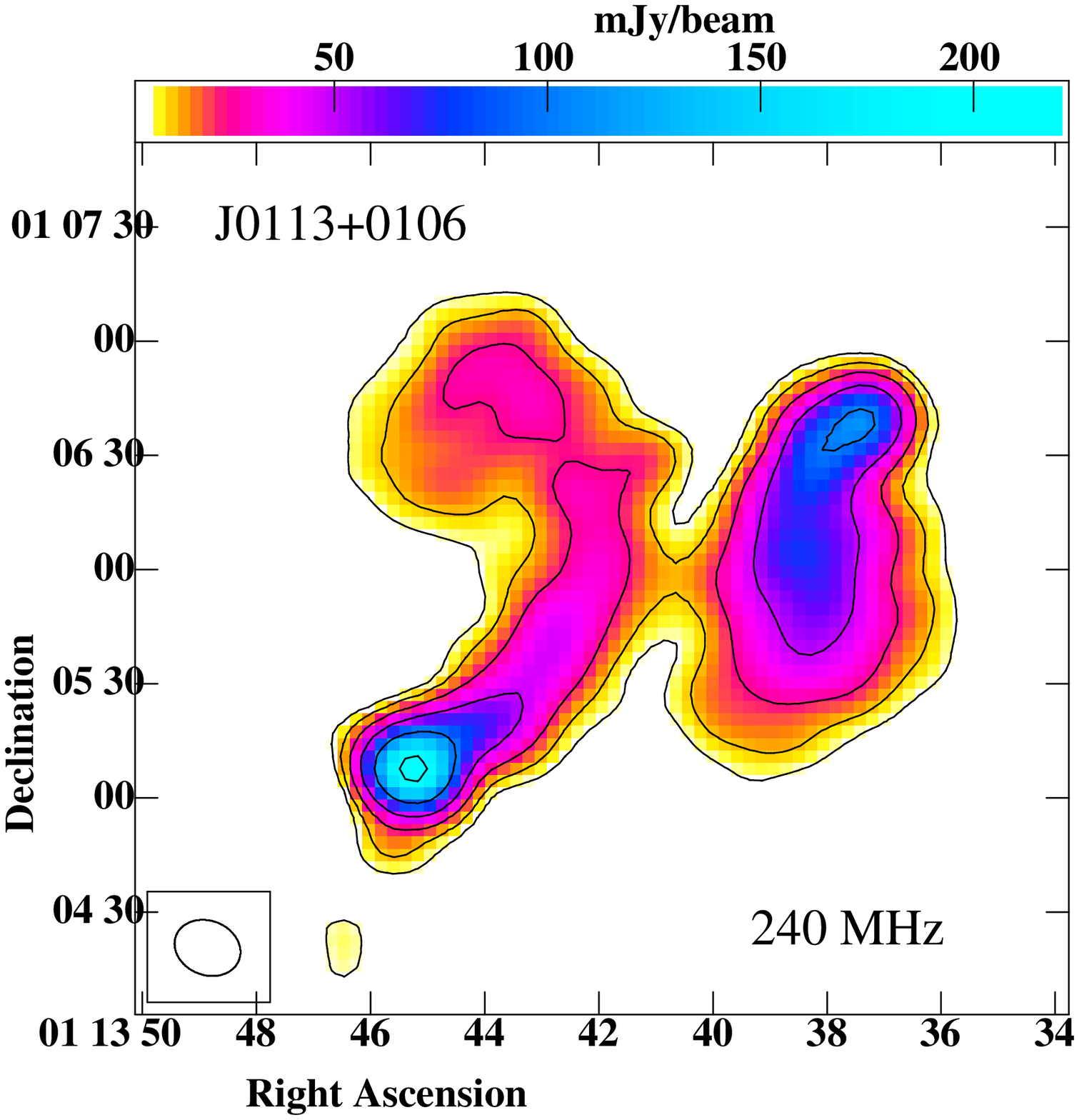} &
\includegraphics[height=4cm]{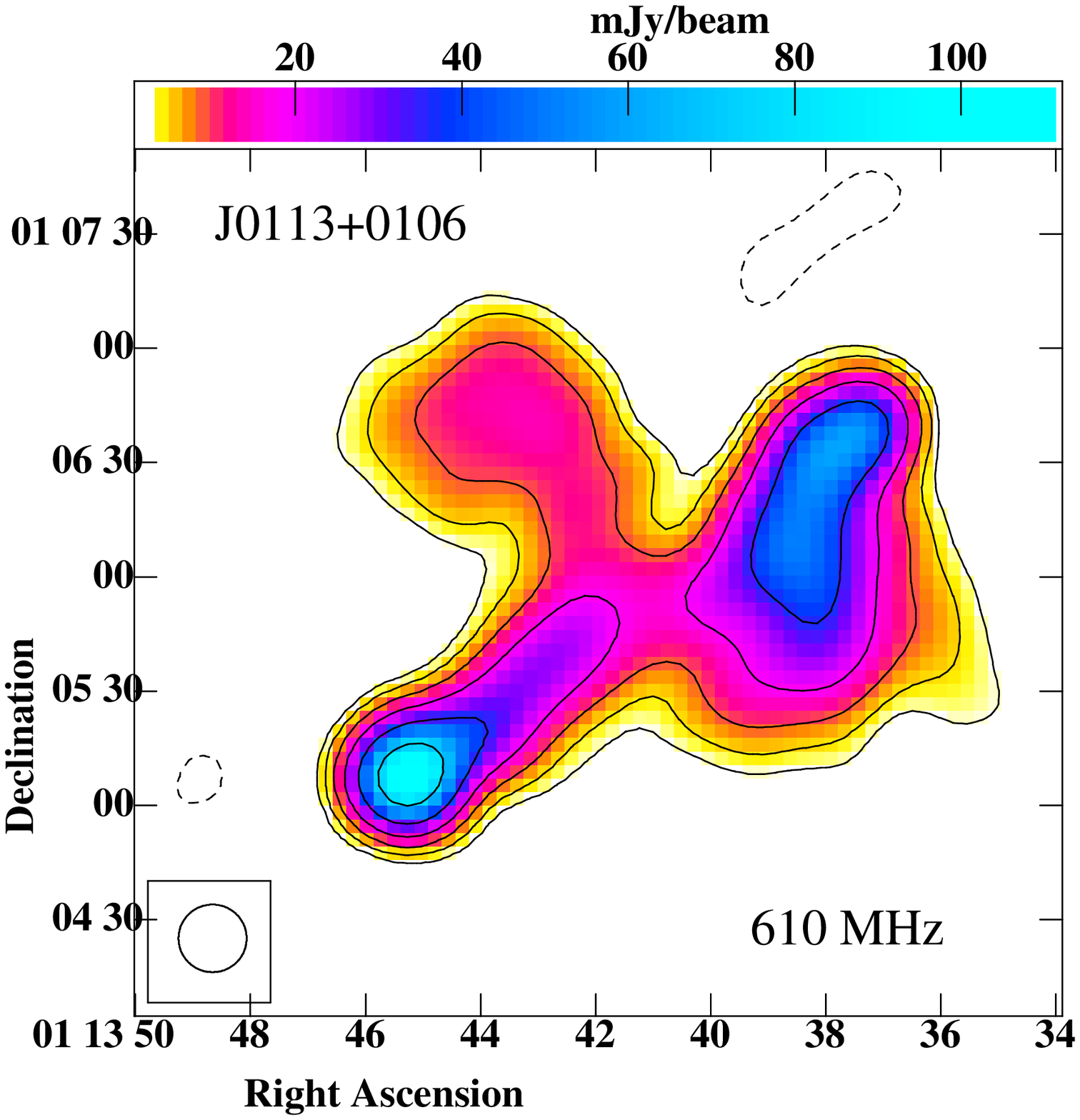} &
\includegraphics[height=5.1cm]{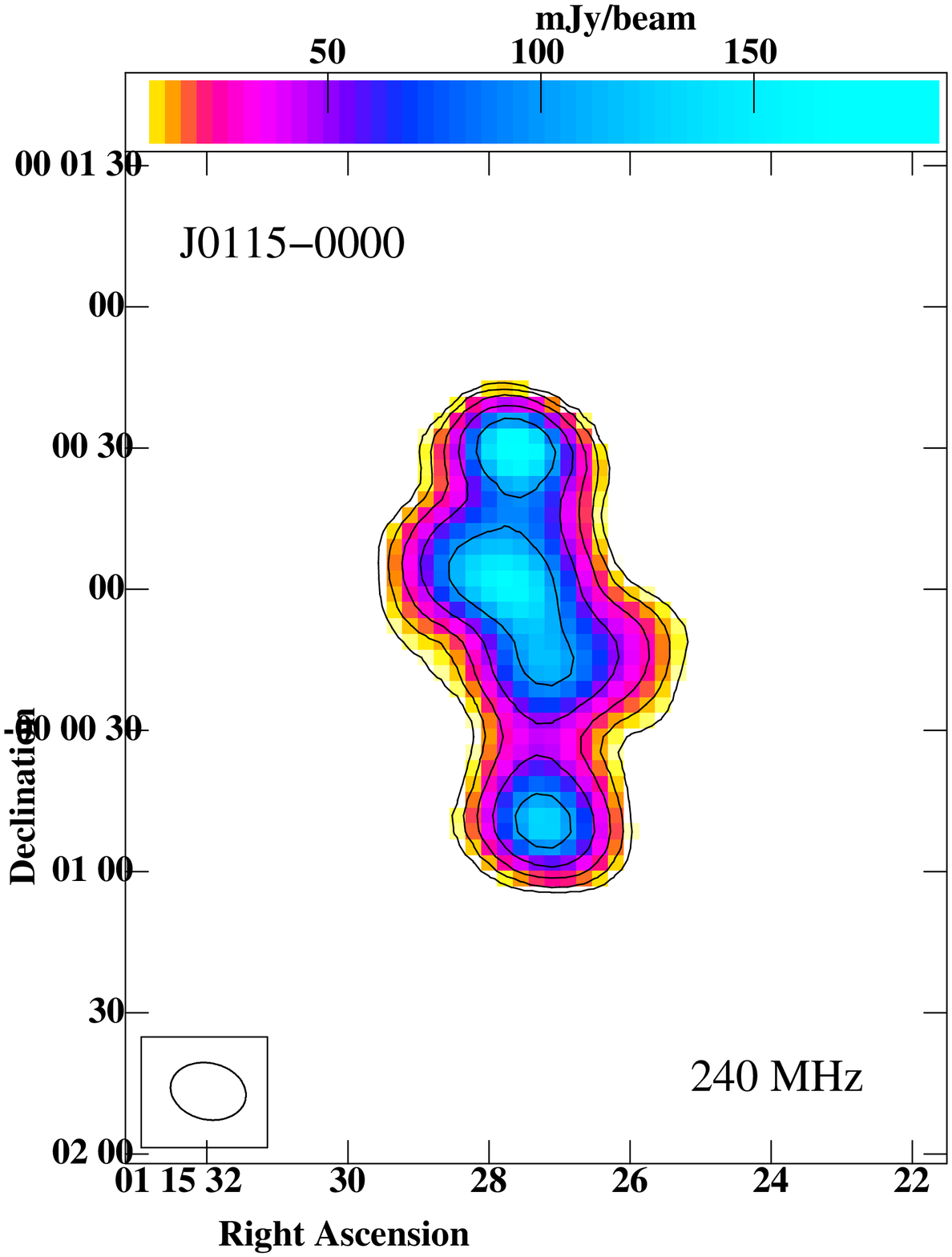} &
\includegraphics[height=5.1cm]{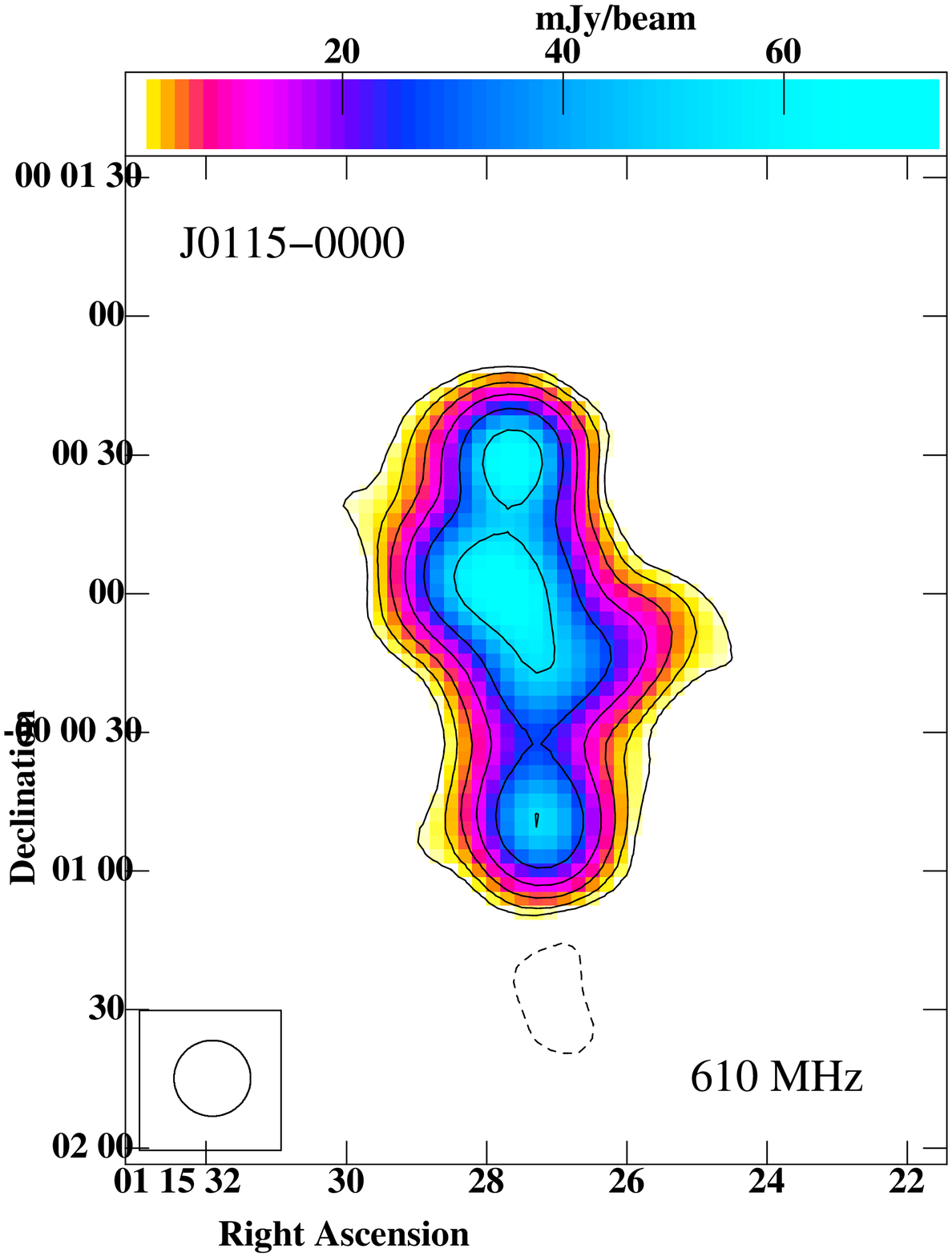} \\
\includegraphics[height=4.15cm]{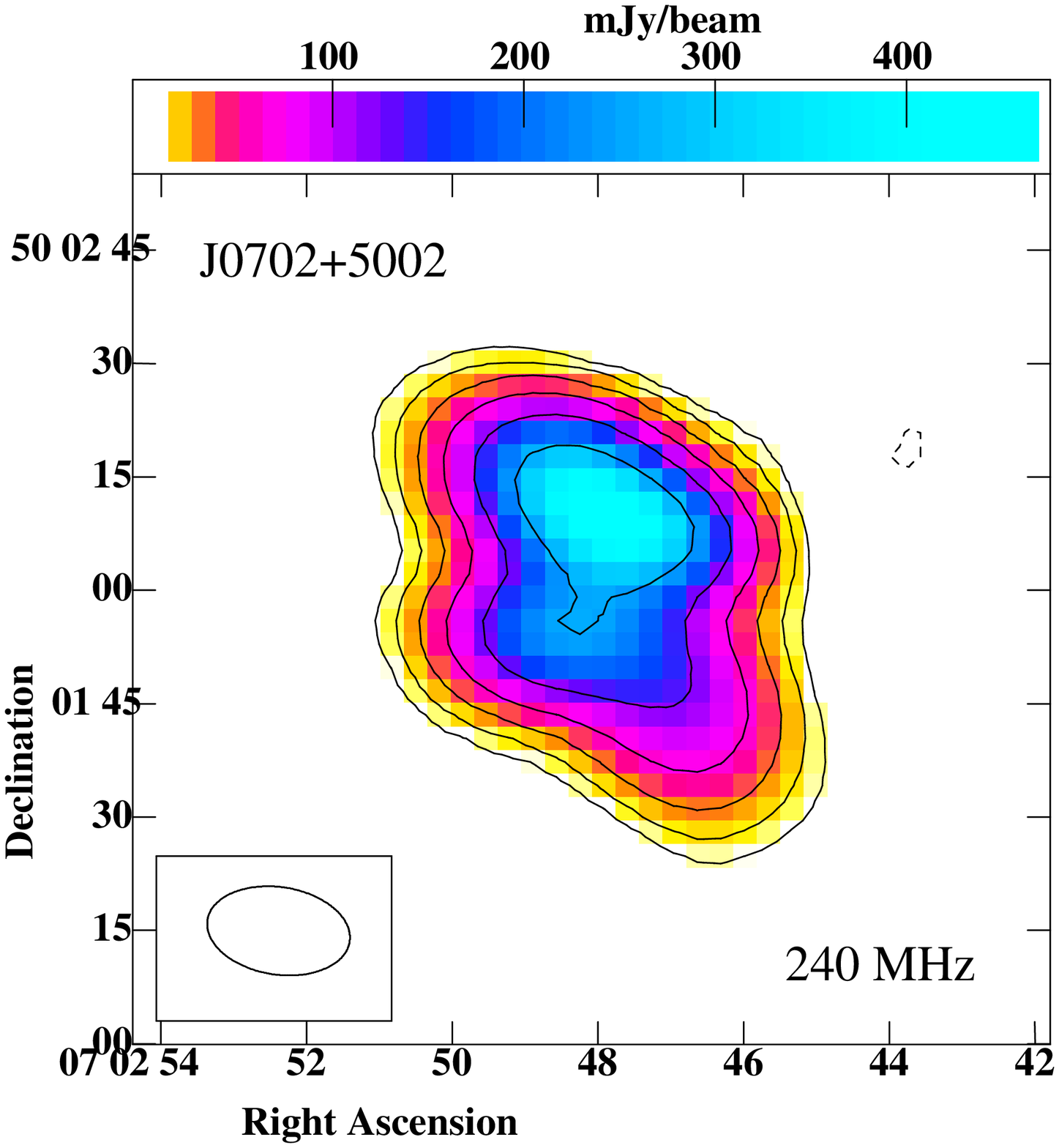} &
\includegraphics[height=4.15cm]{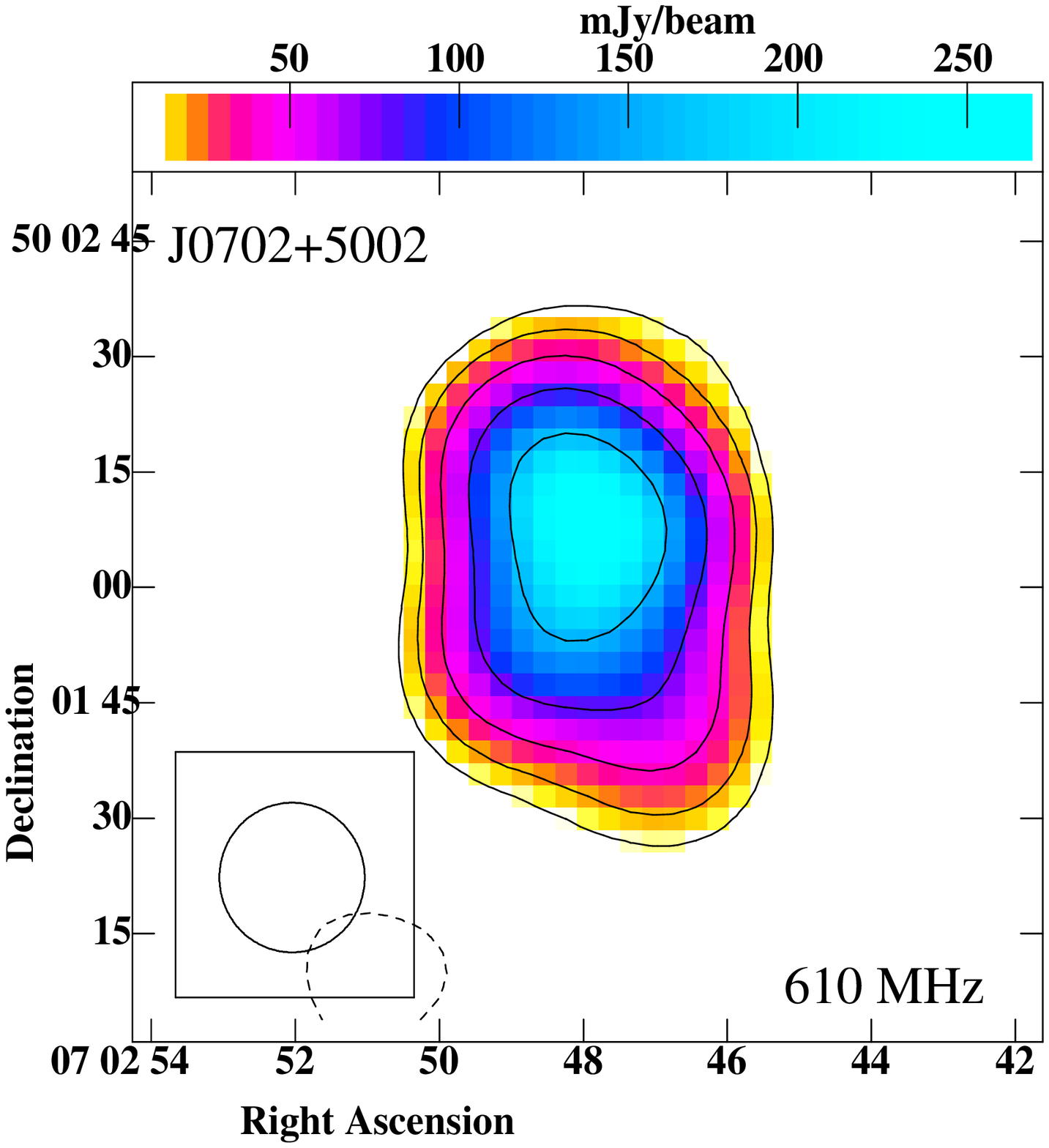} &
\includegraphics[height=4.25cm]{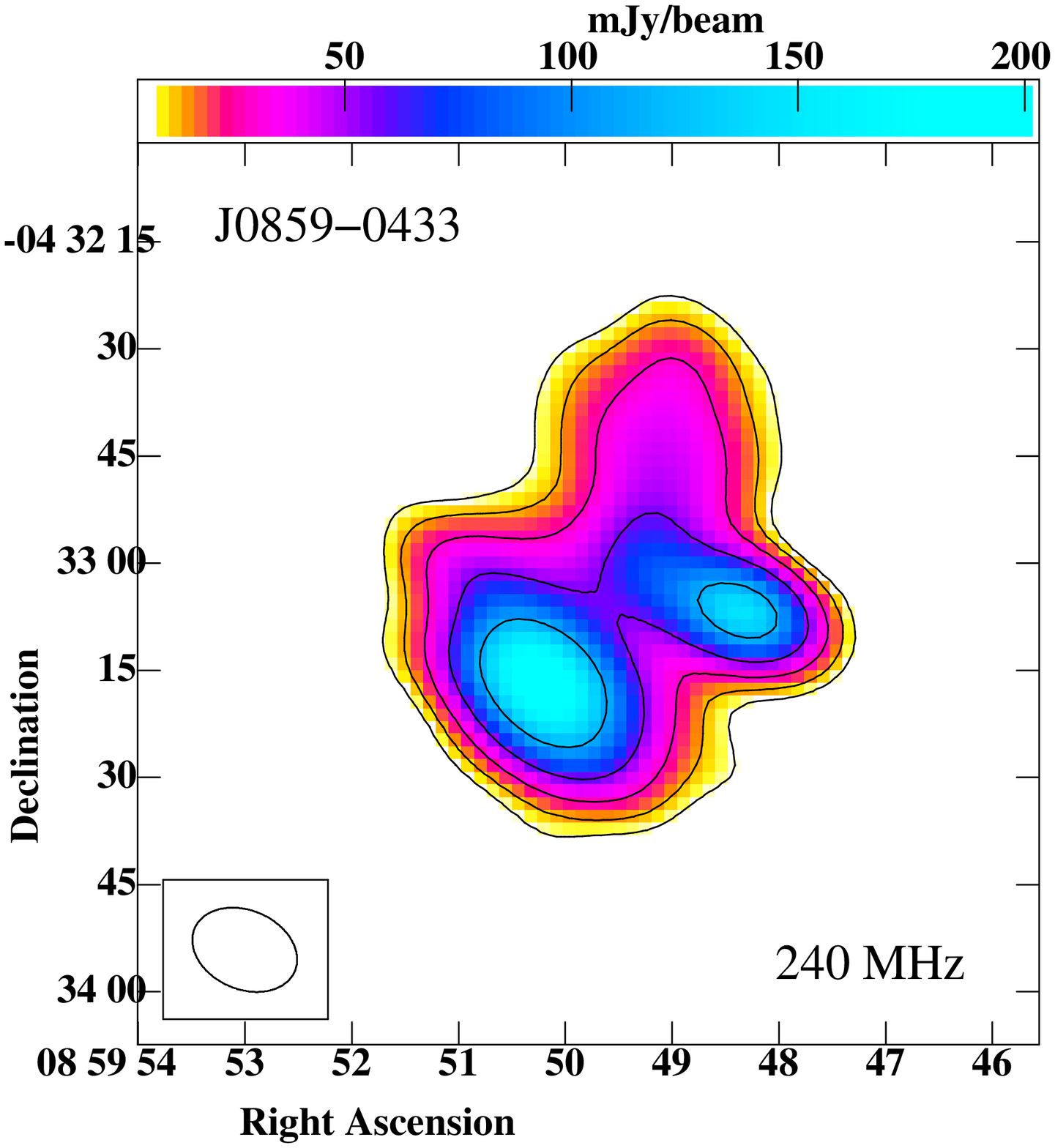} &
\includegraphics[height=4.25cm]{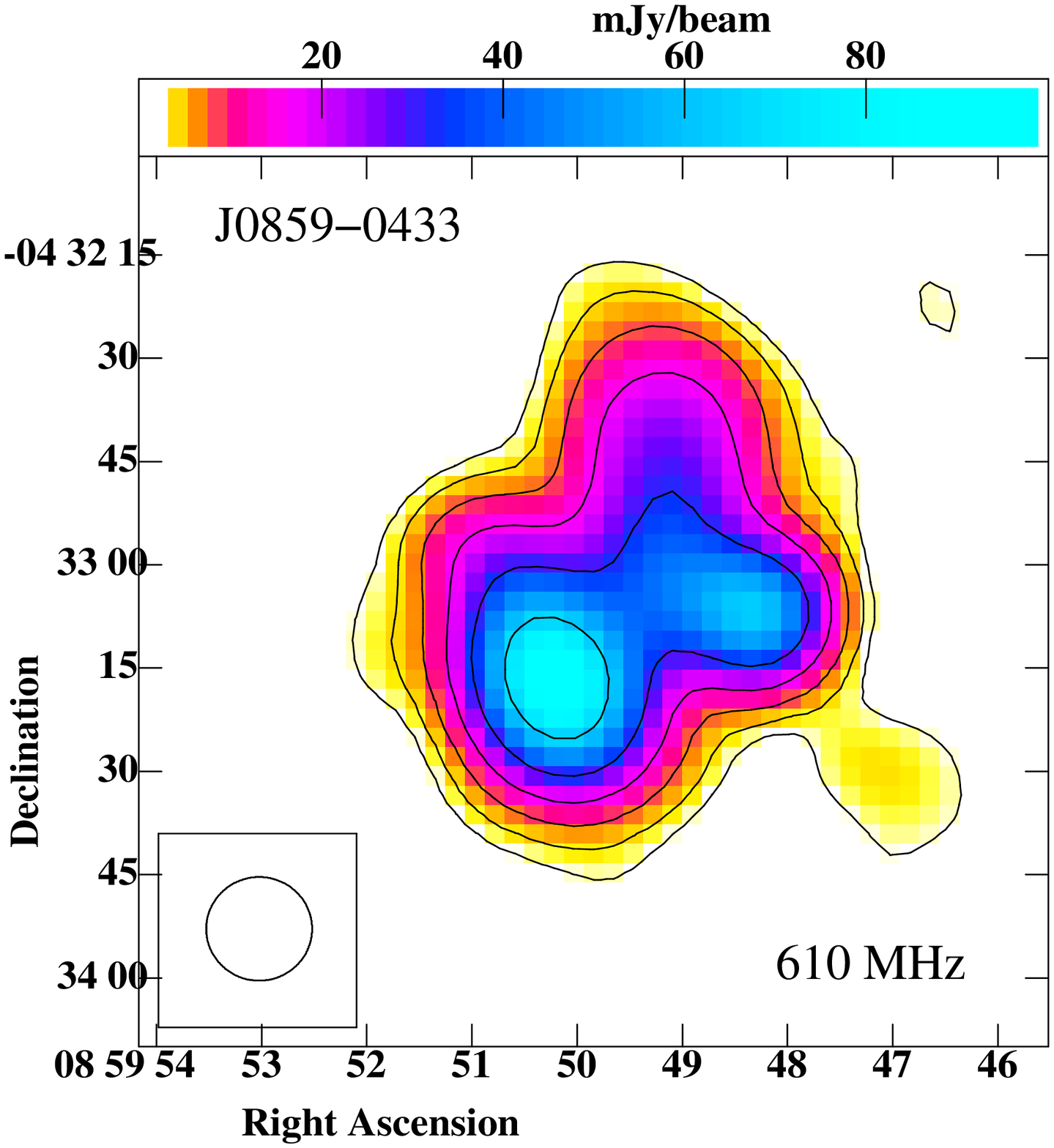} \\
\includegraphics[height=4.1cm]{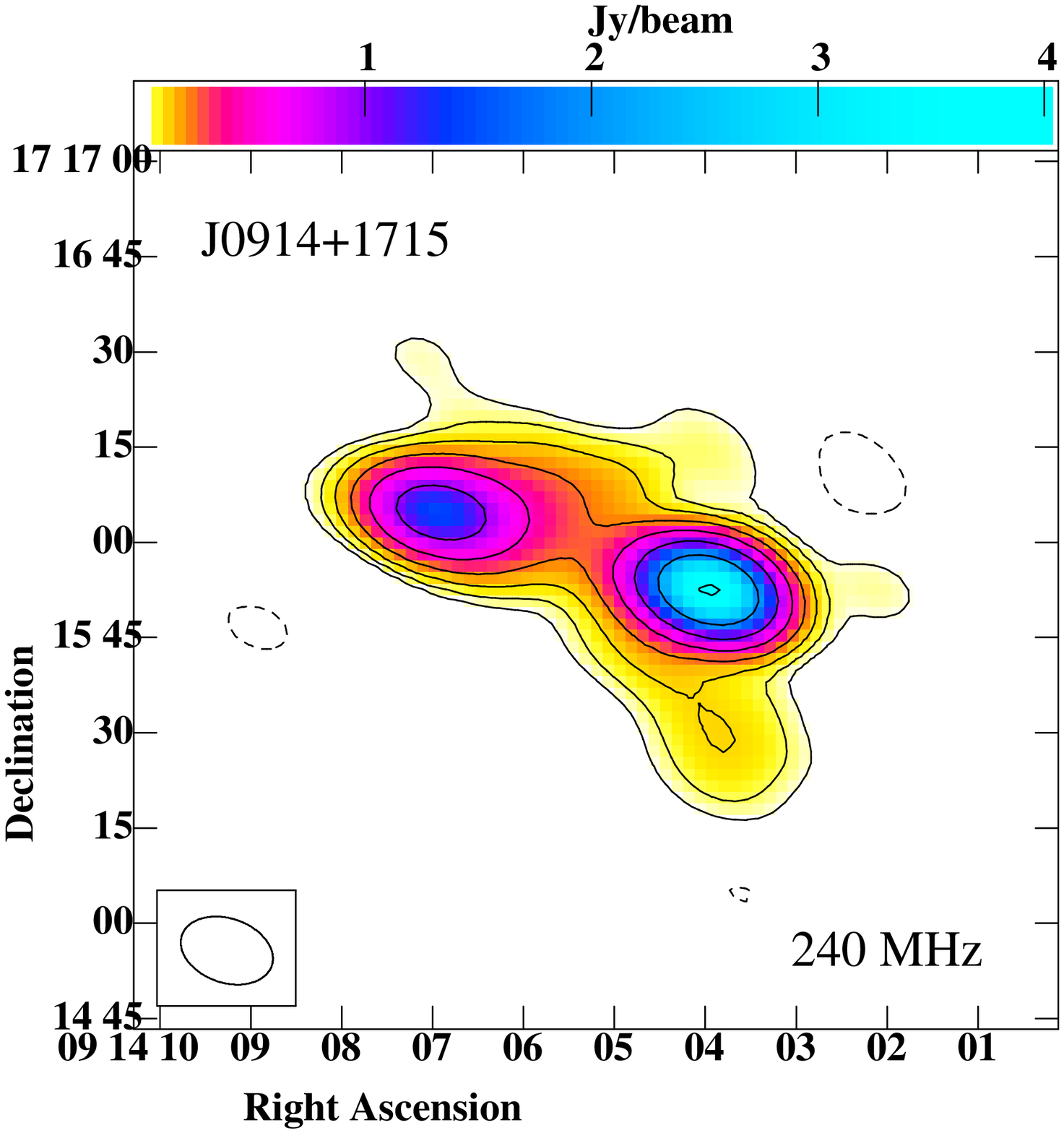} &
\includegraphics[height=4cm]{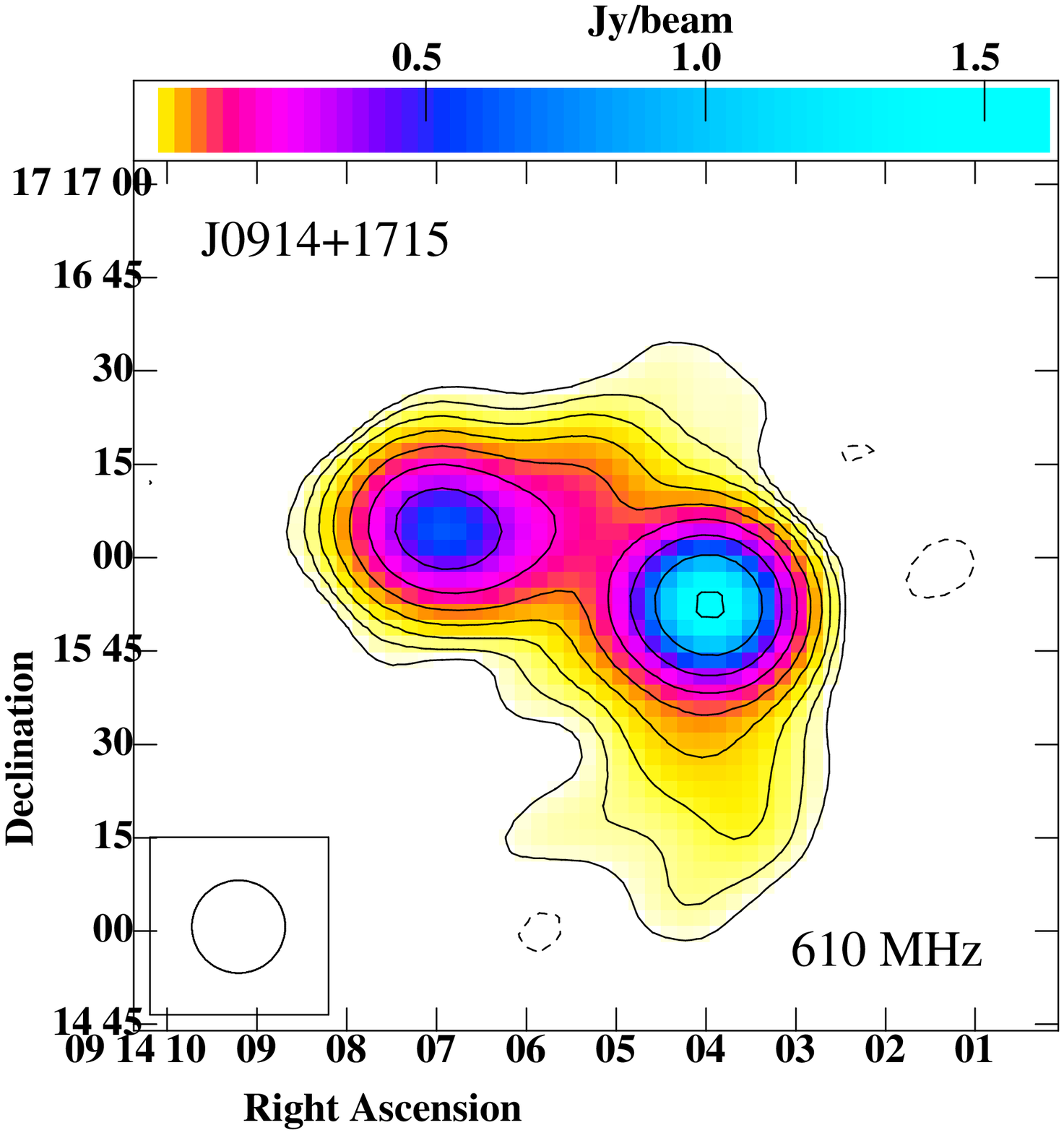} &
\includegraphics[height=4cm]{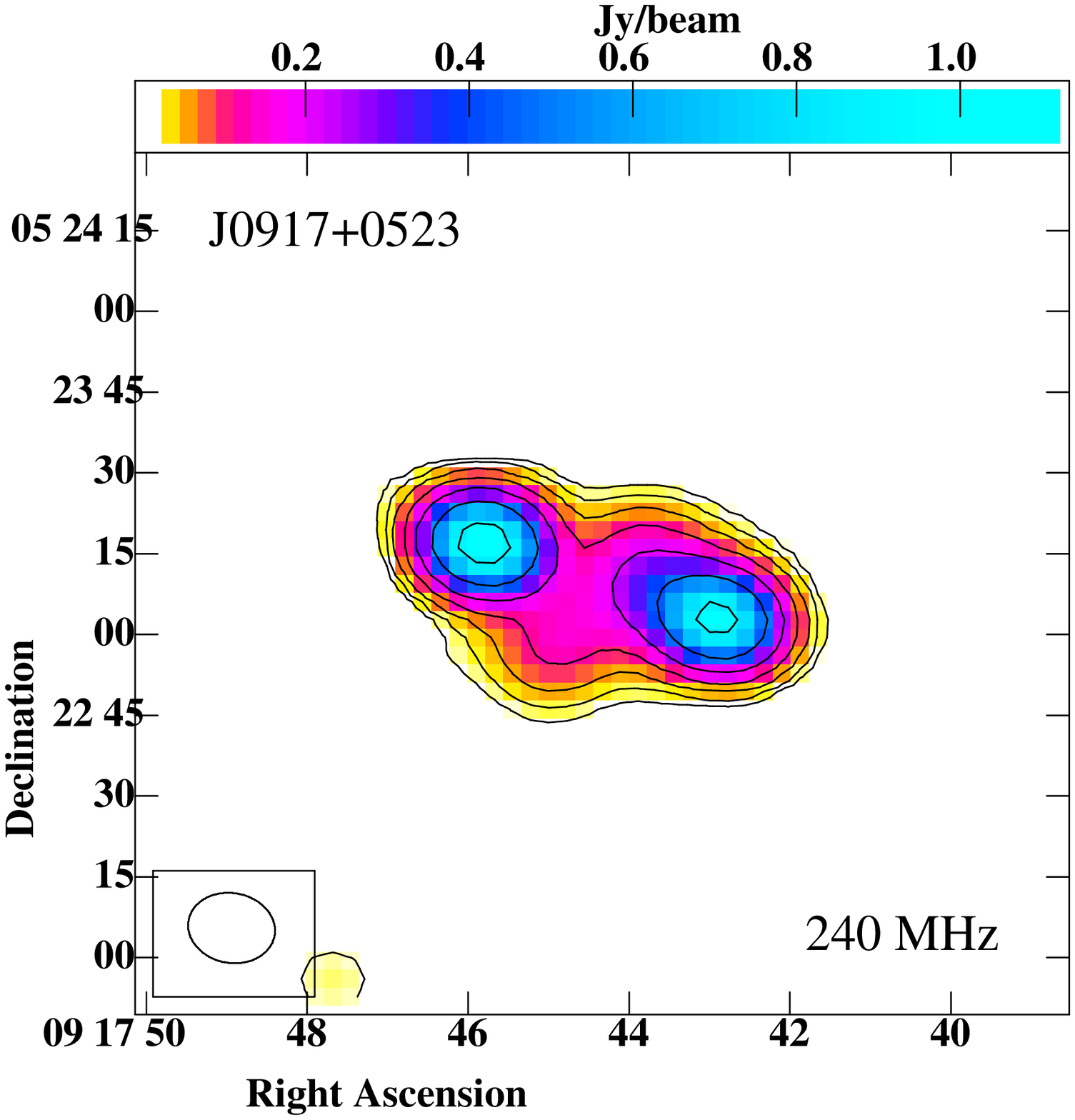} &
\includegraphics[height=4cm]{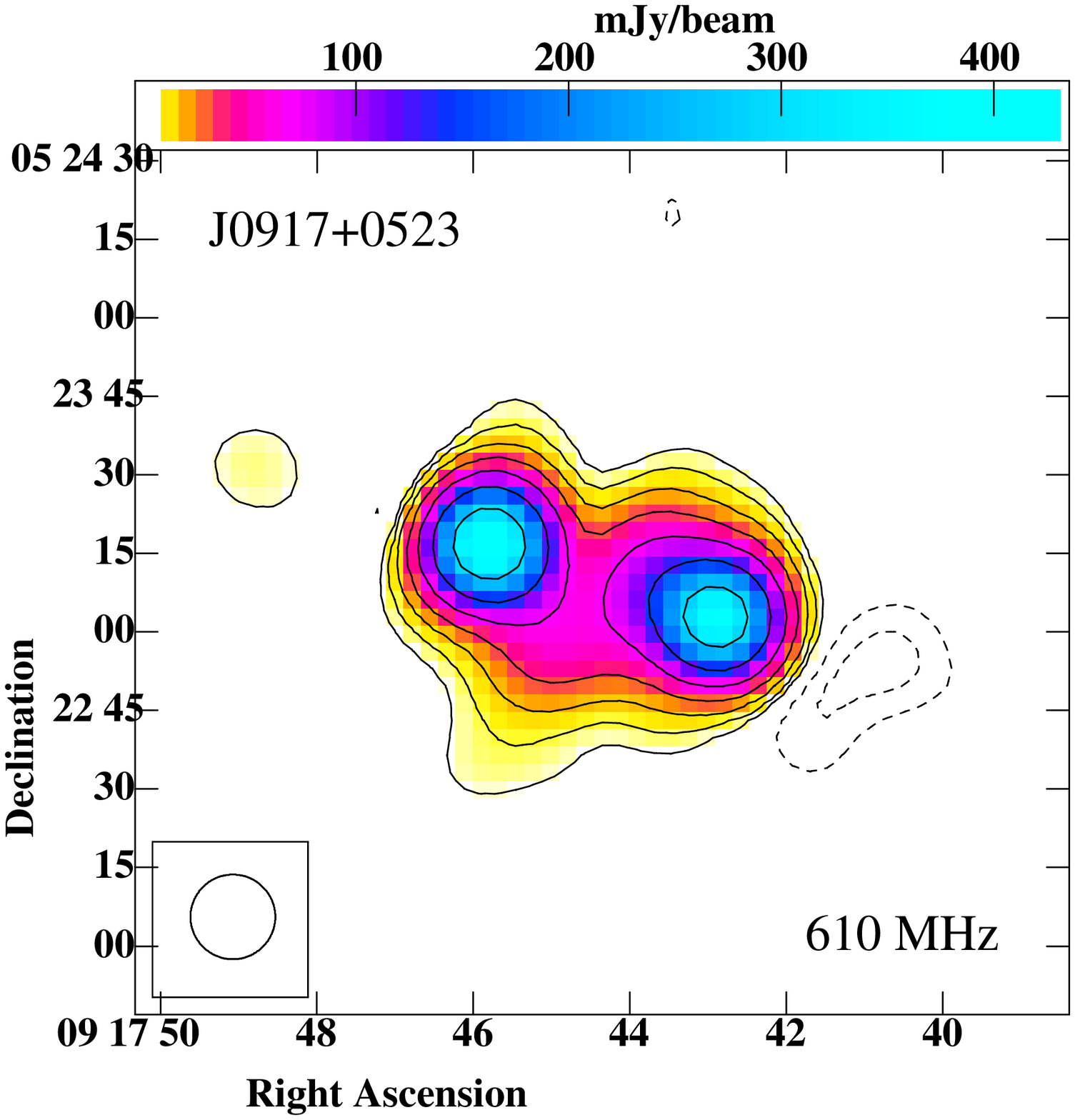} \\
\includegraphics[height=4cm]{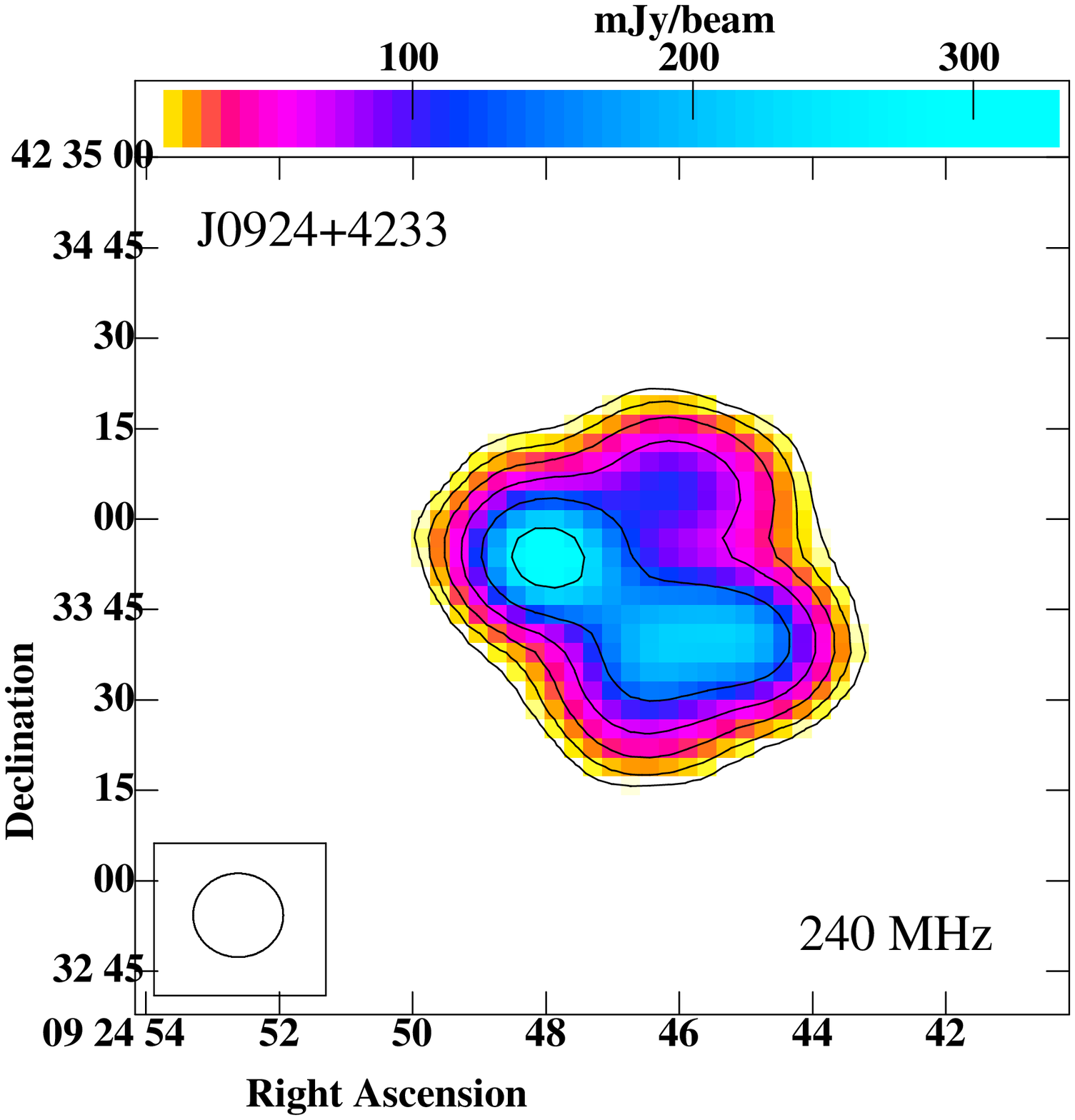} &
\includegraphics[height=4cm]{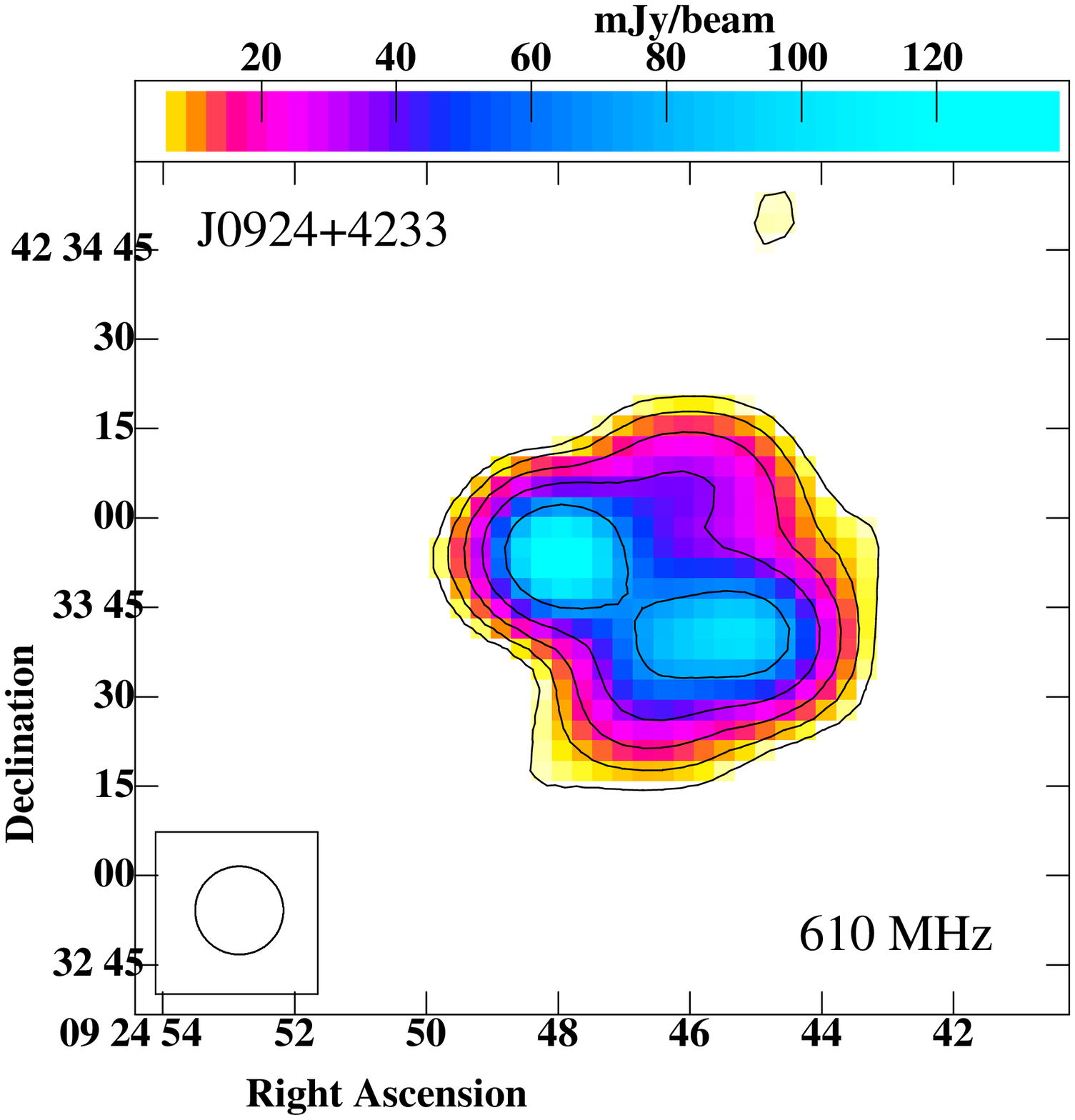} &
\includegraphics[height=4.1cm]{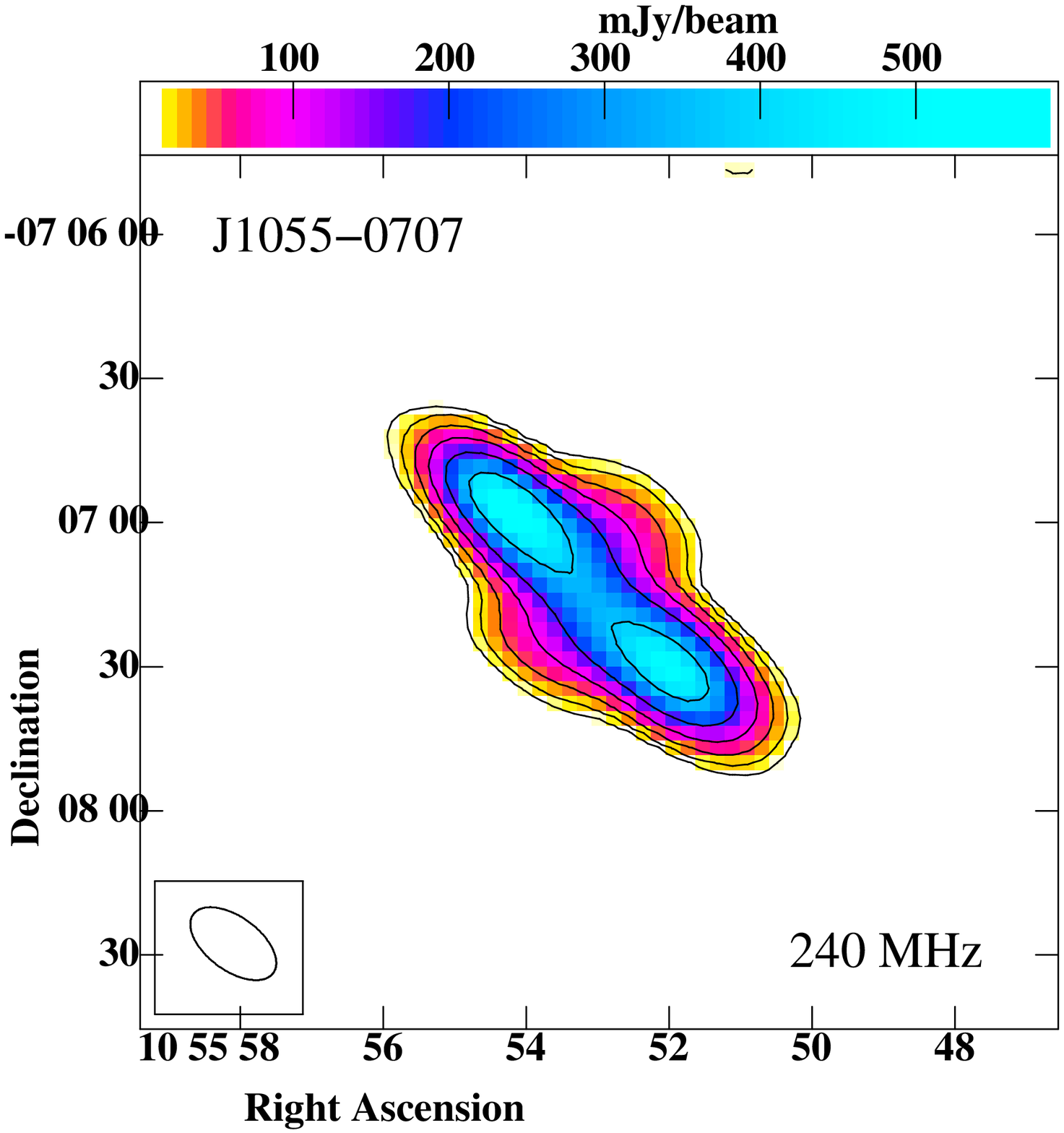} &
\includegraphics[height=4.1cm]{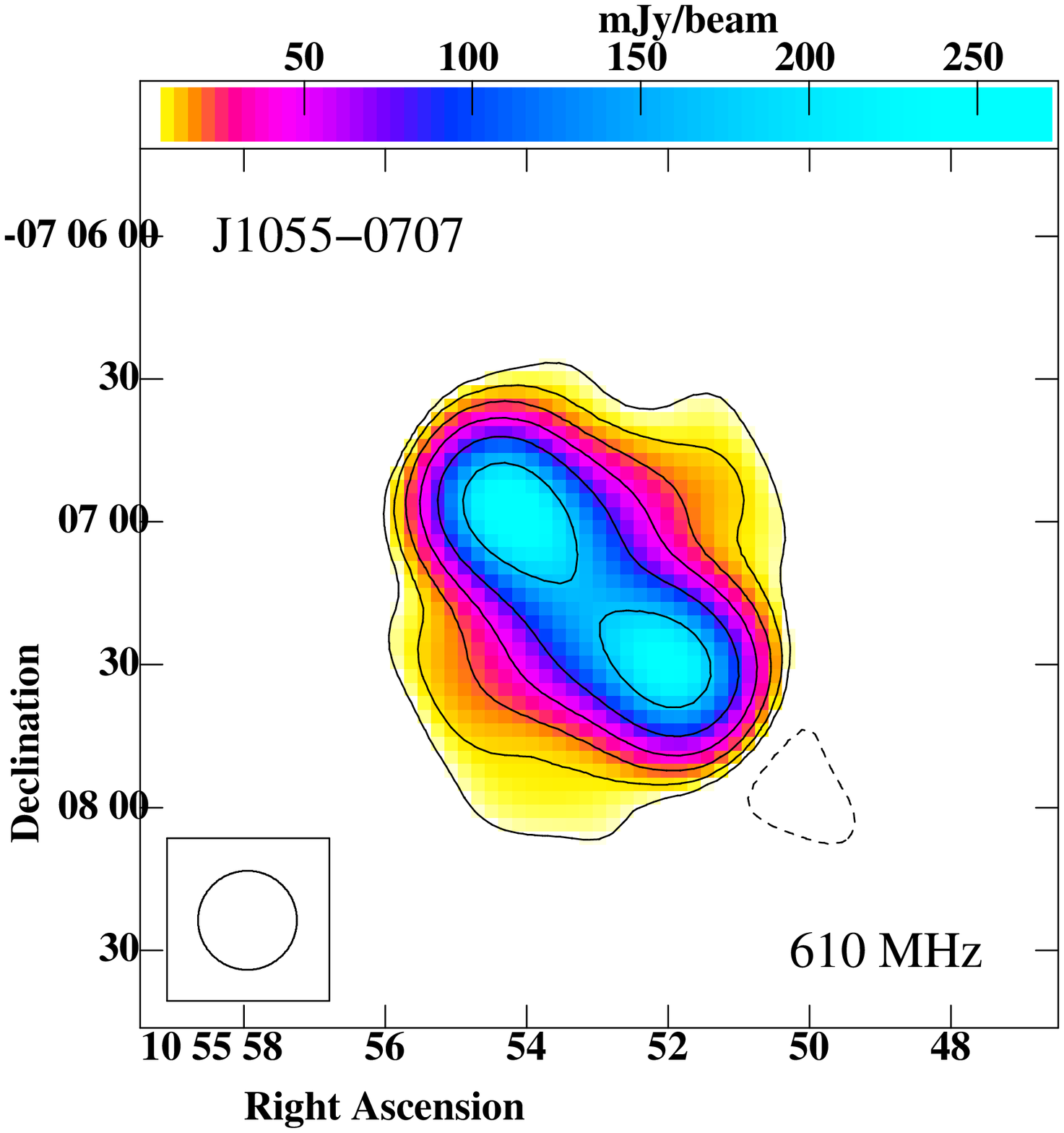} \\
\includegraphics[height=4.2cm]{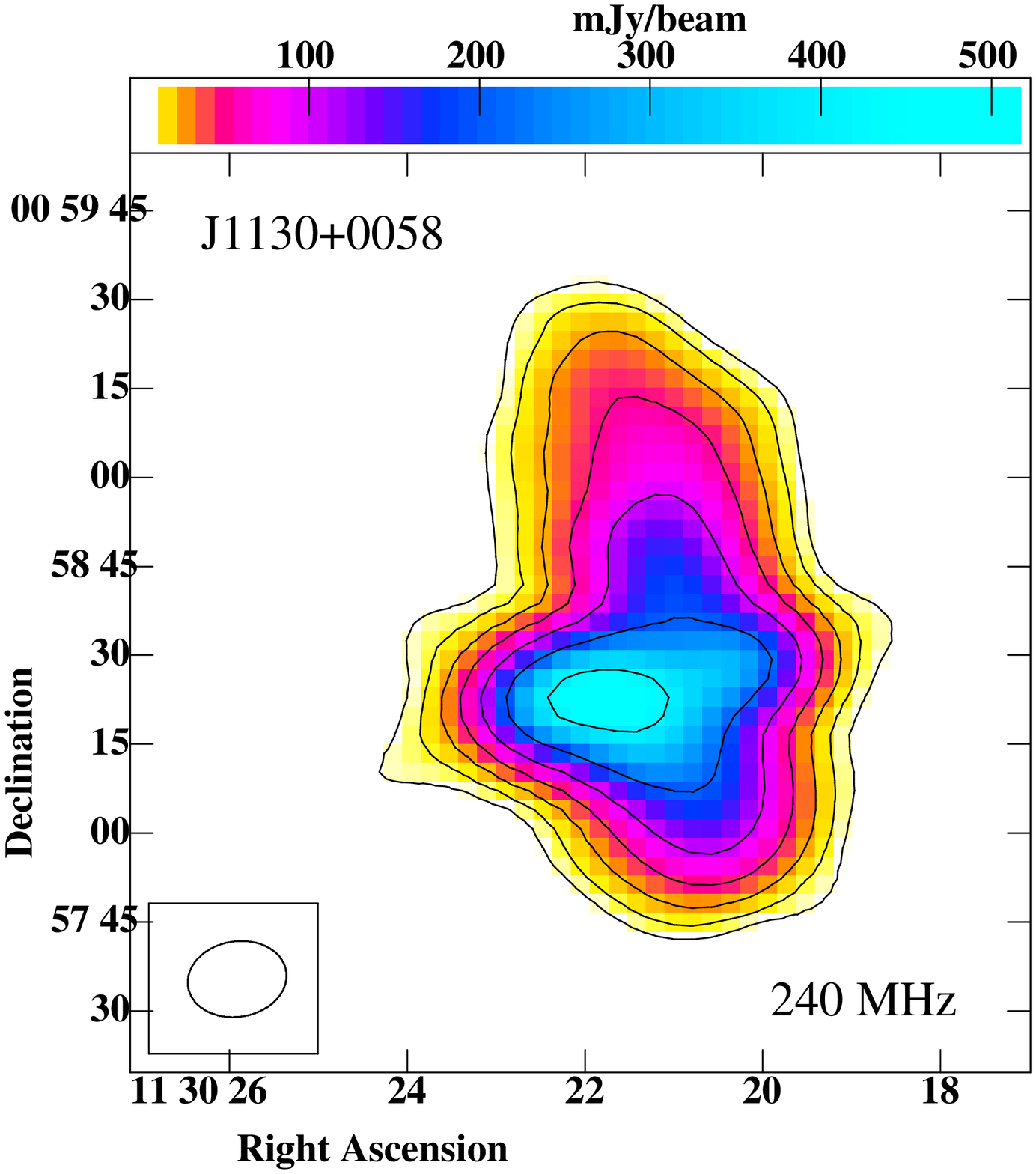} &
\includegraphics[height=4.2cm]{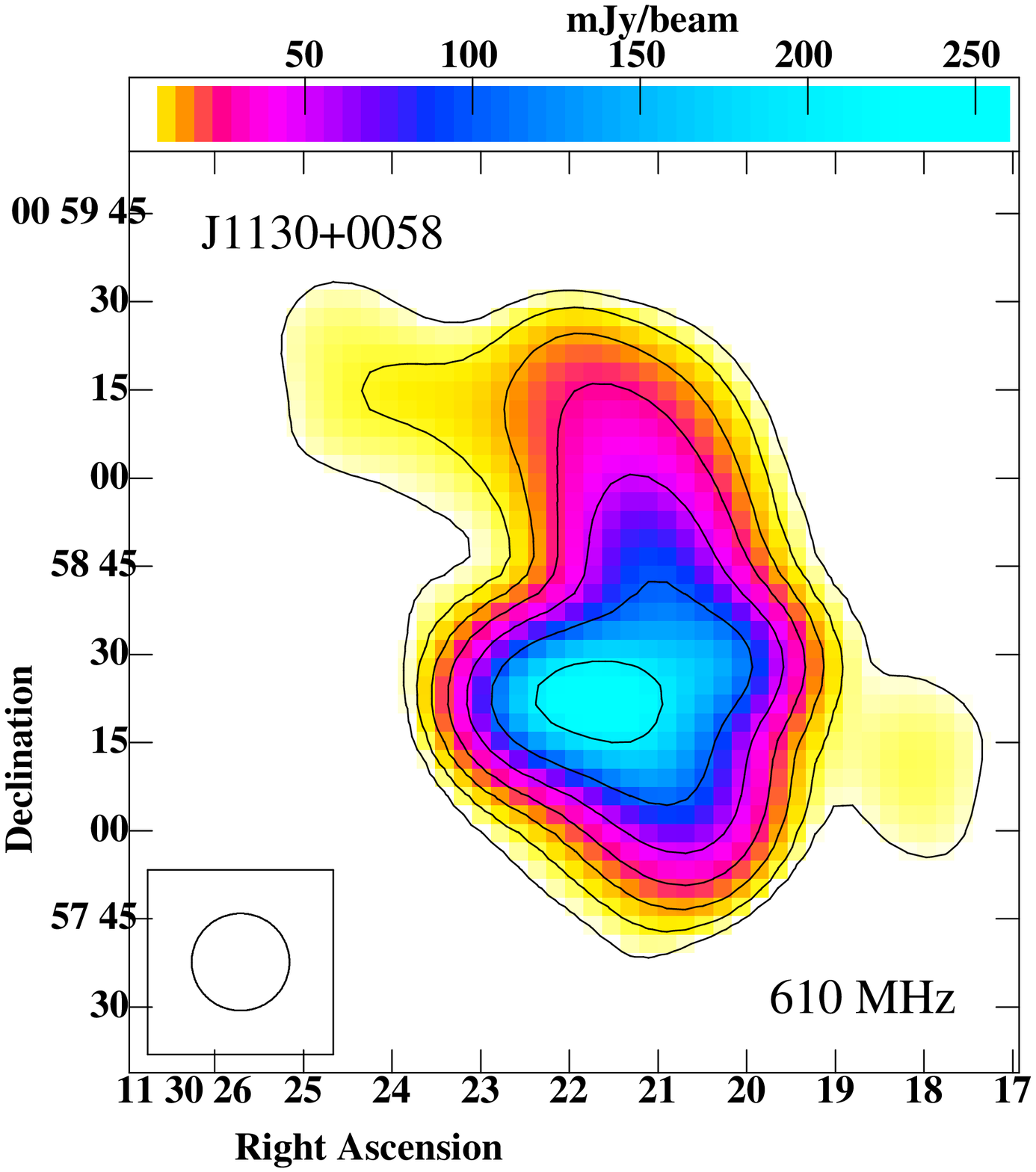} &
\includegraphics[height=4.1cm]{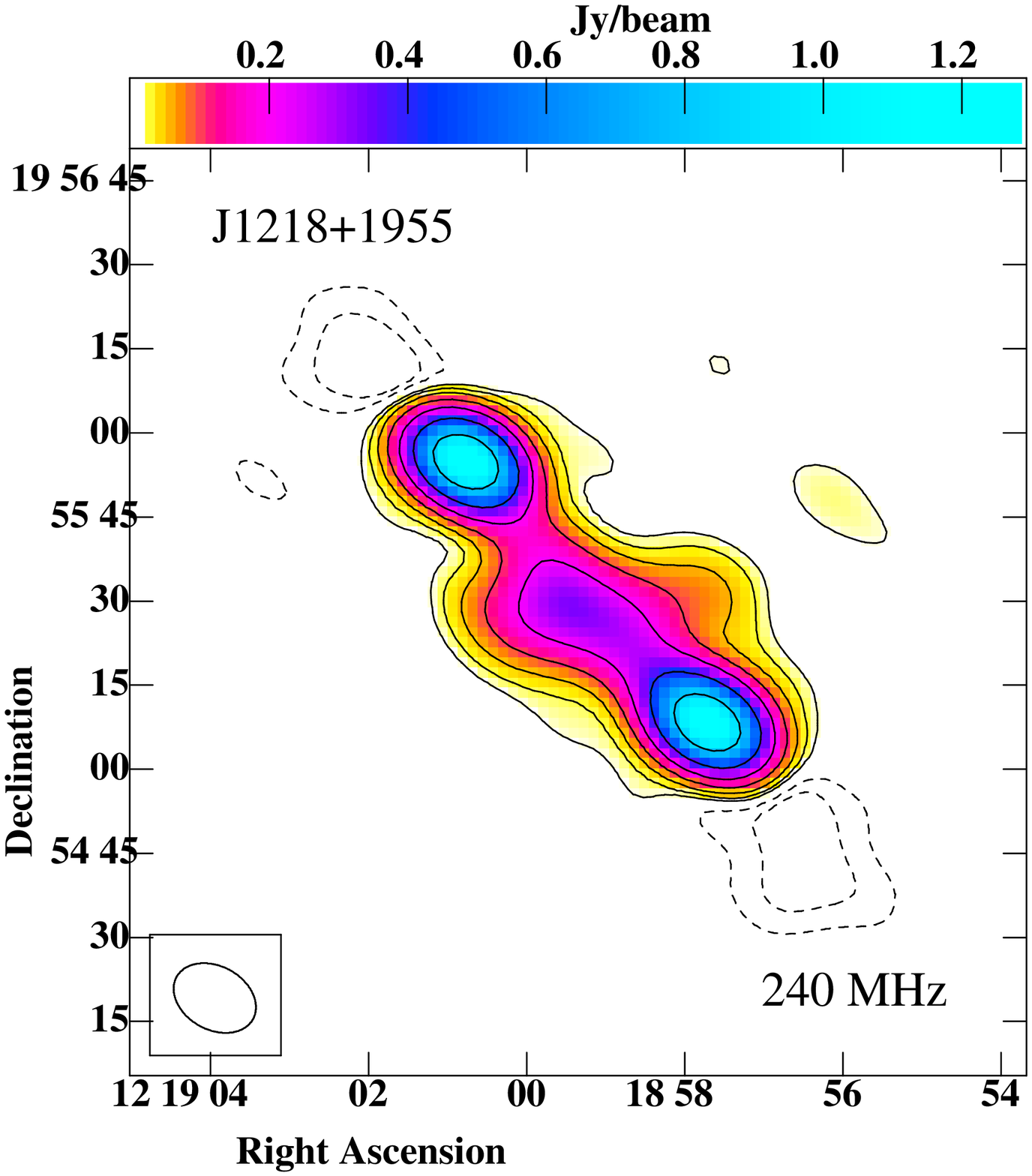} &
\includegraphics[height=4.1cm]{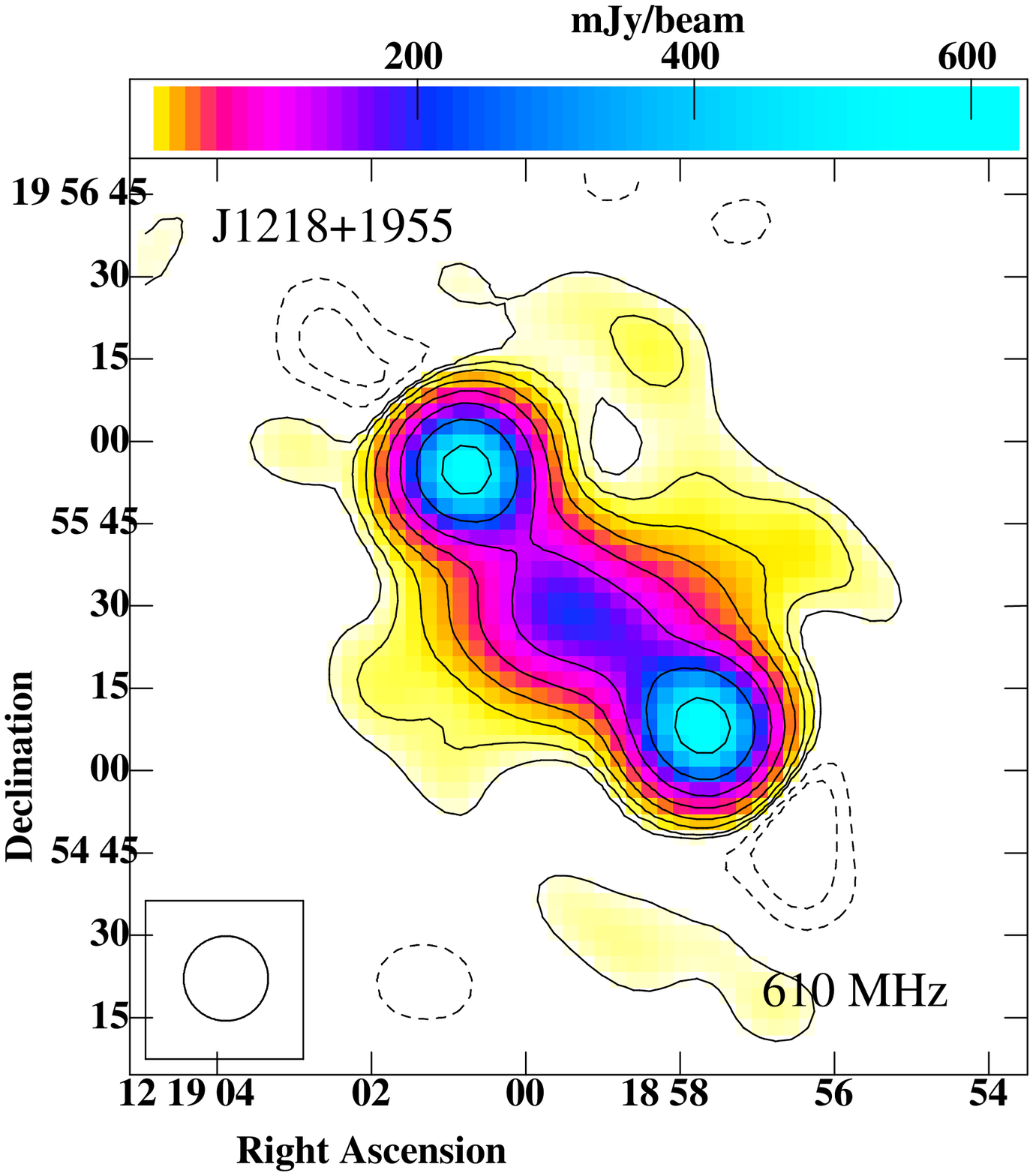}
\end{tabular}
\caption{The full synthesis GMRT images at 240 MHz (first and third columns)) and lower resolution versions of the 610 MHz data (second and fourth columns).
Map parameters are summarized in Table~\ref{maps-appendix}.}
    \label{low-res-collage}
\end{center}
\end{figure*}

\addtocounter{figure}{-1}
\begin{figure*}[ht]
\begin{center}
\begin{tabular}{cccc}
\includegraphics[height=4cm]{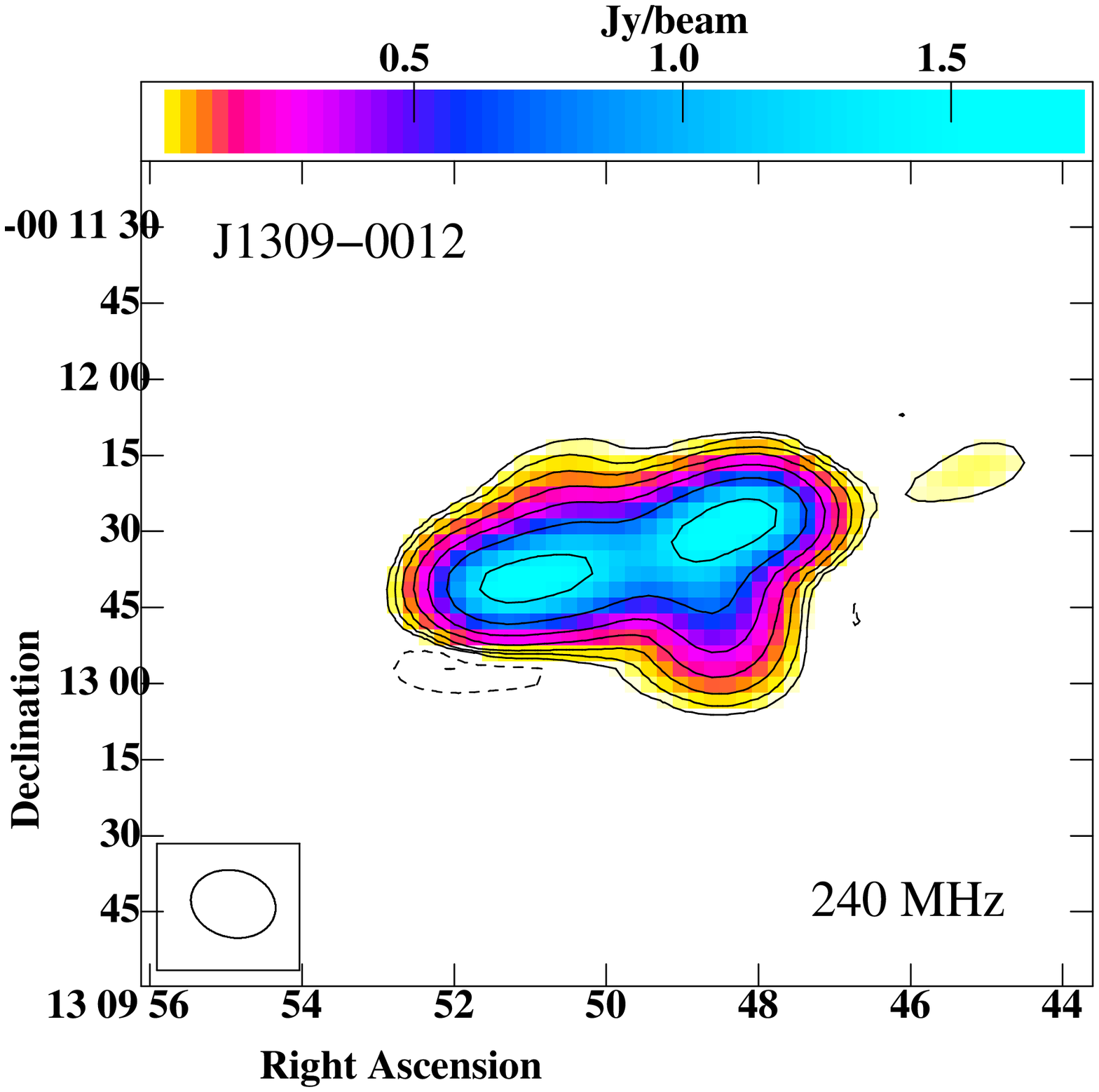} &
\includegraphics[height=4cm]{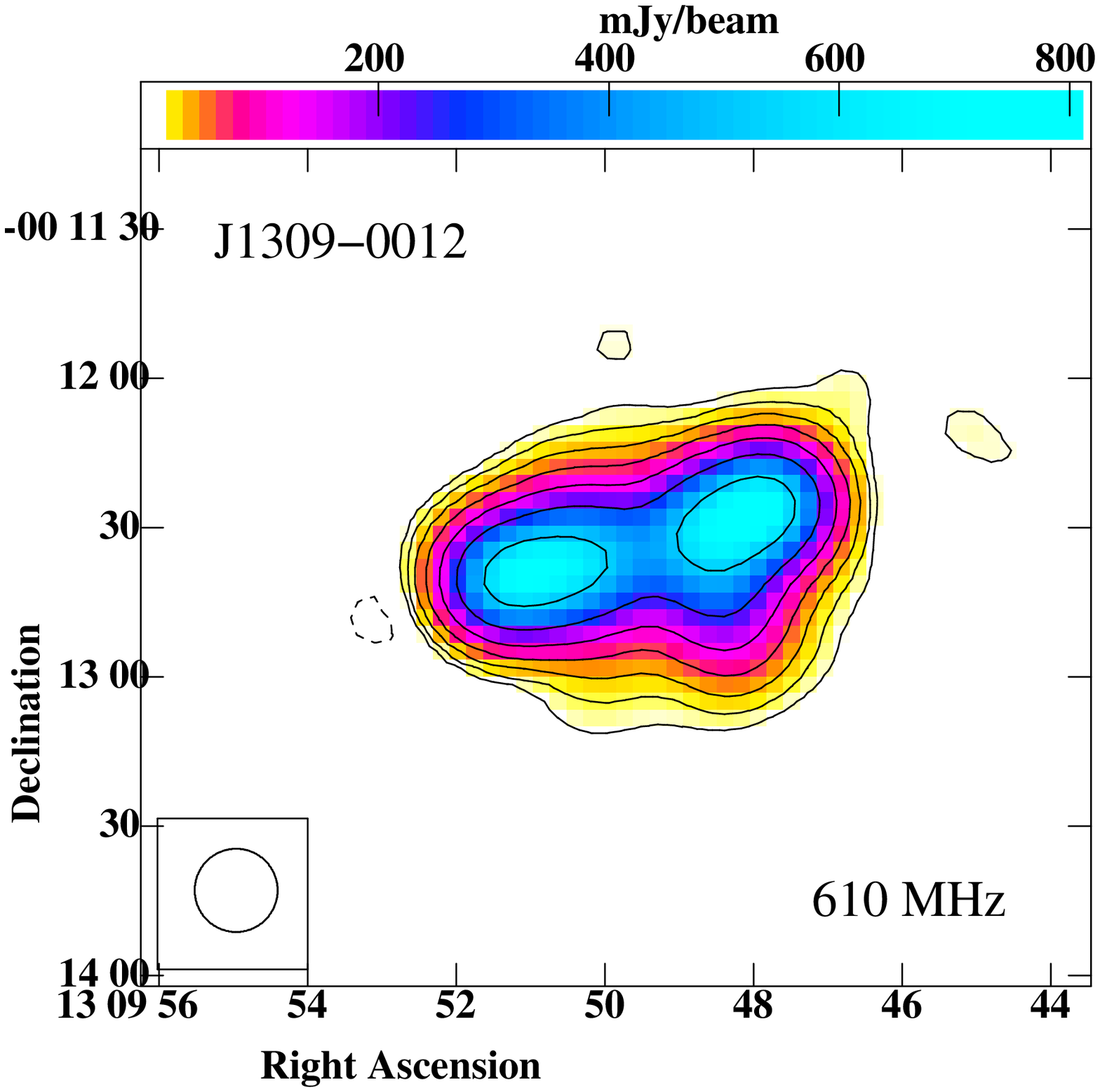} &
\includegraphics[height=3.9cm]{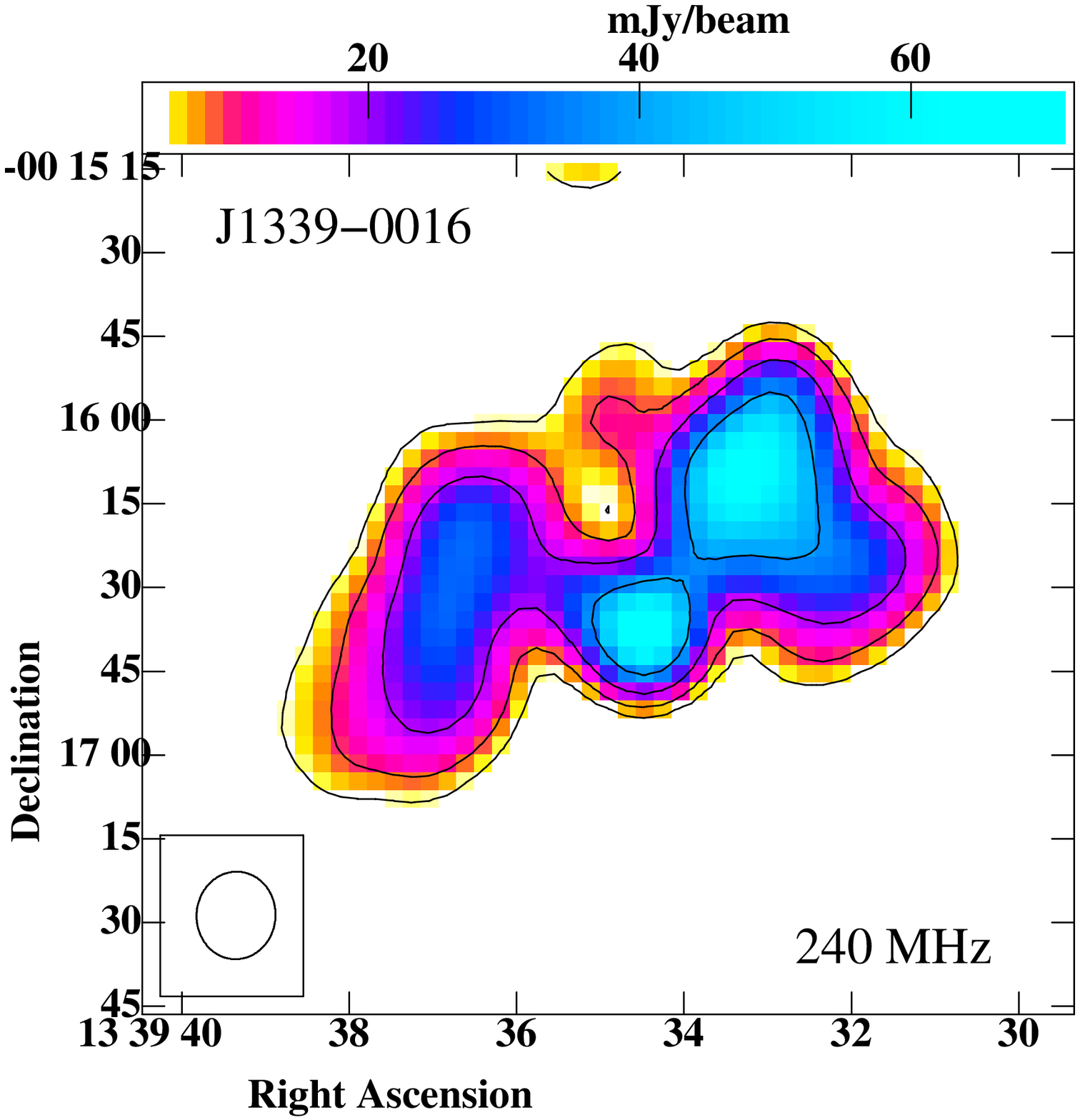} &
\includegraphics[height=3.9cm]{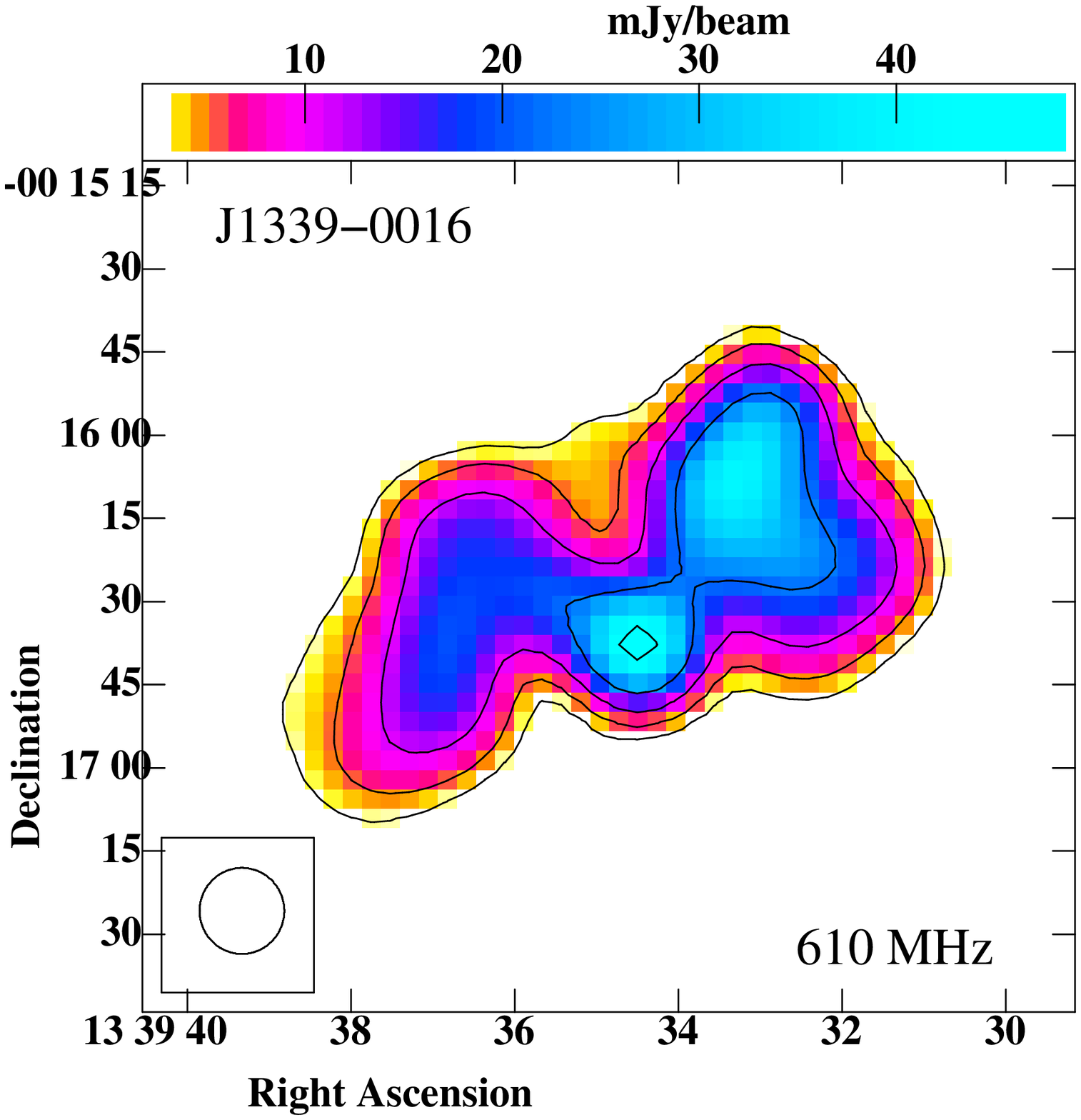} \\
\includegraphics[height=4.1cm]{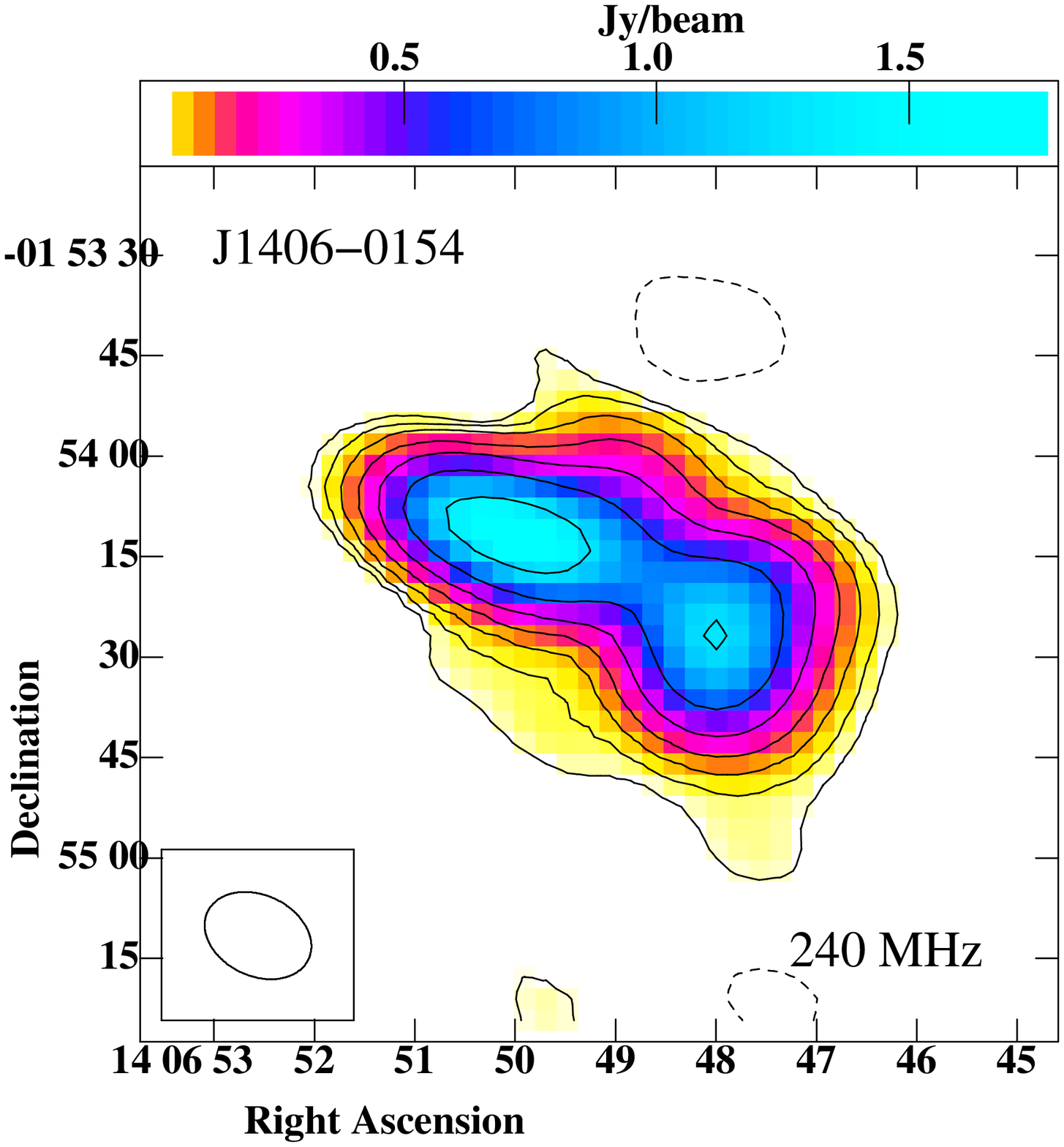} &
\includegraphics[height=4.1cm]{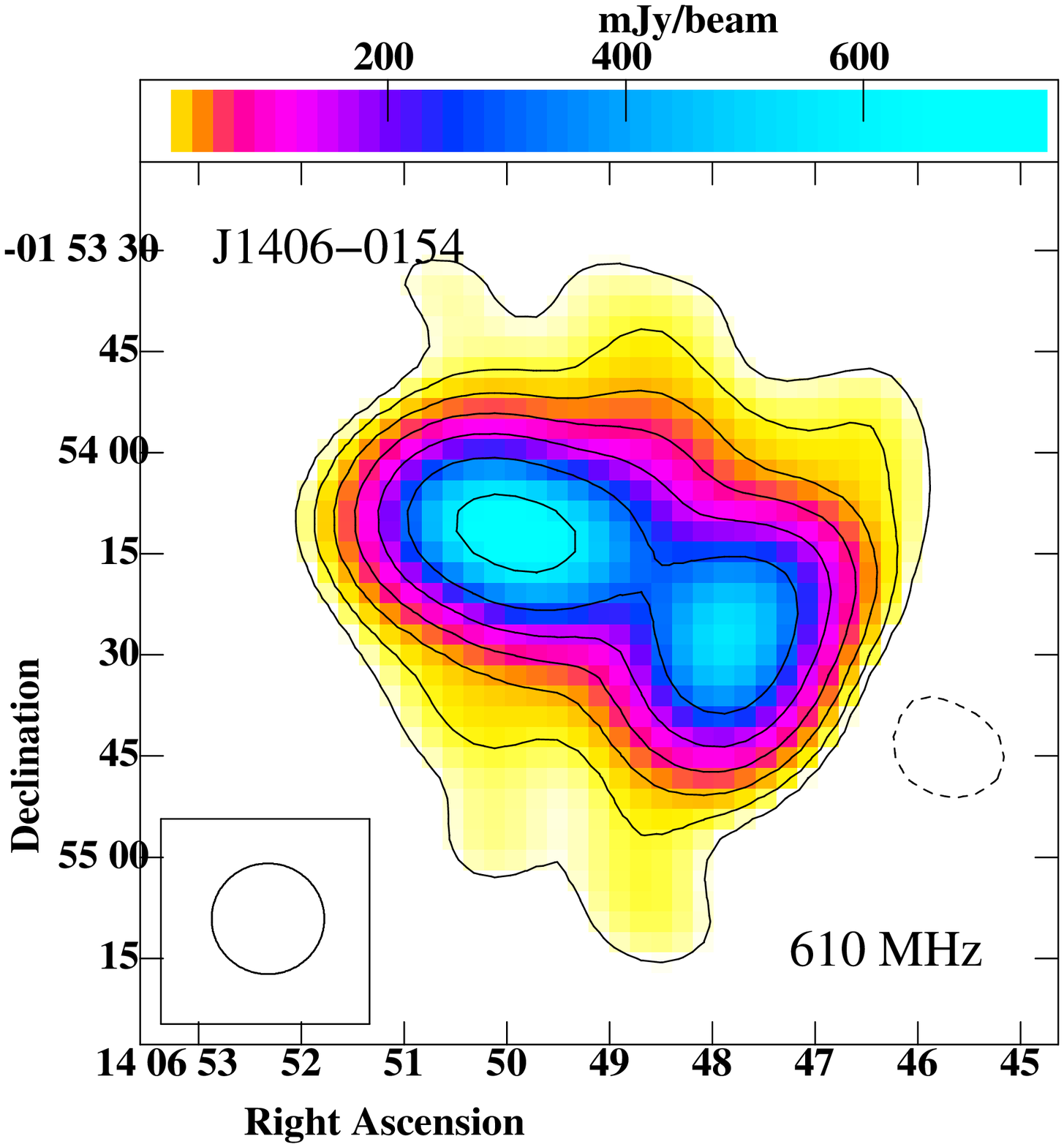} &
\includegraphics[height=4cm]{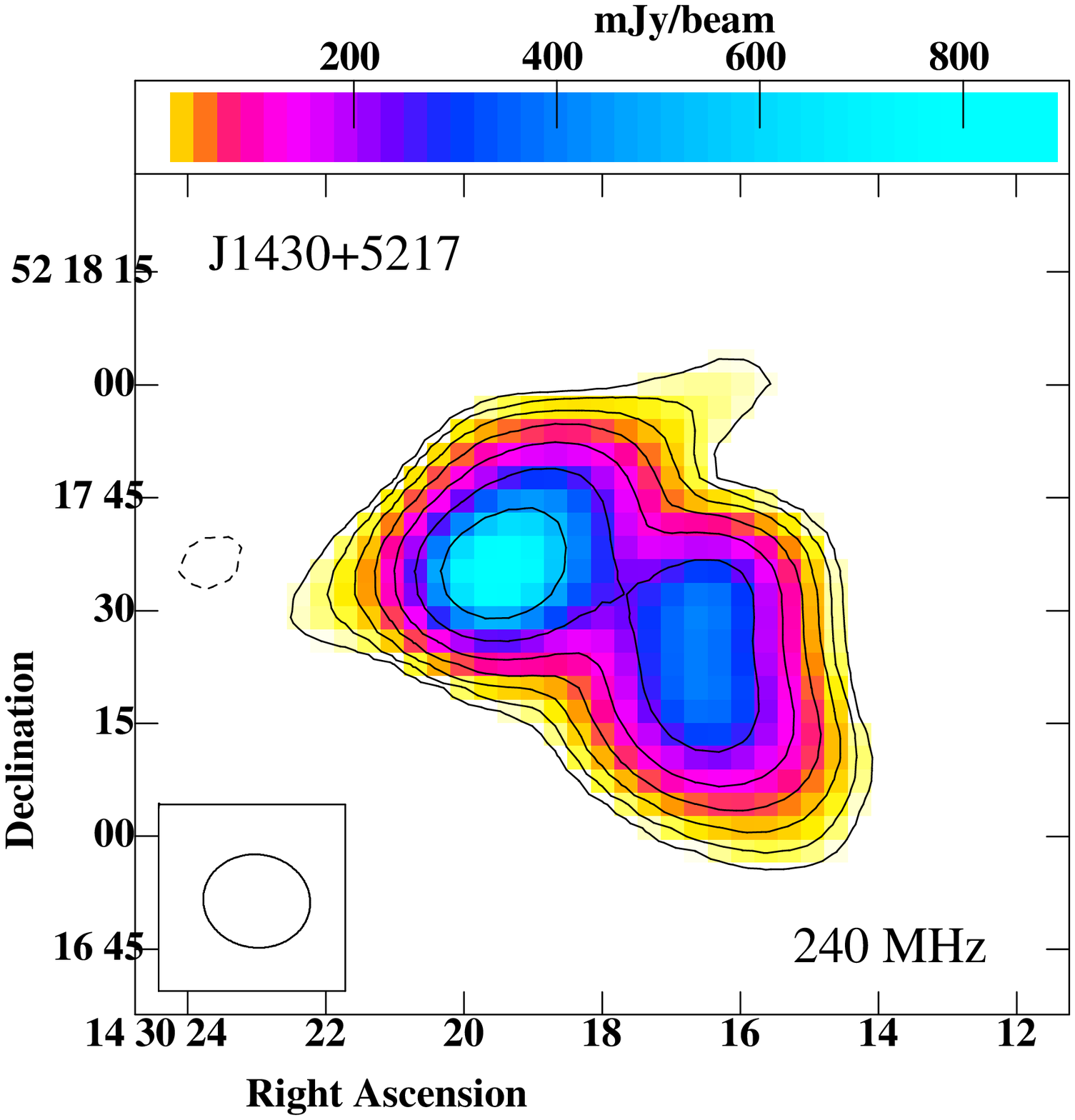} &
\includegraphics[height=4cm]{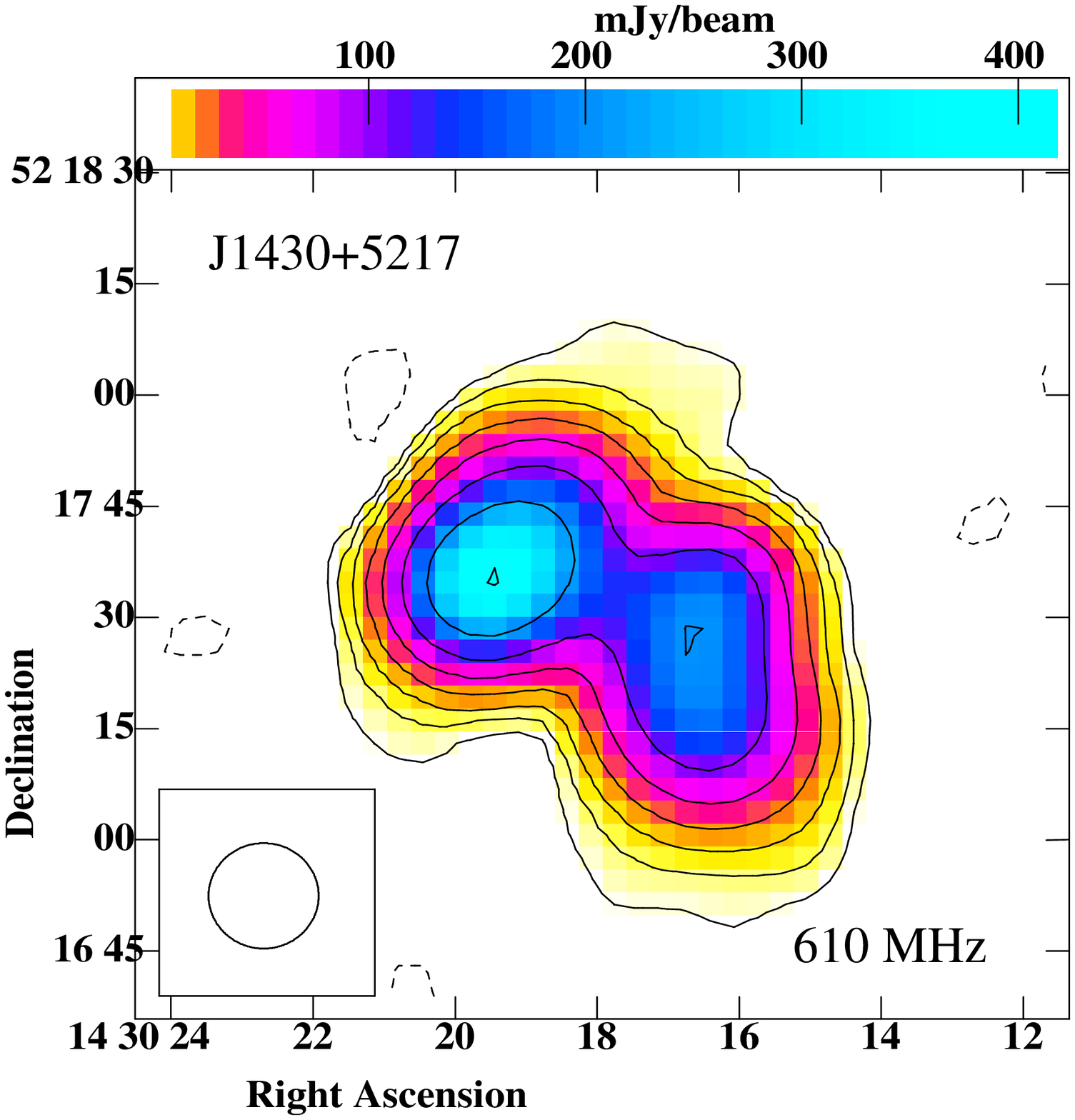} \\
\includegraphics[height=4.2cm]{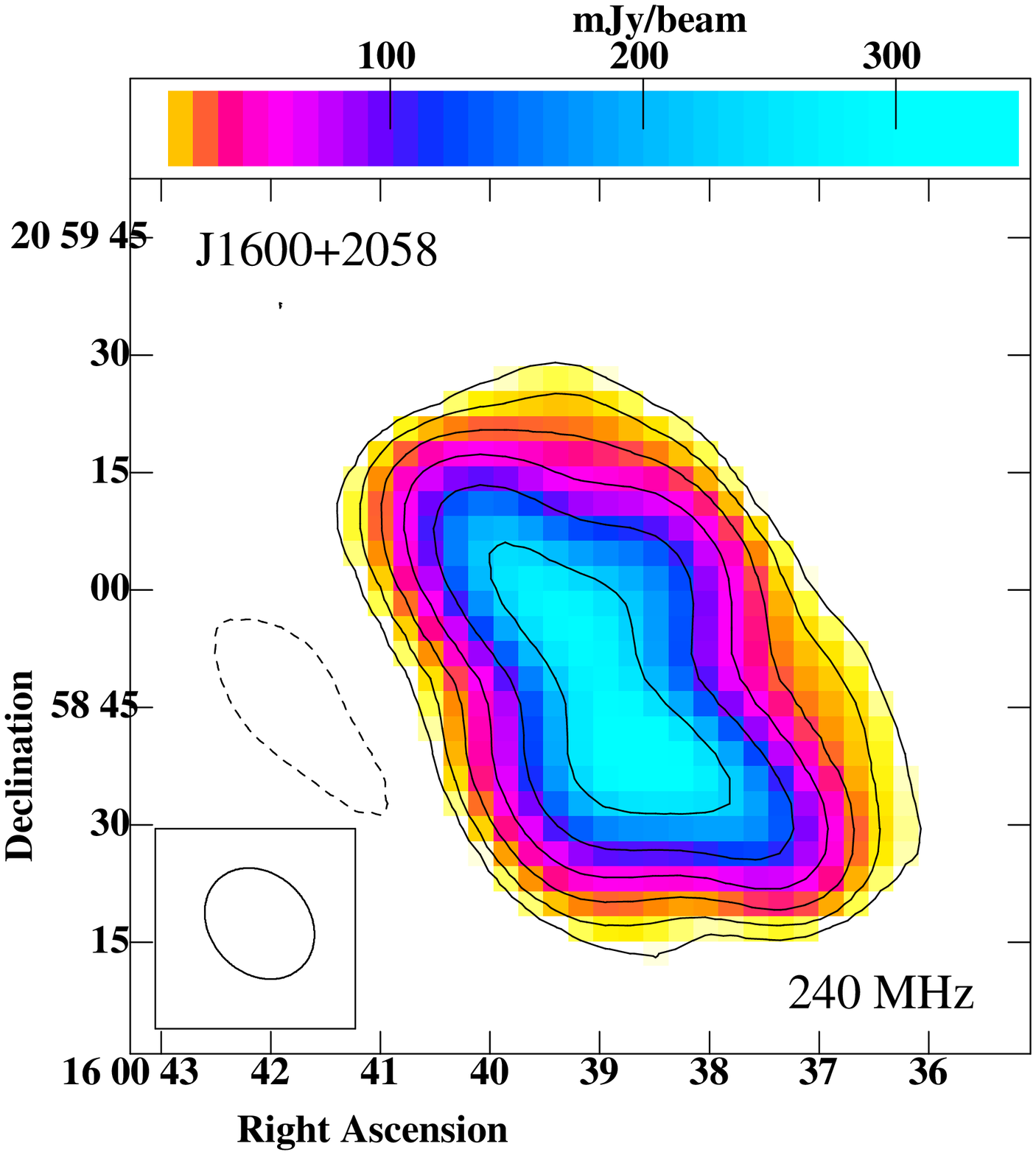} &
\includegraphics[height=4.2cm]{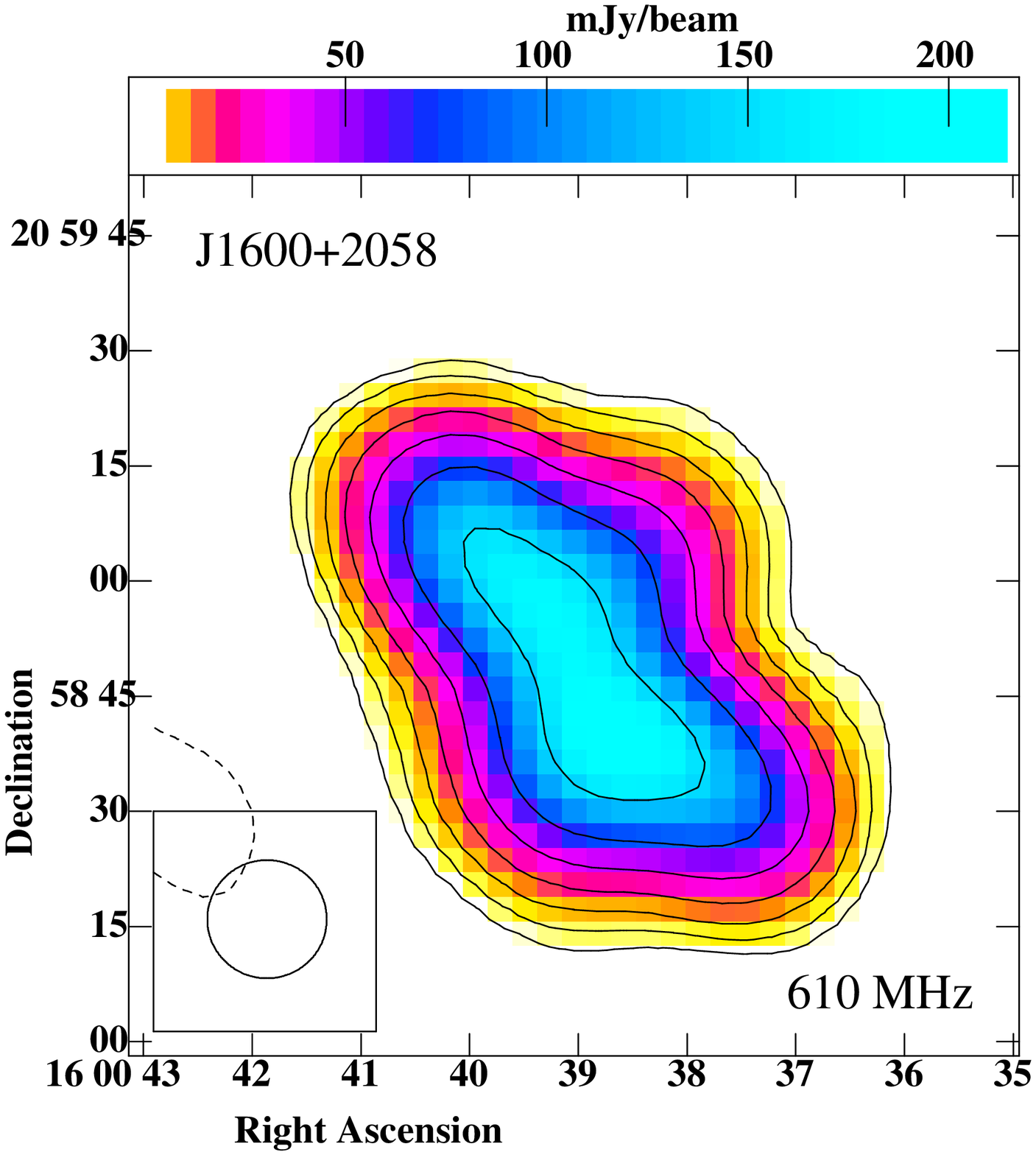} &
\includegraphics[height=4.1cm]{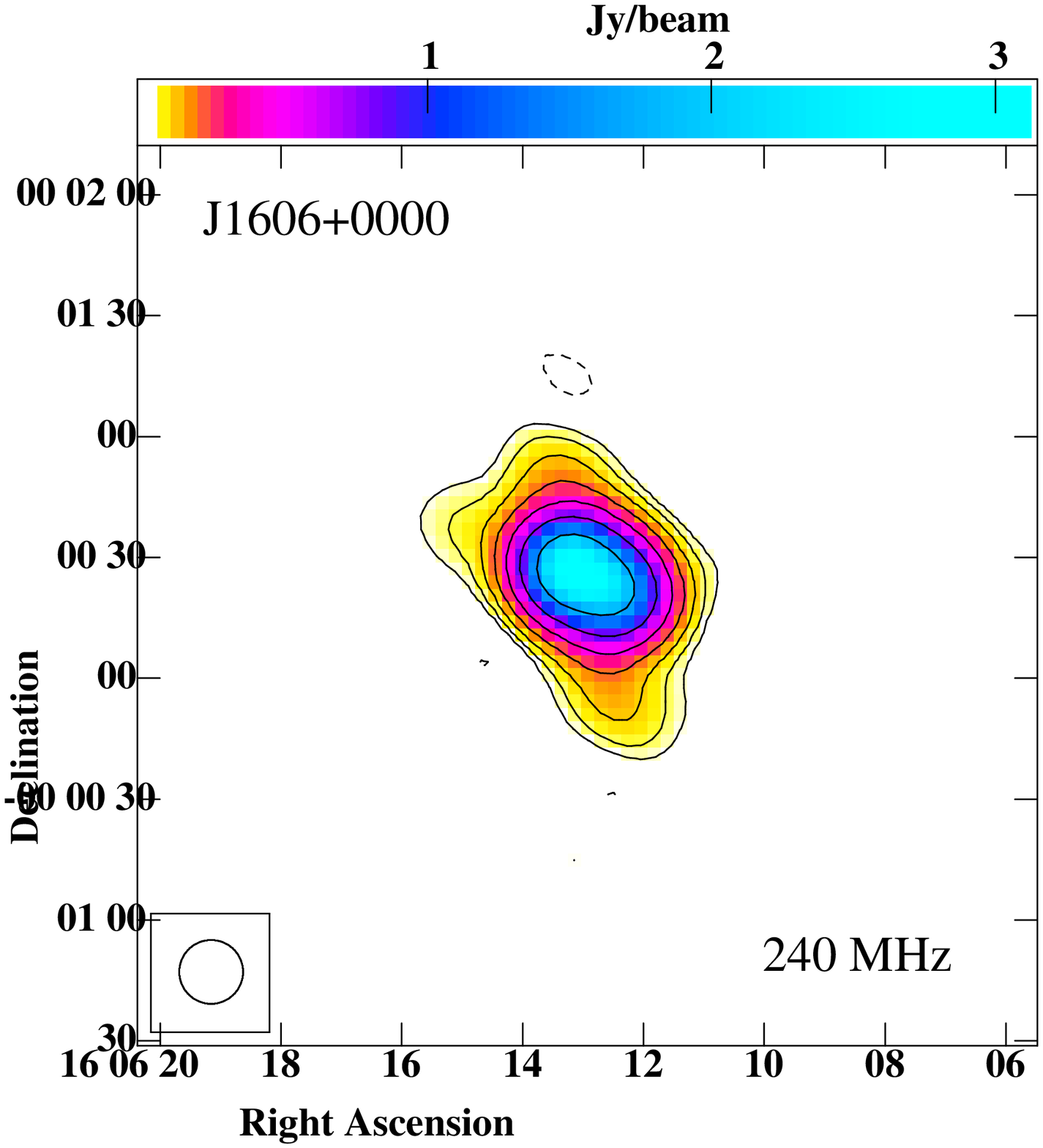} &
\includegraphics[height=4.1cm]{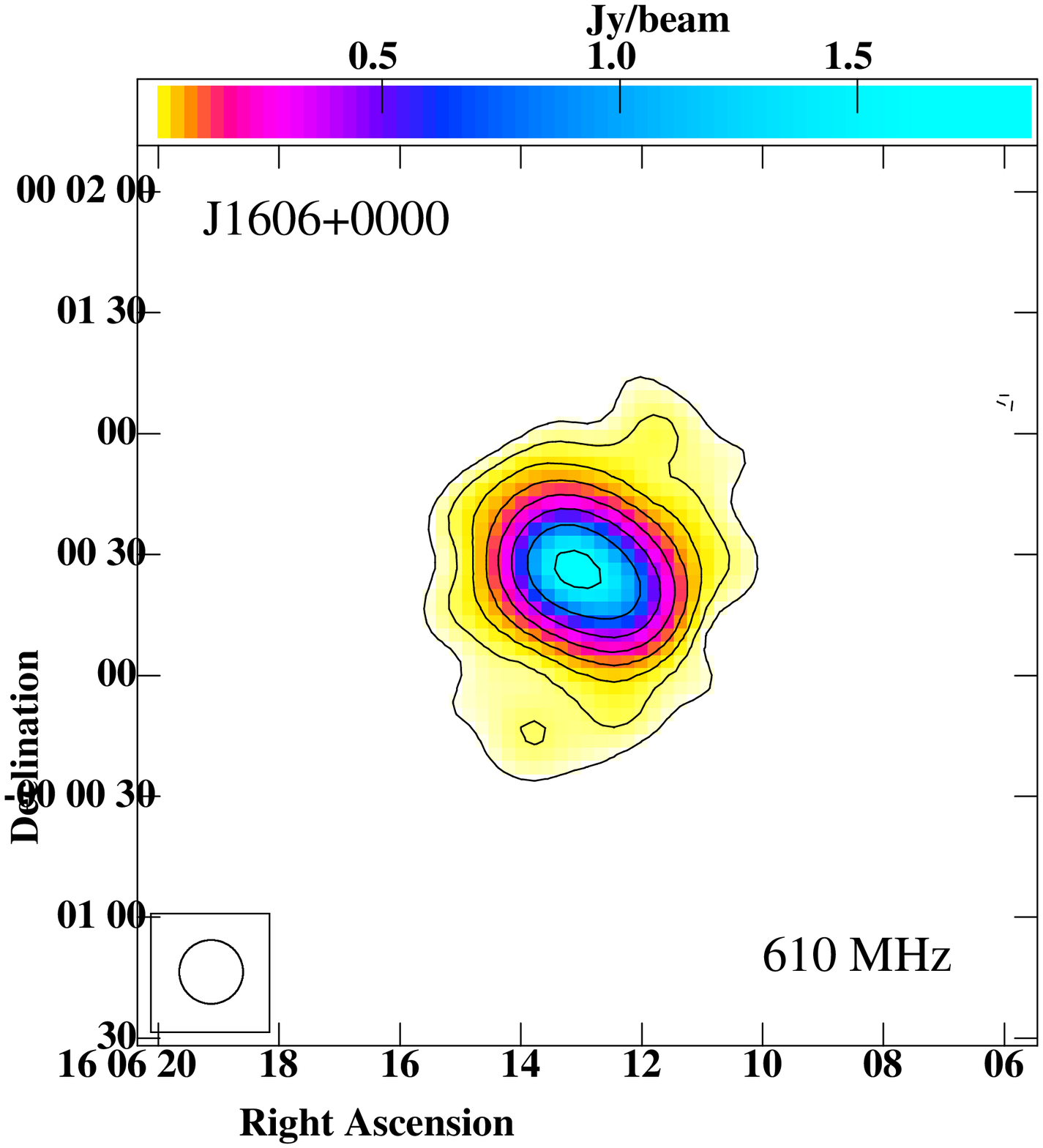}
\end{tabular}
\caption{{\it continued.}}
\end{center}
\end{figure*}

\clearpage

\end{document}